\documentclass[structabstract]{aa}
\usepackage{graphicx}
\usepackage{txfonts}
\usepackage{natbib}
\usepackage{threeparttable}

\defcitealias{Mikolaitis18}{Paper~I}
\defcitealias{Mikolaitis19}{Paper~II}
\defcitealias{Tautvaisiene20}{Paper~III}

\begin{document}

   \title{Abundances of neutron-capture elements in thin- and thick-disc stars in the solar neighbourhood
   \thanks{Based on observations collected with the 1.65~m telescope and VUES spectrograph at the Mol\.{e}tai Astronomical Observatory, Institute of Theoretical Physics and Astronomy, Vilnius University, for the SPFOT Survey.}
   }

   \author{
            G. Tautvai\v{s}ien\.{e}
          \inst{1},
             C. Viscasillas V\'azquez
          \inst{1}, 
            \v{S}. Mikolaitis
          \inst{1},
          E. Stonkut\.{e}
          \inst{1},\\
        R. Minkevi\v{c}i\={u}t\.{e}
          \inst{1},
         A. Drazdauskas
          \inst{1},
\and
            V. Bagdonas
          \inst{1}          
                 }

   \institute{Astronomical Observatory, Institute of Theoretical Physics and Astronomy, Vilnius University, Saul\.{e}tekio av. 3, 10257 Vilnius, Lithuania \\
              \email{grazina.tautvaisiene@tfai.vu.lt} 
        }

   \date{Received 24 November 2020 / Accepted 5 March 2021}

\titlerunning{Abundances of ten neutron-capture elements in thin- and thick-disc solar neighbourhood stars } 
\authorrunning{{G}. Tautvai{\v s}ien{\. e} et al.}

  \abstract
{}
{
The aim of this work is to determine abundances of neutron-capture elements for thin- and thick-disc F, G, and K stars in several selected sky fields near the north ecliptic pole and to compare the results with the Galactic chemical evolution models, to explore elemental gradients according to stellar ages, mean galactocentric distances, and maximum heights above the Galactic plane. 
}
{
The observational data were obtained with the 1.65~meter telescope at the Mol{\. e}tai Astronomical Observatory and a fibre-fed high-resolution spectrograph covering a full visible wavelength range (4000--8500~$\AA$). Elemental abundances were determined using a  differential line-by-line spectrum synthesis using the TURBOSPECTRUM code with the MARCS stellar model atmospheres and accounting for the hyperfine-structure effects. 
} 
{We determined abundances of Sr, Y, Zr, Ba, La, Ce, Pr, Nd, Sm, and Eu for 424 thin- and 82 thick-disc stars. The sample of thick-disc stars shows a clearly visible decrease in [Eu/Mg] with increasing metallicity compared to the thin-disc stars, bringing more evidence of a different chemical evolution in these two Galactic components. Abundance correlation with age slopes  for the investigated thin-disc stars are slightly negative for the majority of $s$-process dominated elements, while $r$-process dominated elements have positive correlations. 
Our sample of thin-disc stars with ages spanning from 0.1 to 9~Gyrs gives the ${\rm [Y/Mg]}=0.022\,(\pm0.015)-0.027\,(\pm 0.003)\cdot{\rm age}$~[Gyr] relation. However, for the thick-disc stars, when we also took data from other studies into account, we found that [Y/Mg] cannot serve as an age indicator. 
The radial abundance-to-iron gradients in the thin disc are negligible for the $s$-process dominated elements and become positive for the $r$-process dominated elements. The vertical gradients are negative for the light $s$-process dominated elements and become positive for the $r$-process dominated elements. In the thick disc, the radial abundance-to-iron slopes are negligible, and the vertical slopes are predominantly negative.}
% Conclusions
{}
   \keywords{Galaxy: evolution -- stars: abundances -- Galaxy: disk -- solar neighbourhood }
   \maketitle

\section{Introduction}
\label{sec:intro}

To gain a better understanding of the evolution of the Galaxy and therefore the Universe, it is necessary to know the evolution of the most basic units that comprise it: the chemical elements.  The variety of chemical elements found in the spectra of stars is a result of multiple chemical processes and specific conditions. A study of elemental abundances and their distribution in various Galactic components therefore allows us to create Galactic chemical enrichment scenarios and trace back the cosmic events that shaped the current state of the Galaxy.

Most of the observable Universe is formed from hydrogen and helium, which originated during the Big Bang, whereas the remaining elements were formed later in stars \citep[e.g.][]{Coc17}. Nuclear fusion reactions are responsible for the production of light chemical elements, while elements heavier than $A=56$ can only be formed during the capture of free neutrons \citep{Burbidge57}. 
Based on timescales of neutron capture versus $\beta^-$-decay ratio, the neutron capture processes are divided into slow ($s$-process) and rapid ($r$-process). The $s$-process takes place at low neutron fluxes (about $10^5$ to $10^{11}$ neutrons per cm$^2$ per second) when there is enough time for radioactive $\beta^-$ decay to occur before the next neutron is captured. The $r$-process occurs when the atomic nucleus captures several neutrons due to the high neutron density (about $10^{24}$ neutrons per cm$^3$) and high temperature \citep[e.g.][]{Burbidge57,Sneden08}. 

The $s$-processes can occur at various sites and are subdivided into three components: weak, main, and strong \citep[e.g.][]{Kappeler89}. The weak component prevails in massive stars ($>10\,M_{\odot}$) where neutrons are provided by the $^{22}{\rm Ne}(\alpha,{\rm n})^{25}{\rm Mg}$  reaction. This component is partly responsible for the production of the elements from iron to strontium \citep[e.g.][]{Peters68, Lamb77,Pignatari10}. Low-mass asymptotic giant branch (AGB) stars are mostly responsible for the main $s$-process component, which receives free neutrons from the $^{13}{\rm C}(\alpha,{\rm n})^{16}{\rm O}$ reaction.  This reaction contributes mostly to production of elements beyond $A=90$ \citep[e.g.][]{Busso99,Kappeler11,Bisterzo15}. The strong $s$-process component is believed to occur in low-metallicity and low initial mass  AGB stars and produces about a half of the solar Pb abundance \citep[e.g.][]{Clayton67,Gallino98,Travaglio01}.
Estimates of $s$- and $r$-process contributions to producing specific elements in various sites vary from study to study \citep[e.g.][and references therein]{Arlandini99,Travaglio04,Sneden08,Bisterzo14,Shen15,Thielemann17,Cowan19,Haynes19,Mishenina19,Siegel19}. 

It is known that the $r$-process is typical in violent conditions with a high density of free neutrons, but environments in which this process takes place are still poorly understood. Several possible sites of the $r$-process have been proposed: neutrino-induced winds from core-collapse supernovae \citep{Woosley94}, polar jets from rotating  core-collapse supernova \citep{Nishimura06}, ejecta from neutron star mergers \citep{Freiburghaus99} or from neutron star  and black hole mergers \citep{Surman08}. The neutron star merging events were revealed in the LIGO/Virgo experiment by detecting gravitational waves (\citet[e.g.][]{Abbott17}). The neutron star mergers could be a dominant source of $r$-process dominated elements \citep{Cote18}, but $s$-process dominated elements also can be produced, as shown by the recent identification of strontium in the merger of two neutron stars by \citet{Watson19}. 
The current Galactic chemical evolution (GCE) models also possess uncertainties in accounting for a contribution by the most massive AGB stars \citep[cf.][]{Karakas14}. 
\citet{Travaglio04} have invoked an additional so-called light element primary process (LEPP) in their GCE model in order to solve the deficit of Sr, Y, and Zr  as well as the Solar System abundances of $s$-only isotopes with $90<A<130$.  Recently, the LEPP mechanism was also involved in  GCE models by \citet{Bisterzo17}. However, some studies have argued that GCE models could avoid the additional LEPP contribution \citep[cf.][]{Cristallo15, Trippella16, Prantzos18, Kobayashi20}. The additional source of neutron-capture (n-capture) elements required to explain the solar abundances that led \citet{Travaglio04}  to introduce the additional LEPP process could for instance apparently be explained by the contribution from rotating massive stars \citep{Prantzos18}. We compared observational results with this new model by \citet{Prantzos18} and with the semi-empirical model by \citet{Pagel97}, which was the first attempt to model the evolution of n-capture chemical elements in the Galaxy.    

Many studies have been dedicated to n-capture element abundance determinations in the Galactic disc field stars \citep[][and references therein]{dasilva12,Mishenina13,Bensby14,Mishenina15, Mishenina19,Nissen15,Nissen16,Nissen20,Zhao16,Battistini16,Spina16,TucciMaia16,Delgado17,Delgado19,Luck17,Luck18a,Nissen17,Reddy17,Slumstrup17,Adibekyan18,Guiglion18,Magrini18, Forsberg19, Griffith19, Mishenina19,Titarenko19, Liu20} with more or fewer chemical elements investigated. The ages of stars, and even less frequently, Galactic radial and vertical locations of stars in the investigated samples are not always determined, however.      
It is very important to interpret observational results by taking into account as many stellar characteristics as possible. We used an advantage of the new possibilities that were opened by the $Gaia$ space mission \citep{Gaia16,Gaia18} in determining accurate  stellar locations in the Galaxy and computed slopes of n-capture element abundances in respect to the mean galactocentric distances and distances from the Galactic plane. We would also like to emphasise that the mean galactocentric distances of stars are much more informative than the {\it \textup{in situ}} galactocentric distances that are often used in other studies.  

 Recently, it was found that abundances of $s$-process dominated chemical elements are higher in young Galactic open clusters than was predicted by the chemical evolution models. This was raised by \citet{D'Orazi09}, who reported a significant barium overabundance in young open clusters, reaching [Ba/Fe]$>0.6$. 
The enhancement of Ba abundances in young open clusters was confirmed by later studies, but to a lesser degree (see \citealt{Reddy17} and \citealt{Spina20} for a discussion and possible influence of stellar activity), while abundances of other $s$-process dominated elements displayed negative or flat abundance trends with increasing age that varied with studies \citep[e.g.][]{Yong12,Jacobson13,Mishenina13,Mishenina15}.  
The investigations were extended to the Galactic field stars, 
especially solar twins, even though the accuracy of the age determinations is less precise \citep[e.g.][]{dasilva12,Nissen17,Feltzing17}. \citet{Battistini16} suggested that the trends of abundance ratios [El/Fe] as a function of stellar ages have different slopes before and after 8~Gyr. 
An increasing trend for solar twins through the entire age range of 10~Gyr was found by \citet{Reddy17} for La, Ce, Nd and Sm, and higher for barium. \citet{Spina18} also found a strong age-dependent trend for all investigated $s$-process dominated chemical elements in solar twins. 
A recent homogeneous study of cluster and field stars in the $Gaia$-ESO Survey by \citet{Magrini18}, encompassing stars in an age range larger than 10~Gyr, found that an increase in abundances with decreasing object age is more clearly visible for Y, Zr, and Ba, but is less prominent for La and Ce. We compared the n-capture element -- age slopes in our sample of stars with those of the solar twins \citep{Spina18} and with results by \citet{Magrini18}, as well as with slopes computed for stars investigated by \citet{Battistini16}. In order to contribute to solving the so-called barium puzzle, we compared our results with the model by \citet{Maiorca12}, which was computed with the aim to explain the high abundances of $s$-process dominated elements in young open clusters. 

In this study, we extend and complement the results of published studies with a homogeneous abundance analysis of ten neutron capture elements of diverse $s/r$-process inputs in their production (Sr, Y, Zr, Ba, La, Ce, Nd, Pr, Sm, and Eu) in a sample of 506 FGK dwarfs, subgiants, and giants in the solar neighbourhood. The aim is to investigate how their abundances depend on metallicity, stellar age, the mean galactocentric distance, and the maximum height from the Galactic plane.

The work is structured as follows. In Section~\ref{sect:2} we describe the stellar sample and analysis methods, in Section~\ref{sect:3} we discuss the elemental abundance trends relative to metallicity and compare with evolutionary models, in Section~\ref{sect:age} we present the abundance gradients with age and evaluate elemental abundance ratios such as stellar age indicators, while radial and vertical abundance gradients are presented in Section~\ref{sect:5}. We conclude this work with a summary and conclusions in Section~\ref{sect:summary}.

\section{Stellar sample and analysis} 
\label{sect:2}

\begin{figure}
        \graphicspath{ {fig1.pdf} }
        \includegraphics[width=\columnwidth]{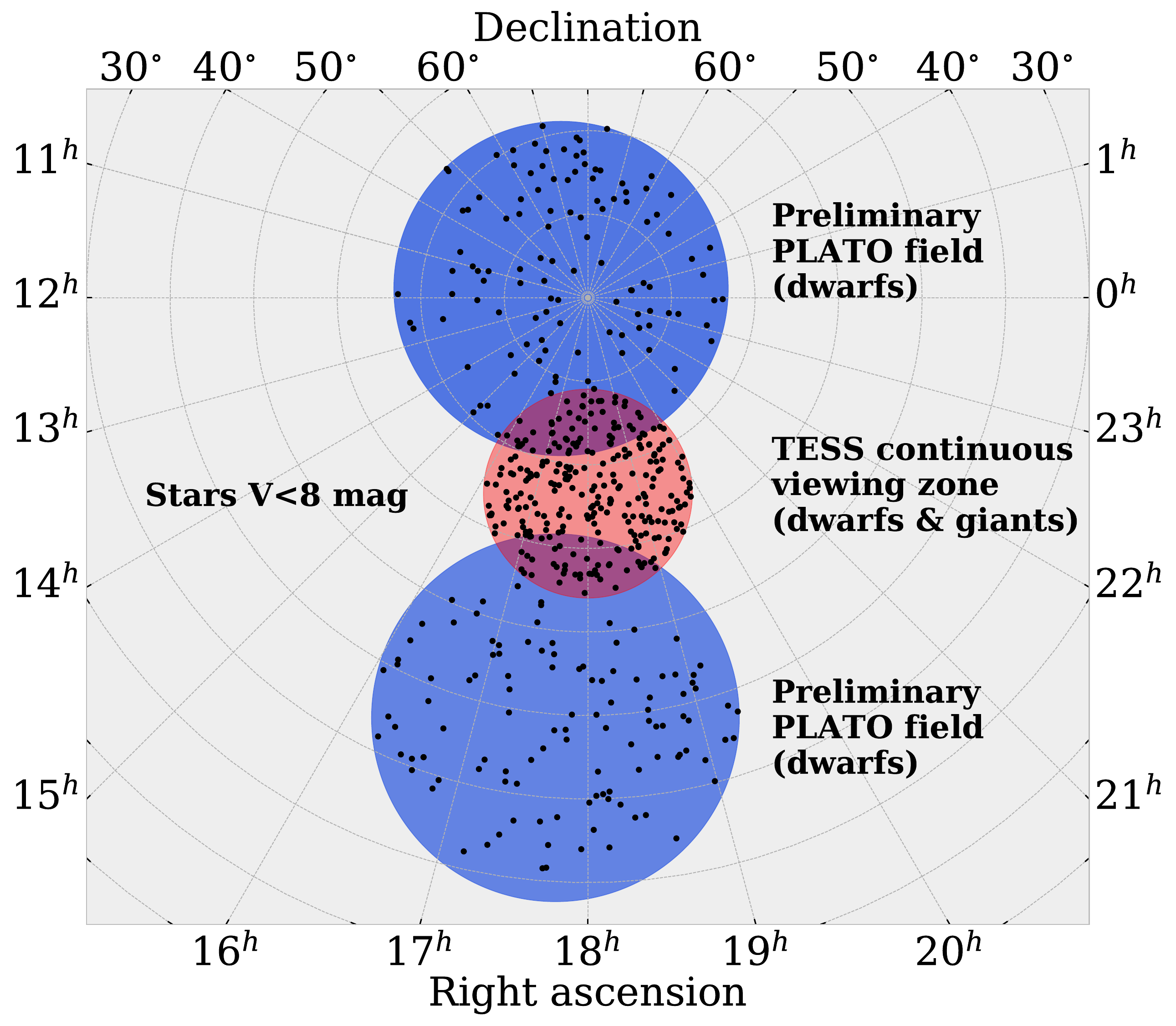}
    \caption{Positions of all selected programme stars (black dots) observed from the Mol\.{e}tai Astronomical Observatory in Lithuania. The violet areas  denote the preliminary PLATO fields and all observed dwarf stars, while the red area marks the continuous viewing zone of the TESS telescope.
    }
    \label{fig:plato and tess fields}
\end{figure}

We used archival stellar spectra of the Spectroscopic and Photometric Survey of the Northern Sky (SPFOT) from the Mol\.{e}tai Astronomical Observatory (MAO, Lithuania) observed by \citet{Mikolaitis18, Mikolaitis19} and \citet{Tautvaisiene20}
(hereinafter called \citetalias{Mikolaitis18}, \citetalias{Mikolaitis19}, and \citetalias{Tautvaisiene20}). 
In \citetalias{Mikolaitis18} and \citetalias{Mikolaitis19}, all F, G, and K spectral type dwarf and subgiant stars with  $V < 8$~mag were observed within two fields of approximately $20^{\circ}$ radii centred at $\alpha(2000) = 161^{\circ}.03552$, $\delta(2000) = 86^{\circ}.60225$ 
and at $\alpha(2000) = 265^{\circ}.08003$, $\delta(2000) = 39^{\circ}.58370$, respectively. These fields are preliminary targets of the upcoming ESA PLATO 
space mission (\citealt{2014ExA....38..249R, Miglio17}). In \citetalias{Tautvaisiene20}, all F5 and cooler stars with $V < 8$~mag were observed within a field of approximately $12^{\circ}$ radii centred at $\alpha(2000) = 270^{\circ}$, $\delta(2000) = 66^{\circ}$. This field corresponds to the continuous viewing zone of the NASA TESS space mission \citep{Ricker15,Sullivan15}. Our final list contained 247 stars in the PLATO fields (138 in the first and 109 in the second preliminary field) and 275 in the TESS field. Sixteen stars are in common in the two fields. The list contains no double-line spectroscopic binaries and no stars rotating faster than 25~km\,s$^{-1}$. The positions of all programme stars are presented in Fig.~\ref{fig:plato and tess fields}.

The observations were carried out with the Vilnius University Echelle Spectrograph (VUES) \citep{2014SPIE.9147E..7FJ, Jurgenson16} mounted on the MAO $f/12$ 1.65 meter Ritchey-Chretien telescope. The VUES is a multimode spectrograph with three spectral resolutions ($R = 36\,000, 51\,000$, and 68\,000) and covers a wavelength range from 4000 to 8800 {\AA}. 
Stars in \citetalias{Mikolaitis18} were observed using the $R=68\,000$ mode, in \citetalias{Mikolaitis19} we used the $R=36\,000$ mode, and in \citetalias{Tautvaisiene20} the $R= 68\,000$ mode for the M spectral type stars and the $R= 36\,000$ mode for other objects. 

\begin{figure}
        \graphicspath{ {fig2.pdf} }
        \includegraphics[width=\columnwidth]{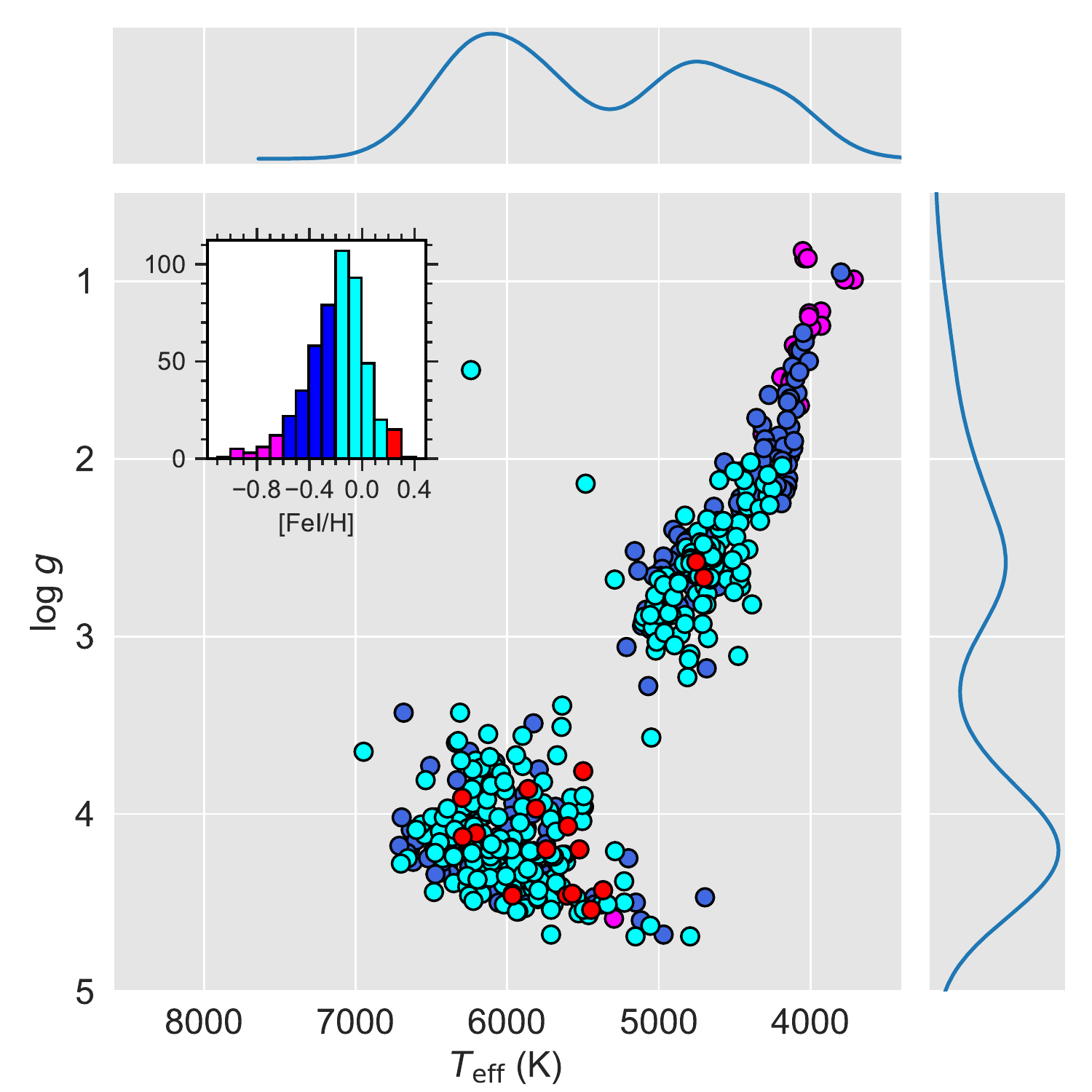}
    \caption{Investigated stars in a surface gravity and effective temperature diagram showing a density estimation  and bivariate distribution.}
    \label{fig:KD}
\end{figure}

\subsection {Stellar atmospheric parameters} 
\label{sec:kinematic}
The main atmospheric parameters (effective temperature $T_{\rm eff}$, surface gravity log\,$g$, metallicity [Fe/H], and microturbulence velocity $v_{\rm t}$) of the investigated stars were taken from \citetalias{Mikolaitis18},  \citetalias{Mikolaitis19}, and \citetalias{Tautvaisiene20}. Fig.~\ref{fig:KD} displays the investigated stars in a surface gravity and effective temperature diagram.

The stellar atmospheric parameters in these papers were determined from equivalent widths of Fe\,{\sc i} and Fe\,{\sc ii} lines using standard spectroscopic techniques. Effective temperatures were derived by minimising a slope of abundances determined from Fe\,{\sc i} lines with different excitation potentials. Surface gravities were found by requiring the Fe\,{\sc i} and Fe\,{\sc ii} lines to give the same iron abundances. The microturbulence velocity values were attributed by requiring the Fe\,{\sc i} lines to give the same iron abundances regardless of their equivalent widths. In total, 299 Fe\,{\sc i} and Fe\,{\sc ii} lines were used to compute the stellar atmospheric parameters with the tenth version of the MOOG code \citep{1973PhDT.......180S} and the MARCS grid of stellar atmosphere models \citep{Gustafsson08}.

\subsection{Line list of neutron-capture chemical elements}

\begin{table}
\centering
\footnotesize
        \caption{Solar abundances of the investigated chemical elements as determined using two different resolution modes of the VUES spectrograph and values by \citet{Asplund09}}.
        \label{tab:solar-abund}
        \begin{threeparttable}
        \begin{tabular}{llccc} 
        \hline
        \hline
El. & $\lambda$ & VUES & VUES & Asplund et al.\\ &({\AA}) & $R=68\,000$ & $R=36\,000$ & 2009\\
        \hline

        Sr\,{\sc        i}      &       4607.33 &       2.86    &       2.85    &       $2.87\pm 0.07$   \\
%       \hline                                                                          
        Y\,{\sc ii}     &       4883.69 &       1.98    &       2.05    &               \\
        Y\,{\sc ii}     &       4900.12 &       2.18    &       2.22    &               \\
        Y\,{\sc ii}     &       4982.13 &       1.96    &       2.05    &               \\
        Y\,{\sc ii}     &       5087.42 &       1.98    &       2.00    &               \\
        Y\,{\sc ii}     &       5200.42 &       2.28    &       2.34    &               \\
        Y\,{\sc ii}     &       5289.81 &       2.22    &       2.26    &               \\
        Y\,{\sc ii}     &       5402.78 &       2.27    &       2.31    &               \\
        Y\,{\sc ii}     &       5728.87 &       2.28    &       2.25    &               \\
    Y$^\ast$   &           & $2.18\pm0.13$ & $2.19\pm0.12$ & $2.21\pm 0.05$ \\
%       \hline                                                                          
        Zr\,{\sc        i}      &       5384.96 &       2.52    &       2.50    &               \\
        Zr\,{\sc        i}      &       6127.44 &       2.60    &       2.62    &               \\
        Zr\,{\sc        i}      &       6134.20 &       2.75    &       2.81    &               \\
        Zr\,{\sc        i}      &       6143.18 &       2.62    &       2.63    &               \\
%       Zr\,{\sc        i}      &       6445.75 &       atmos$^*$       &       atmos$^*$       & \\
    Zr$^\ast$              &           & $2.62\pm 0.08$ & $2.63\pm 0.10$ & $2.58\pm 0.04$ \\
%       \hline                                                                          
        Zr\,{\sc        ii}     &       5350.09 &       2.50    &       2.53    &       $2.58\pm 0.04$   \\
%       \hline                                                                          
        Ba\,{\sc        ii}     &       5853.67 &       2.12    &       2.19    &               \\
        Ba\,{\sc        ii}     &       6141.71 &       2.21    &       2.26    &               \\
        Ba\,{\sc        ii}     &       6496.91 &       2.33    &       2.37    &               \\
    Ba$^\ast$              &           & $2.22\pm 0.09$ & $2.27\pm 0.07$ & $2.18\pm 0.09$ \\
%       \hline                                                                          
        La\,{\sc        ii}     &       4748.72 &       1.00    &       1.04    &               \\
        La\,{\sc        ii}     &       5123.01 &       1.16    &       1.18    &               \\
        La\,{\sc        ii}     &       5303.53 &       0.99    &       1.02    &               \\
        La\,{\sc        ii}     &       6320.41 &       1.03    &       1.06    &               \\
        La\,{\sc        ii}     &       6390.48 &       1.15    &       1.15    &               \\
    La$^\ast$              &           & $1.07\pm 0.07$ & $1.09\pm 0.06$ & $1.10\pm 0.04$ \\
%       \hline                                                                          
        Ce\,{\sc        ii}     &       5274.22 &       1.50    &       1.53    &       \\
        Ce\,{\sc        ii}     &       5512.06 &       1.76    &       1.77    &               \\
        Ce\,{\sc        ii}     &       5975.82 &       1.32    &       1.31    &               \\
        Ce\,{\sc        ii}     &       6043.38 &       1.62    &       1.63    &               \\
    Ce$^\ast$              &           & $1.55\pm 0.16$ & $1.56\pm 0.17$ & $1.58\pm 0.04$ \\
%       \hline                                                                          
        Pr\,{\sc        ii}     &       5219.02 &       0.82    &       0.86    &               \\
        Pr\,{\sc        ii}     &       5259.72 &        0.63   &       0.60    &               \\
        Pr\,{\sc        ii}     &       5322.77 &        0.88   &       0.89    &               \\
    Pr$^\ast$              &           & $0.78\pm  0.13$ & $ 0.78\pm 0.16$ & $0.72\pm 0.04$ \\
%       \hline                                                                          
        Nd\,{\sc        ii}     &       5092.80 &       1.29    &       1.27    &       \\
        Nd\,{\sc        ii}     &       5276.86 &       1.25    &       1.26    &               \\
        Nd\,{\sc        ii}     &       5356.97 &       1.61    &       1.58    &               \\
        Nd\,{\sc        ii}     &       5740.86 &       1.45    &       1.50    &               \\
    Nd$^\ast$              &           & $1.40\pm 0.14$ & $1.40\pm 0.14$ & $1.42\pm 0.04$ \\
%       \hline                                                                          
        Sm\,{\sc        ii}     &       4854.37 &       0.87    &       0.90    &       $0.96\pm 0.04$   \\
%       \hline                                                                          
        Eu\,{\sc        ii}     &       6645.13 &       0.49    &       0.49    &       $0.52\pm 0.04$   \\
    \hline
        \end{tabular}
        
        \begin{tablenotes}
        \item[$^\ast$] The mean abundance and r.m.s.
        \end{tablenotes}
        \label{tab:lines}
        \end{threeparttable}
\end{table}

The line list selection was made based on line quality in terms of their purity and depth, considering only lines that were relatively unblended. 
The blue region of visible spectra contains many lines of n-capture elements, but this region is disadvantageous because of rather strong blending and lower sensitivity of the CCD detector. 

Our main line list was therefore confined to the range $4740-6650$~{\AA}. An exception was made just for the Sr\,{\sc i} line, which is at 4607~{\AA}. This line is relatively clean and deep, and we did not have other good candidates in the redder spectral region. In total, 35 spectral lines were selected (Table~\ref{tab:lines}). 

We used the $Gaia$-ESO Survey line-list version~5  \citep{Heiter2021}, and took the lines with flags "Yes" or "Undecided". These labels indicate that the line is "relatively unblended" or "may be useful in some stars". 

Here we also give a summary of the references for transition probabilities (oscillator strengths, log\,$gf$), and where applicable, the information for the hyperfine structure and isotopic splitting (hereafter HFS and IS).

The log\,$gf$ value for the selected line of Sr\,{\sc i} was taken from \citet{Parkinson76}, and for all Y\,{\sc ii} lines we used
from \citet{Biemont11}. The values of log\,$gf$ for all Zr\,{\sc i} lines were taken from \citet{Biemont81}, and for the Zr\,{\sc ii} line, we took the values from \citet{Cowley83}. 
For all Ba\,{\sc ii} lines the log\,$gf$ values were taken from \citet{Davidson92}. 
The strong barium lines have a background line list containing Ba\,{\sc ii} data and are from \citet{Miles69}.
 The log\,$gf$ value for the La\,{\sc ii} 6320.4\,{\AA} line was taken from \citet{Corliss62} and the values for the remaining La\,{\sc ii} lines from \citet{Lawler01La}. We took log\,$gf$ from the high-quality experimental data by \citet{Lawler09} for all Ce\,{\sc ii} lines. Experimental log\,$gf$ values for two Pr\,{\sc ii} lines at 5219 and 5322~\AA\, were taken from \citet{Li07}, and for the third line at 5259~\AA,\,no data were provided. We therefore adopted a value from \citet{Ivarsson01}. All Nd\,{\sc ii} lines were taken from \citet{DenHartog03}. The oscillator strength experimental value for the Sm\,{\sc ii} line  was taken from \citet{Lawler06} and from \citet{Lawler01Eu} for the Eu\,{\sc ii} line.
 
 The HFS effects were taken into account to investigate the lines of Ba, La, Nd, Pr, Sm, and Eu.
 The HFS and IS values were taken for
 all Ba\,{\sc ii} lines from \citet{Davidson92}; for the La\,{\sc ii} 5123, 5303, and 6390~{\AA}\, lines from \citet{Lawler01La}; for the Nd\,{\sc ii} 5092 and 5740~{\AA}\, lines from \citet{DenHartog03}; for the 5276~{\AA}\, line from \citet{Meggers75};  for all Pr\,{\sc ii} lines from \citet{Sneden2009}; for the Sm\,{\sc ii} line from \citet{Lawler06}; and for the  Eu\,{\sc ii} line from \citet{Lawler01Eu}.
 
For the remaining lines, for example, La\,{\sc ii} 4748.7 and 6320.4~{\AA}, the HFS were not provided. These lines are relatively weak, however, and give abundances that are similar to those of the three other La\,{\sc ii} lines used with HFS. We therefore assume that the HFS effect on our sample stars is small regardless of the spectral resolution. 

\subsection{Abundance determination}

The abundances of all investigated chemical elements were determined using spectral synthesis. 
To model the synthetic spectra, we used the spectrum synthesis code TURBOSPECTRUM (v12.1.1, \citealt{1998A&A...330.1109A}). 
To compute synthetic spectra, we used a set of plane-parallel, one-dimensional, hydrostatic, and constant-flux local thermodynamic equilibrium (LTE) model atmospheres taken from the MARCS\footnote{http://marcs.astro.uu.se/} stellar model atmosphere and flux library  described by \citet{Gustafsson08}. This is the same as assuming that the atmosphere is in statistical equilibrium, homogeneous in abundances, constant in time, $R_{\rm atm} << R_{\rm star} $, and that the energy transport takes place only by radiation and convection. For the barium abundances, we applied  the non-local thermodynamic equilibrium (NLTE) corrections taken from \citet{Korotin15} as the strongest  Ba\,{\sc ii} lines are affected by deviations from LTE. In our sample, for stars with [Fe/H]>0, the corrections are about $-0.027\pm 0.01$~dex, in the interval $-0.5<{\rm [Fe/H]}<0$ the corrections are about $-0.033\pm 0.016$~dex, and for [Fe/H]<$-0.5$ the corrections are about $-0.06\pm 0.02$~dex. In this work, we provide both the LTE and NLTE barium abundances.  

The analysis was carried out differentially with respect to the Sun by applying a line-by-line investigation. 
As our target stars were observed using two different resolutions (36\,000 and 68\,000), 
their spectra were investigated differentially to the solar spectra observed in the corresponding resolution mode. In order to increase precision in determining the solar elemental abundances, we averaged the abundances derived from several observations.  Table~\ref{tab:solar-abund} presents the solar abundance values for each investigated line and the corresponding values by \citet{Asplund09} for comparison.
For some stars, due to shallowness of lines, the abundances of some elements were not determined. This was encountered most frequently for the Zr\,{\sc i} and Sm\,{\sc ii} lines. 

\subsection{Mean galactocentric distances, maximum heights from the Galactic plane, and ages} \label{sec:distances and ages}

The mean galactocentric distances ($R_{\rm mean}$) and maximum heights from the Galactic plane $|z_{\rm max}|$ 
were taken from \citetalias{Mikolaitis19} and \citetalias{Tautvaisiene20}, where they were computed using input data (parallaxes, proper motions, and coordinates) from the {\it Gaia} DR2 catalogue \citep{Luri18, Katz18, Gaia16, Gaia18} and the python-based package for galactic-dynamics calculations {\it galpy} by \citet{Bovy15}. Our stellar sample is very well confined in the proximity of the Sun up to 0.14~kpc. We  therefore applied a standard distance definition (1/parallax). As recommended, the parallax shift of  $-0.029$~mas from \citet{Luri18} was adopted, although the effect of this shift on our stellar sample is very small. We did not apply any cut in parallax uncertainties either. We took them into account when we calculated the $R_{\rm mean}$ and $z_{\rm max}$ uncertainties.
We are aware that for more distant stars in the Gaia DR2, reliable distances should be delivered using a more correct inference procedure that accounts for the  nonlinearity  of the transformation and the  asymmetry of the resulting probability  distribution (cf. \citealt{Bailer-Jones2018}). Our estimates of stellar distances compared with those provided by \citet{Bailer-Jones2018} differ by just $0.0015\pm0.0048$~kpc, which is much smaller than the uncertainties on the distance estimates.

The ages of the investigated stars were taken from \citetalias{Mikolaitis18}, \citetalias{Mikolaitis19}, and \citetalias{Tautvaisiene20}, as determined with the UniDAM code by \citet{2017A&A...604A.108M}.
This code combines spectroscopic data with infrared photometry from 2MASS (\citealt{Skrutskie06}) and AllWISE (\citealt{Cutri14}) and compares them with PARSEC isochrones to derive probability density functions (PDFs) for ages.  

The maximum uncertainty allowed for age is 3~Gyr. The mean uncertainty of the age determination as calculated from the uncertainties of individual stars is 2~Gyr, the mean uncertainty of the mean galactocentric distances is 0.2~kpc, and the mean uncertainty of the maximum height from the Galactic plane is 0.05~kpc.

\subsection{Evaluation of elemental abundance uncertainties}
\label{sec:uncertainties}

 Systematic and random uncertainties should be taken into consideration. 
 The systematic uncertainties may occur due to uncertainties in atomic data, but they were mostly eliminated because of the differential analysis  relative to the Sun.  
 
The random uncertainties may appear because of a local continuum placement and a specific line fitting. A scatter that occurs in averaging abundances from multiple lines of the same element represents the random uncertainty quite well. 
The average random uncertainties ($\sigma_{\rm rand}$) for every investigated element are presented in Table~\ref{tab:uncertainties}  (the last column). 
Because we used one Sr\,{\sc i}, Zr\,{\sc ii}, and Eu\,{\sc ii} line for the abundance determination, we attributed the standard deviation value of 0.07~dex, which was the average of all standard deviations from all the stars and all elements with more than one measured line.  
    
Abundance uncertainties may also appear due to  uncertainties in atmospheric parameters. 
The uncertainties in abundances can be measured by calculating the abundance changes that are caused by the error of each individual atmospheric parameter, keeping other parameters fixed. The uncertainties of atmospheric parameters for every star are presented in \citetalias{Mikolaitis18}, \citetalias{Mikolaitis19}, and \citetalias{Tautvaisiene20}. The calculated medians of atmospheric parameter determination  uncertainties from  all stars in our sample are the following:  $\sigma T_{\rm eff}=48$~K, $\sigma {\rm log}\,g=0.30$, $\sigma {\rm [Fe/H]}=0.11$, and $\sigma v_{\rm t}=0.28$~km\,s$^{-1}$ for dwarfs (${\rm log}\,g>3.5$) and $\sigma T_{\rm eff}=57$~K, $\sigma {\rm log}\,g=0.21$, $\sigma {\rm [Fe/H]}=0.11$, and $\sigma v_{\rm t}=0.22$~km\,s$^{-1}$ for giants (${\rm log}\,g\le3.5$). 
The median dwarf star (${\rm log}\,g>3.5$) in our sample has $T_{\rm eff}=6057$~K, ${\rm log} g=4.20$, ${\rm [Fe/H]}=-0.10$, and $v_{\rm t}=1.21$~km\,s$^{-1}$ and the median giant has (${\rm log}\,g\le3.5$) $T_{\rm eff}=4646$~K, ${\rm log} g=2.50$, ${\rm [Fe/H]}=-0.14$, and $v_{\rm t}=1.49$~km\,s$^{-1}$.  The sensitivity of the abundances to the median values of uncertainties in atmospheric parameters for stars with log\,$g$ smaller or larger than 3.5 as well as for solar twins is presented in Table~\ref{tab:uncertainties}. Because most of the investigated chemical elements are ionised, the sensitivity to the uncertainties in log\,$g$ is highest. The Ba lines, which are relatively strong, are also quite sensitive to $v_{\rm t}$.

Finally, the total uncertainty budget was estimated for every elemental abundance determination by taking the uncertainties from stellar parameters as well as random uncertainties into account. 
We thus calculated $\sigma_{\rm El}$ for every given element of every given star as
\\ \\
$\sigma_{\rm El} = \sqrt{\sigma^2_{\rm atm}+\left(\frac{\sigma_{\rm rand}}{\sqrt{N}}\right)^2} $,\\ \\
where $\sigma_{\rm rand}$ is the line-to-line scatter and $N$ is the number of analysed atomic lines. If the number of lines is $N=1$, we adopted 0.07~dex for a given element as $\sigma_{\rm rand}$. The uncertainties from four stellar parameters were quadratically summed for every star and gave us $\sigma_{\rm atm}$). 
The final combined uncertainty $\sigma_{\rm El}$ is provided for every element of every star in Table~\ref{tab:results}.

\begin{table}
        \caption{Sensitivity of abundances to the median values of uncertainties in atmospheric parameters and random uncertainties.}
        \label{tab:uncertainties}
        \begin{tabular}{lccccc} 
                    \hline
                    \hline
El. & $\Delta{T_{\rm eff}}$ & $\Delta{\rm log}\,g$ & $\Delta{\rm [Fe/H]}$ & $\Delta v_{\rm t}$ & Random \\ % & Total \\
% & $\pm46$~K & $\pm0.30$ & $\pm0.11$ & $\pm0.27$ & error \\ % & \\
% & & & & km\,s$^{-1}$ & \\ % & \\
\hline
\multicolumn{6}{c}{${\rm log}\,g>3.5$}\\

        Sr\,{\sc i} & $\pm$0.05 & $\mp$0.01 & $\mp$0.01 & $\mp$0.07 & $\pm$0.07 \\ % & 0.11\\ 
                Y\,{\sc ii} & $\pm$0.01 & $\pm$0.10 & $\mp$0.02 & $\mp$0.07 & $\pm$0.06 \\ % & 0.15\\
                Zr\,{\sc i} & $\pm$0.06 & $\mp$0.05 & $\mp$0.01 & $\mp$0.01 & $\pm$0.10 \\ % & 0.13 \\
                Zr\,{\sc ii} & $\pm$0.01 & $\pm$0.11 & $\mp$0.02 & $\mp$0.01 & $\pm$0.07 \\ % & 0.13 \\
        Ba\,{\sc ii} & $\pm$0.03 & $\pm$0.04 & $\mp$0.01 & $\mp$0.18 & $\pm$0.06 \\ % & 0.19  \\
        La\,{\sc ii} & $\pm$0.02 & $\pm$0.12 & $\mp$0.02 & $\mp$0.01 & $\pm$0.10 \\ % & 0.16 \\
        Ce\,{\sc ii} & $\pm$0.02 & $\pm$0.11 & $\mp$0.02 & $\mp$0.01 & $\pm$0.11 \\ % & 0.16 \\
        Pr\,{\sc ii} & $\pm$0.02 & $\pm$0.09 & $\mp$0.02 & $\mp$0.01 & $\pm$0.08 \\ % & 0.12 \\
        Nd\,{\sc ii} & $\pm$0.01 & $\pm$0.13 & $\mp$0.03 & $\mp$0.01 & $\pm$0.08 \\ % & 0.15  \\
        Sm\,{\sc ii} & $\pm$0.03 & $\pm$0.11 & $\mp$0.03 & $\mp$0.01 & $\pm$0.07 \\ % & 0.14  \\
        Eu\,{\sc ii} & $\pm$0.01 & $\pm$0.12 & $\mp$0.02 & $\mp$0.01 & $\pm$0.07 \\ % & 0.14 \\
\hline
\multicolumn{6}{c}{${\rm log}\,g\le3.5$}\\
        Sr\,{\sc i} & $\pm$0.05 & $\mp$0.01 & $\mp$0.01 & $\mp$0.08 & $\pm$0.08 \\ % & 0.11\\ 
                Y\,{\sc ii} & $\pm$0.01 & $\pm$0.11 & $\mp$0.06 & $\mp$0.07 & $\pm$0.08 \\ % & 0.15\\
                Zr\,{\sc i} & $\pm$0.08 & $\mp$0.06 & $\mp$0.01 & $\mp$0.01 & $\pm$0.11 \\ % & 0.13 \\
                Zr\,{\sc ii} & $\pm$0.01 & $\pm$0.12 & $\mp$0.04 & $\mp$0.01 & $\pm$0.08 \\ % & 0.13 \\
        Ba\,{\sc ii} & $\pm$0.04 & $\pm$0.05 & $\mp$0.01 & $\mp$0.17 & $\pm$0.07 \\ % & 0.19  \\
        La\,{\sc ii} & $\pm$0.03 & $\pm$0.13 & $\mp$0.02 & $\mp$0.01 & $\pm$0.11 \\ % & 0.16 \\
        Ce\,{\sc ii} & $\pm$0.02 & $\pm$0.12 & $\mp$0.04 & $\mp$0.01 & $\pm$0.12 \\ % & 0.16 \\
        Pr\,{\sc ii} & $\pm$0.03 & $\pm$0.09 & $\mp$0.03 & $\mp$0.02 & $\pm$0.08 \\ % & 0.12 \\
        Nd\,{\sc ii} & $\pm$0.02 & $\pm$0.11 & $\mp$0.04 & $\mp$0.01 & $\pm$0.10 \\ % & 0.15  \\
        Sm\,{\sc ii} & $\pm$0.03 & $\pm$0.11 & $\mp$0.04 & $\mp$0.01 & $\pm$0.08 \\ % & 0.14  \\
        Eu\,{\sc ii} & $\pm$0.01 & $\pm$0.13 & $\mp$0.03 & $\mp$0.01 & $\pm$0.09 \\ % & 0.14 \\
\hline
\multicolumn{6}{c}{Solar twins}\\
        Sr\,{\sc i}     &       $\pm$0.04       &       $\mp$0.01       &       $\mp$0.01       &       $\mp$0.07       &       $\pm$0.08       \\
  Y\,{\sc ii}   &       $\pm$0.01       &       $\pm$0.02       &       $\mp$0.02       &       $\mp$0.04       &       $\pm$0.04       \\
  Zr\,{\sc i}   &       $\pm$0.07       &       $\mp$0.06       &       $\mp$0.01       &       $\mp$0.01       &       $\pm$0.11       \\
  Zr\,{\sc ii}  &       $\pm$0.01       &       $\pm$0.10       &       $\mp$0.04       &       $\mp$0.01       &       $\pm$0.07       \\
        Ba\,{\sc ii}    &       $\pm$0.01       &       $\pm$0.05       &       $\mp$0.01       &       $\mp$0.04       &       $\pm$0.04       \\
        La\,{\sc ii}    &       $\pm$0.03       &       $\pm$0.09       &       $\mp$0.02       &       $\mp$0.01       &       $\pm$0.09       \\
        Ce\,{\sc ii}    &       $\pm$0.02       &       $\pm$0.09       &       $\mp$0.04       &       $\mp$0.01       &       $\pm$0.12       \\
        Pr\,{\sc ii}    &       $\pm$0.02       &       $\pm$0.07       &       $\mp$0.03       &       $\mp$0.02       &       $\pm$0.07       \\
        Nd\,{\sc ii}    &       $\pm$0.01       &       $\pm$0.08       &       $\mp$0.04       &       $\mp$0.01       &       $\pm$0.07       \\
        Sm\,{\sc ii}    &       $\pm$0.01       &       $\pm$0.04       &       $\mp$0.01       &       $\mp$0.01       &       $\pm$0.05       \\
        Eu\,{\sc ii}    &       $\pm$0.01       &       $\pm$0.09       &       $\mp$0.03       &       $\mp$0.01       &       $\pm$0.06       \\
        \hline        
        \end{tabular}
        
\end{table}

\begin{figure}
        \graphicspath{ {fig3.pdf} }
        \includegraphics[width=\columnwidth]{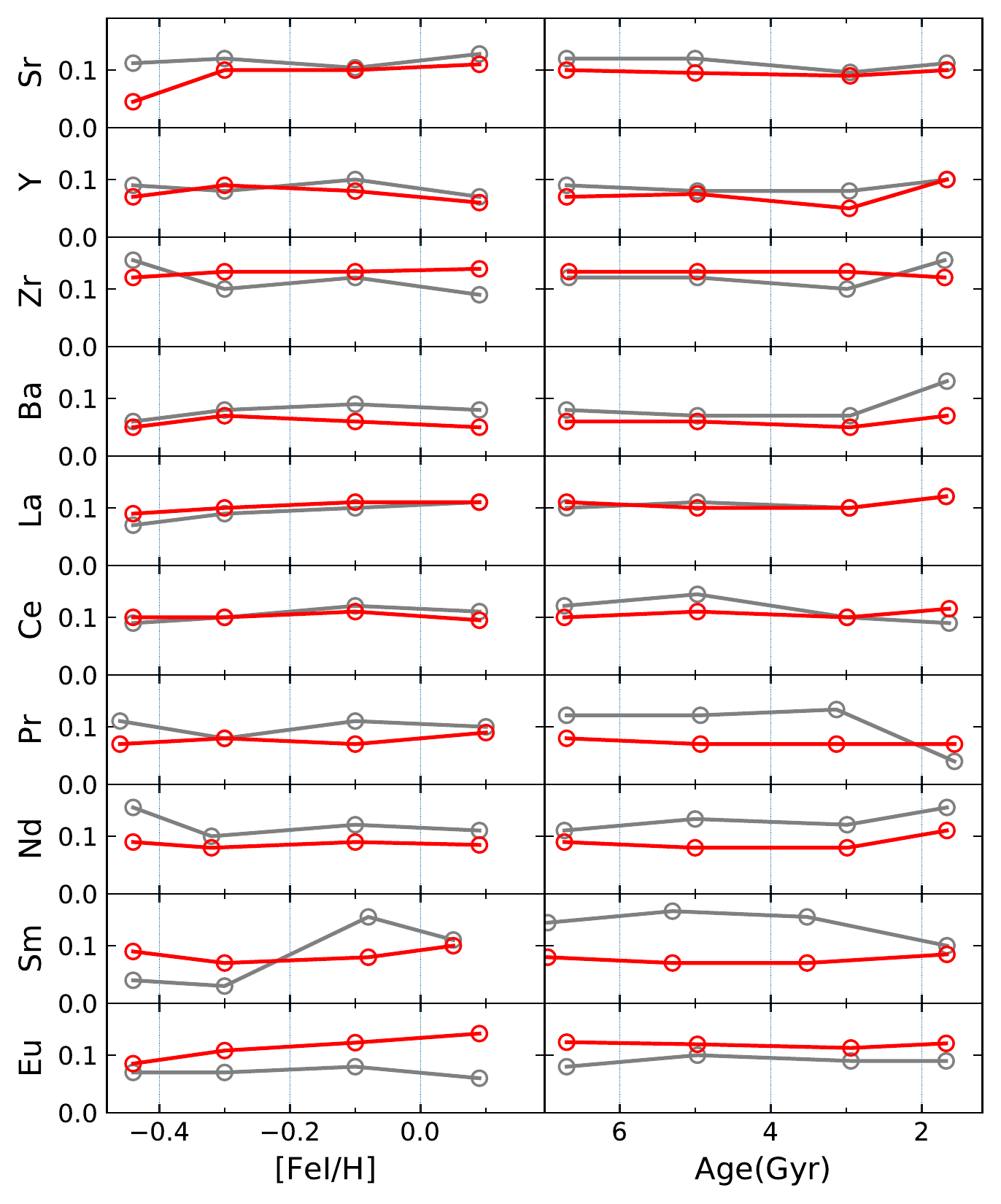}
    \caption{Observed rms scatter around the median [El/Fe] abundance (grey lines) and expected uncertainties (red lines) in four metallicity and age bins (thin vertical lines). The open circles at every bin are placed at the median value of the [Fe/H] or age of the corresponding bin.}
    \label{fig:err-scat}
\end{figure}

\begin{figure*}
    \graphicspath{ {fig4.pdf} }
        \includegraphics[width=\textwidth]{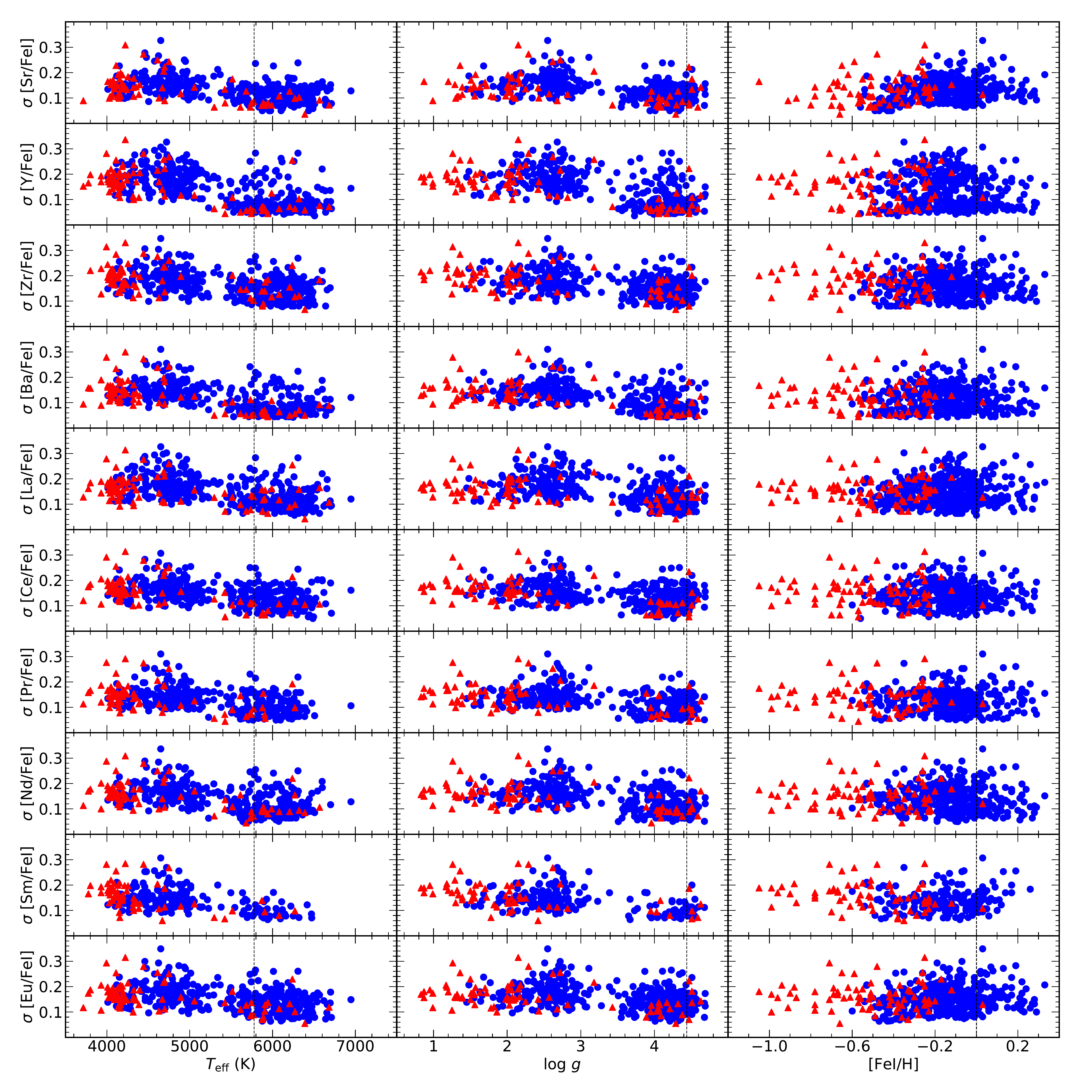}
            \caption{Uncertainties on the abundance ratios as a function of $T_{\rm eff}$, log\,$g$, and [Fe/H]. The blue dots represent the thin-disc and the red triangles represent the thick-disc stars. See Sect.~\ref{sec:uncertainties} for details of how the uncertainties have been evaluated.}
    \label{fig:uncertainties}
\end{figure*}

In Fig.~\ref{fig:err-scat} we show a comparison between the expected uncertainties (red line) and the measured dispersion (black line) in metallicity and age bins, where the bin limits are indicated by thin dashed vertical lines. The bin sizes were selected in order to have enough points (at leat five) for the statistical analysis. Because there are fewer the thick-disc stars, the comparison was made for the thin-disc component alone. If the thin disc is well mixed, homogeneous, and well defined, we expect that the scatter of the results in every bin is comparable to the expected uncertainties.  In Fig.~\ref{fig:uncertainties} we show the individual  uncertainties ($\sigma_{\rm El}$) as a function of $T_{\rm eff}$, log\,$g$, and [Fe/H]. Stars of the thin and thick discs (their attribution to the discs is described in the next section) are indicated by different symbols.

%%%%%%%%%%%%%%%%%%
\section{Elemental abundance trends relative to metallicity}
%\footnote{We refer to [Fe\,{\sc i}/H] when talking about metallicity.}}
\label{sect:3}
%%%%%%%%%%%%%%%%%%

\begin{table*}
        \caption{Atmospheric parameters and abundances of neutron-capture elements.}      
        \label{tab:results}
             %\begin{threeparttable}
        \begin{tabular}{lccrrccrccccl} % seven columns, alignment for each
                    \hline
                    \hline
Star & $T_{\rm eff}$ & ${\rm log}\,g$ & [Fe\,{\sc i}/H] & [Sr\,{\sc i}/H] & $\sigma$Sr & ... & [Eu\,{\sc ii}/H] & $\sigma$Eu & Age & $R_{\rm mean}$ & $|z_{\rm max}|$ & Disc \\
  &  K &   &   &  & & & &  & Gyr & kpc & kpc & \\
       \hline
    TYC 1547-367-1 & 6221 & 3.94 & $-0.19$ &  $-0.20$ & 0.13 & ... & $-0.14$ & 0.16 & 3.78 & 8.78 & 0.07 & thin \\
    TYC 1563-3551-1 & 5423 & 4.51 & $-0.38$ & $-0.62$ & 0.12 & ... & $-0.16$ & 0.12 & 5.39 & 7.58 & 0.10 & thick \\
    TYC 1563-3552-1 & 6199 & 3.83 & $-0.17$ & $-0.20$ & 0.09 & ... & $-0.11$ & 0.11 & 3.85 & 8.07 & 0.15 & thin\\
    TYC 2057-709-1 & 5505 & 4.04 & $ 0.00$ & $ 0.06$ & 0.09 & ... & $ 0.09$ & 0.07 & 7.58 & 7.70 & 0.40 & thin \\
    TYC 2070-1061-1 & 5973 & 4.20 & $-0.01$ & $-0.01$ & 0.12 & ... & $ 0.02$ & 0.15 & 4.92 & 7.91 & 0.68 & thin \\
\hline
        \end{tabular}
                 %\begin{tablenotes} 
                 \tablefoot{Only a portion of this table is shown here for guidance regarding its form and content. A machine-readable version of the full table is available with the online version of this paper. The atmospheric parameters, age, and kinematic data are from \citetalias{Mikolaitis18}, \citetalias{Mikolaitis19}, and \citetalias{Tautvaisiene20}.}
                %\end{tablenotes}
                 %\end{threeparttable}
\end{table*}

 \begin{table}
 \centering
        \caption{Solar percentage of $s$-process input for the n-capture elements.}
        \label{tab:percentages}
        \begin{tabular}{lcccc} % FIVE columns, alignment for each
                \hline
                \hline
                El. & Arlandini  & Sneden &  Bisterzo & Prantzos\\
                & et al.& et al.& et al. & et al. \\
                &  1999 & 2008 & 2014 & 2020\\

                \hline
                Sr & 85 & 89 & 69  & 91 \\ 
                Y & 92 & 72 & 72  & 78 \\
                Zr & 83 & 81 & 66 & 82\\
                Ba & 81 & 85 & 85 & 89\\ 
                La & 62 & 75 & 76  & 80\\
                Ce & 77 & 81 & 84 & 85\\
                Pr & 49 & 49 & 50 & 54\\
                Nd & 56 & 58 & 58 & 62\\
                Sm & 29 & 33 & 31 & 33\\
                Eu & 6 &  2  & 6 & 5\\
        \hline
        \end{tabular}
        \\
\end{table}

We determined the abundances of Sr (453 stars), Y (506 stars), Zr\,{\sc i} (307 stars), Zr\,{\sc ii} (476 stars), Ba (504 stars), La (504 stars), Ce (467 stars), Pr (402 stars), Nd (466 stars), Sm (257 stars), and Eu (488 stars). 
The results are presented in Table~\ref{tab:results} (a full version is available online). The stars in our sample span an approximate range of $T_{\rm eff}$ between 3800 and 6900~K, log\,$g$ between 0.9 and 4.7, [Fe/H] between $-1.0$ and 0.5~dex, ages between 0.1 and 10~Gyr, $R_{\rm mean}$ between 5.5 and 11.8~kpc, and $|z_{\rm max}|$ between 0.03 and 3.73~kpc. These parameters are also presented in Table~\ref{tab:results} for convenience. 
As follows from \citetalias{Mikolaitis18}, \citetalias{Mikolaitis19}, and \citetalias{Tautvaisiene20}, there are 424 stars of the Galactic thin disc and 82 of the thick disc in our sample. The [Mg\,{\sc i}/Fe\,{\sc i}] ratio was adopted to define the thin-disc ($\alpha$-poor) and thick-disc ($\alpha$-rich) stars. The last column in Table~\ref{tab:results} contains information about which of the Galactic discs a star is attributed to. 

We display the elemental abundances versus metallicities\footnote{We refer to [Fe\,{\sc i}/H] when talking about metallicity.} in Fig.~\ref{fig:Abundances}.
For the analysis of the Zr abundances, we used both neutral and ionised lines. The abundance of zirconium was calculated from Zr\,{\sc i} lines in 307 stars and from Zr\,{\sc ii} lines in 476 stars. In the plots, only the [Zr\,{\sc ii}/Fe\,{\sc i}] abundance distribution is depicted, while the [Zr\,{\sc i}/H] abundances can be found in Table~\ref{tab:results}. 

Our results coincide well with the previous studies within the uncertainties.
\citet{Delgado17} investigated equivalent widths of Sr, Y, Zr, Ba, Nd, and Eu lines in a large sample of about 1000 FGK dwarfs using spectra of $R\sim 115\,000$. \citet{Mishenina13} investigated 276 FGK dwarfs and determined abundances of Y, Zr, Ba, La, Ce, Nd, Sm, and Eu (the abundances of Ba and Eu were determined from spectral synthesis). \citet{Forsberg19} investigated abundances of Zr, La, and Eu in 196 thin- and 70 thick-disc stars using spectral synthesis. In these studies, the stars were divided into thin- and thick-disc populations, which is essential when GCE models are evaluated, but no ages and kinematic parameters were determined. \citet{Bensby14} and \citet{Battistini16} analysed the n-capture element abundance and age relations in dwarf stars of the solar neighbourhood (using spectral synthesis, the Y and Ba abundances were determined in the first study, and the Sr, Zr, La, Ce, Nd, Sm, and Eu abundances in the second study).    
In the study by \citet{Magrini18}, the abundances of Y, Zr, Ba, La, and Ce were homogenised from results obtained with several different methods used in the $Gaia$-ESO Survey, and their time evolution was investigated.

 A remark is due about the strontium abundances.  The [Sr/Fe] abundances in the \citet{Battistini16} sample of stars are systematically lower at the solar metallicity by about $-0.2$~dex. 
 \citet{Battistini16} and \citet{Delgado17} used the same single neutral line of strontium (4607.33\,{\AA}) as we did. \citet{Delgado17} employed the equivalent width method, while \citet{Battistini16} and we applied spectral synthesis. As the line possibly does not suffer from the HFS splitting \citep[cf.][]{Gratton94}, the method that is used should not play an important role in this case. 
See \citet{Mishenina19} for a wider discussion of Sr abundance studies. 

Furthermore, we discuss the abundances of n-capture chemical elements divided into groups depending on their origins as well as abundance gradients as a function of stellar age, mean galactocentric radii, and maximum heights from the Galactic plane. We compare the results with model predictions and previous studies if available.

\subsection{Sr, Y, and Zr}

Sr, Y and, Zr belong to the first $s$-process peak and are called the light $s$-process (ls) dominated chemical elements.  While the $s$-process is the key contributor to their abundances, their detailed production is still debated. As displayed in Table~\ref{tab:percentages}, studies sometimes provide different contributions of $s$-process to production of these elements \citep{Arlandini99, Sneden08, Bisterzo14, Prantzos20}. Moreover, these elements are produced not only by the main but also by the weak $s$-processes and by the $r$-process. 

\begin{figure*}
    \graphicspath{ {fig5.pdf} }
        \includegraphics[width=\textwidth]{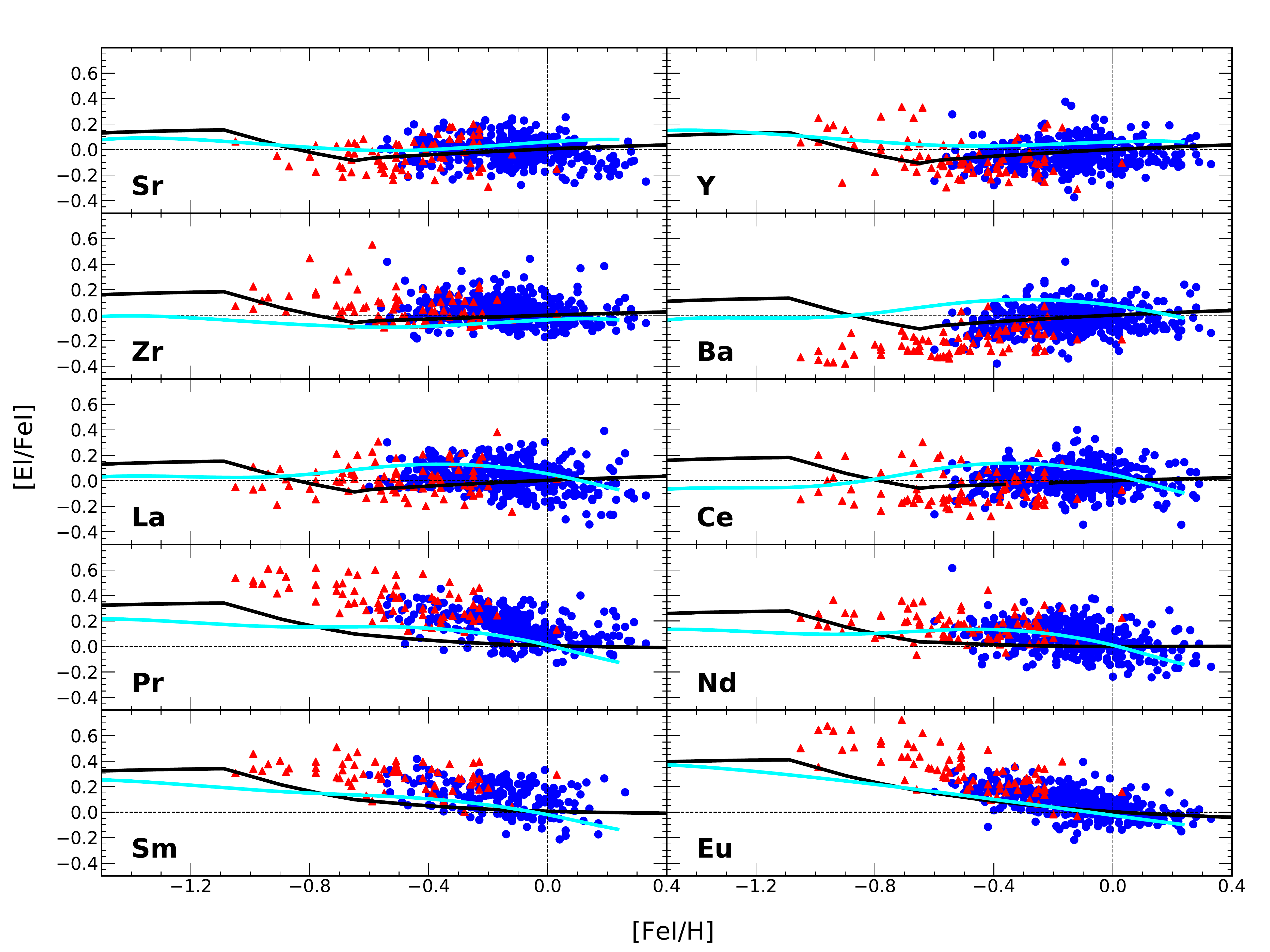}
            \caption{Elemental abundance trends relative to [Fe\,{\sc i}/H]. The blue dots represent the thin-disc stars, and red triangles indicate the thick-disc stars. The continuous lines show the models by \citet{Prantzos18} (cyan) and \citet{Pagel97} (black).}
    \label{fig:Abundances}
\end{figure*}

According to \citet{Bisterzo14}, in addition to the $s$-process contributions, there are pure $r$-process contributions of 12, 8, and 15\%\ to Sr, Y, and Zr, respectively. Hence, the origin of about 8\%\ of Sr and by 18\%\ of Y and Zr is still unclear. As proposed by \citet{Travaglio04}, the missing LEPP process might be present in low-metallicity massive stars. However, this hypothesis was challenged for instance by  \citet{Cristallo15}, \citet{Trippella16}, \citet{Prantzos18}, and \citet{Kobayashi20}. In Fig.~\ref{fig:Abundances} we compare our observational results to the model by \citet{Prantzos18}. In this grid of models, the authors used metallicity-dependent isotopic yields from low- and intermediate-mass stars and yields from massive stars that include the combined effect of metallicity, mass loss, and rotation for a large grid of stellar masses and for all stages of stellar evolution. The yields of massive stars are weighted by a metallicity-dependent function of the rotational velocities, constrained by observations as to obtain a primary-like $^{14}{\rm N}$ behaviour at low metallicity and to avoid overproduction of $s$-elements at intermediate metallicities.  The semi-empirical model by \citet{Pagel97}, the first attempt to model the evolution of the n-capture elements in the Galaxy, is also displayed to show the advancement of modern models.  

 As shown in Fig.~\ref{fig:Abundances}, the [Sr, Y, Zr/Fe] values follow the thin-disc models by \citet{Prantzos18} and \citet{Pagel97} quite well. Regarding the abundance values in the thick-disc stars, in our sample of stars they are quite close to those in the thin-disc stars and do not provide strong evidence to support the finding by \citet{Delgado17} that in metal-deficient thick-disc stars the zirconium abundances are much higher than in the thin-disc stars. A similar pattern to ours was reported in \citet{Mishenina13}.  This question should be explored further.

\subsection{Ba, La, and Ce}

The production for barium, lanthanum, and cerium, the so-called heavy $s$-process (hs) elements belonging to the second $s$-process peak, are clearly dominated by the $s$-process (see Table~\ref{tab:percentages}). 
As indicated in \citet{Kappeler11}, all isotopes beyond  $A=90$, except for Pb, are mainly produced by the $s$-process main component. The remaining element yields are produced via the $r$-process component. Thus, a similar production pattern for all three elements should be reflected in their abundances. 

 Fig.~\ref{fig:Abundances} shows that the chemical evolution of Ba, La, and Ce is indeed very similar.  
 In the \citet{Prantzos18} baseline model, the LIMS contribution to Ba, La, and Ce abundances  reaches its maximum at around ${\rm [Fe/H]}=-0.4$ and then starts to drop, creating a downward trend at sub-solar metallicities.  Fig.~\ref{fig:Abundances} shows that a convex shape of the model should have its maximum at [Fe/H] of about $-0.2$~dex,  and should start accounting for the LIMS input at higher metallicities ($\sim -0.7$~dex), as was modelled by \citet{Pagel97}. This is also very clearly seen in the Ba and Ce abundance trends.  The model by  \citet{Prantzos18} reflects elemental abundances in the stars of super-solar metallicities quite well.   

The [Ba, La, Ce/Fe] ratios in the thick-disc stars are slightly lower than in the thin-disc sample stars of the neighbouring metallicity. They still agree with the  thin-disc model by \citet{Pagel97}, however. 
They also indicate that the time-delay of the main $s$-process contribution should be longer in the model by \citet{Prantzos18}.

\subsection{Pr and Nd}

Praseodymium and neodymium are attributed to the second $s$-process peak as well. They have a more mixed origin than Ba, La, and Ce, however, because about half of their abundances comes from the $s$-process main component and the other half from the $r$-process (see Table~\ref{tab:percentages}). 

For praseodymium, our results agree well with the \citet{Prantzos18} model from slightly sub-solar to super-solar metallicities. [Pr/Fe] are more enhanced at metallicities lower than $-0.2$~dex, however. The lower LIMS effect on Pr compared to Ba, La, or Ce is compensated for by the $r$-process. The model by \citet{Prantzos18} therefore has no expressed peak at around $-0.4$~dex. However, our observational data for Pr indicate that the $r$-process contribution to the praseodymium production may be higher. 
The model underestimation of Pr abundances at lower metallicity stars of our sample could indicate that the $r$-process effect was underestimated both by \citet{Prantzos18} and \citet{Pagel97}. The trend of [Pr/Fe] versus metallicity is quite similar to the [Sm/Fe] and [Eu/Fe] trends. 

The model by \citet{Prantzos18} fits to the distribution of [Nd/Fe] ideally. 
Differently from [Pr/Fe], at sub-solar metallicities [Nd/Fe] abundances show flatter behaviour. This corresponds to the high percentage of $s$-process input in Nd than in Pr. 

The thick-disc stars have higher ratios of Pr and Nd to Fe than in the case of more $s$-process dominated elements, which qualitatively agree well with the GCE predictions for the thin disc. However, in the metal-deficient thick-disc stars, the abundance difference between the $s$- and $r$-process dominated chemical elements is stronger.

\subsection{Sm and Eu}

Samarium and europium are two $r$-process dominated elements we investigated. About 70\%\  of Sm is made by this process, and more than 90\%\  of Eu. The $s$-process main component is responsible for the remaining abundances of these elements. 

Samarium abundances were determined for fewer stars than in the case of other chemical elements.  Nevertheless, the abundances form an increasing trend towards lower metallicities, as expected for the element with a high $r$-process input. Compared with the \citet{Mishenina13} sample, some stars have higher [Sm/Fe] ratios. Compared with the sample from \citet{Battistini16}, however, the overall patterns of the samarium abundance results agree well. The [Eu/Fe] abundance results for the stars in our sample show a complete agreement with other studies discussed in this work.  Recently, \citet{Forsberg19} debated that europium abundances might be super-solar at the solar metallicity. They investigated abundances of Zr, La, Ce, and Eu in a sample of 291 giants in the Galactic thin and thick discs. Their Fig.~5 shows that the abundances of zirconium, which were determined from the neutral lines, agree with the results of other studies, while the abundances of La, Ce, and Eu, which were determined from the ionised lines, are systematically higher. This may be caused by systematic uncertainties in determining a surface gravity of these stars. The authors used the stellar atmospheric parameters from \citet{Jonsson17}, who reported that the  surface gravities are overestimated while comparing the results with the $Gaia$ stellar parameter benchmark values. If this is the case, the abundances of the ionised elements are overestimated as well. 
  
The thin- and thick-disc stars show higher abundances, especially for europium, compared with the $s$-process dominated elements. This agrees with the GCE theory. 

If all the stars in our sample can be correctly attributed to the thin and thick discs, which can hardly be achieved for all studies, the chemical evolution models both by \citet{Prantzos18} and \citet{Pagel97}, which agree with each other quite well, are slightly too low. 
Further developments in modelling the $r$-process yields and time delays in production of these elements are continuing. New production sites are about to be revealed (see e.g. \citealt{Cote19},  \citealt{Schonrich19}). 
The additional origin in Eu production could be related to neutron star mergers and other sources, as discussed in the introduction. \citet{Matteucci14}, for instance, found that the models with only neutron star mergers producing Eu or mixed with type~II supernovae can reproduce the observed Eu abundances in the Galactic disc. However, \citet{Cote19} emphasised 
that the neutron star mergers cannot reproduce the observational constraints at all metallicities on their own. They proposed to consider an input of so-called magnetorotational or other rare types of supernovae acting in the early universe, but fading away with increasing metallicity.

\subsection{Comparison of the light and heavy s-process element abundances}

\begin{figure}
        \graphicspath{ {fig6.pdf} }
        \includegraphics[width=\columnwidth]{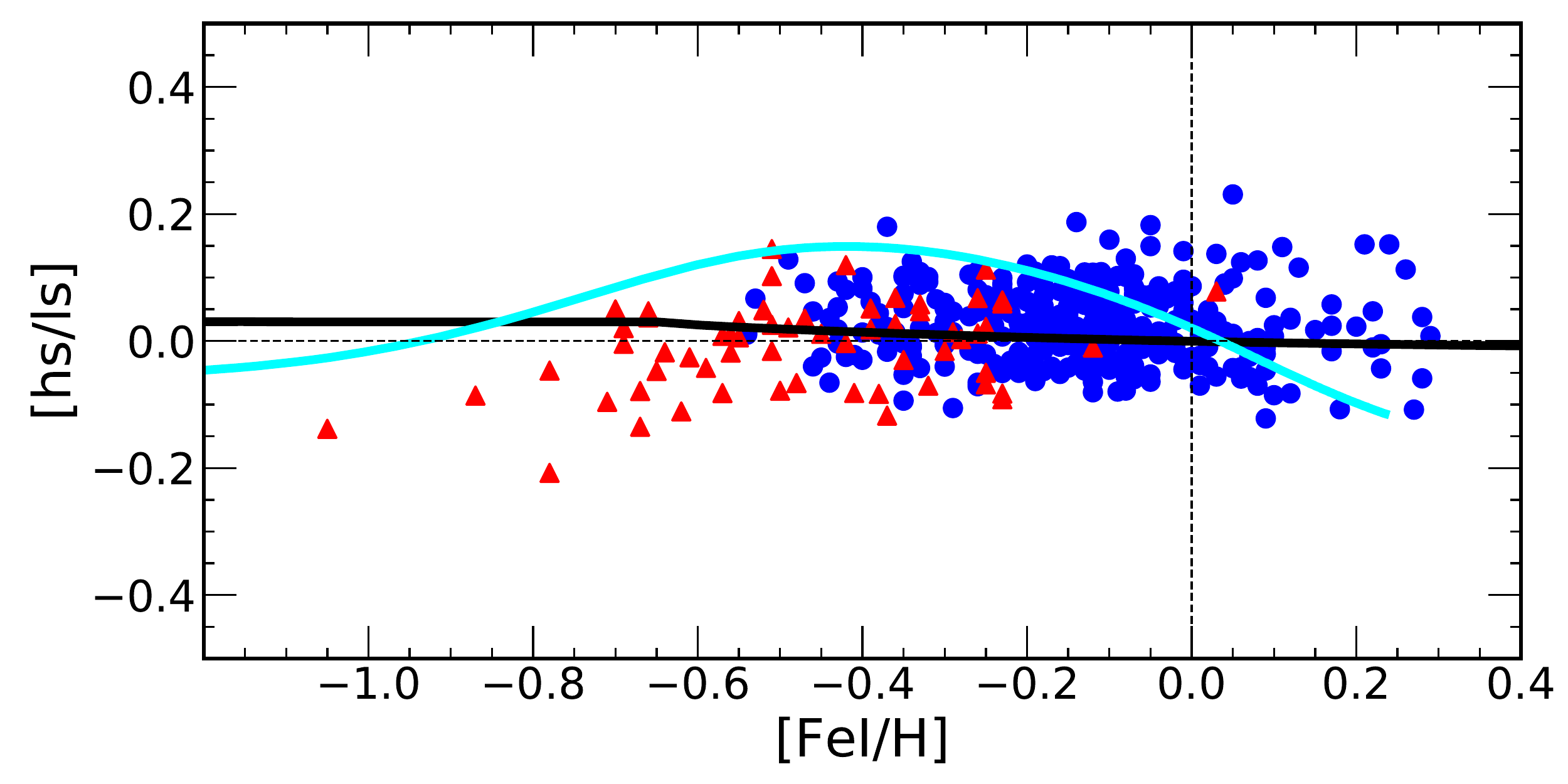}
    \caption{Ratio of heavy (Ba, La, Ce, Nd) and light (Sr, Y, Zr) $s$-process dominated element abundances with respect to [Fe\,{\sc i}/H] compared with the GCE model predictions by \citet{Prantzos18} (cyan line) and the thin-disc model by \citet{Pagel97} (black line). Our results are marked as blue circles for the thin-disc and as red triangles for the thick-disc stars.}
    \label{fig:hs_ls_feh}
\end{figure}

The reliability of $s$-process yields in the GCE models can be verified by comparing the abundances of 
heavy and light $s$-process dominated elements. It is common to monitor the $s$-process efficiency through
the relative abundances of the $s$-elements with neutron magic numbers $N = 50$ at the Zr peak with respect to those with the neutron magic numbers $N = 82$ at the Ba peak. These nuclei are mainly synthesised by the $s$-process and act
as bottlenecks for the $s$-process path because of their low neutron
capture cross-sections. 

\citet{Prantzos18} compared in their Fig.~17 the computed [hs/ls] versus [Fe/H] evolution model with
observations for the unevolved thin- and thick-disc stars by \citet{Delgado17}.  The [hs/ls] ratios were computed   using the average abundance
ratios of Ba, Ce, and Nd for the [hs/Fe] ratios, while for the [ls/Fe] ratios,
the averages of Sr, Y, and Zr were taken. From the comparison, the authors concluded that the yields used in their baseline GCE model are reliable and reproduce observational trends of unevolved stars. However, they did not take into account that measurements of the specific elements considered in the observational
average [hs/Fe] and [ls/Fe] abundances were not present for every star due to the quality of the observed spectra. 
If the authors had taken only the stars from the \citet{Delgado17} sample with the abundances of all the six chemical elements determined, it would be visible that the agreement is poor. This is also seen when the model is compared with our sample stars (see Fig.~\ref{fig:hs_ls_feh}). As the heavy $s$-element, we also included lanthanum. The abundances of this chemical element were absent in \citet{Delgado17}. When the plots with and without La are compared, the systematic difference in the [hs/ls] ratios is just 0.003~dex.

The model agreement with the observations appears to have been better if the convex shape of the model had its maximum not at the [Fe/H] of $-0.4$~dex, but at about $-0.2$~dex,  and would start accounting for the LIMS input at higher metallicities (about $-0.7$~dex).
The GCE model by \citet{Prantzos18} still has to be further developed by investigating the roles of the yields from rotating massive stars and low- and intermediate-mass stars. This is also visible from the comparisons of the model in the [El/Fe] versus age plots that we discussed in Subsect.~\ref{sec:distances and ages}.  

\subsection{Comparison of the s- and r-process dominated element abundances}

\begin{figure}
    \graphicspath{fig7.pdf} 
        \includegraphics[width=\columnwidth]{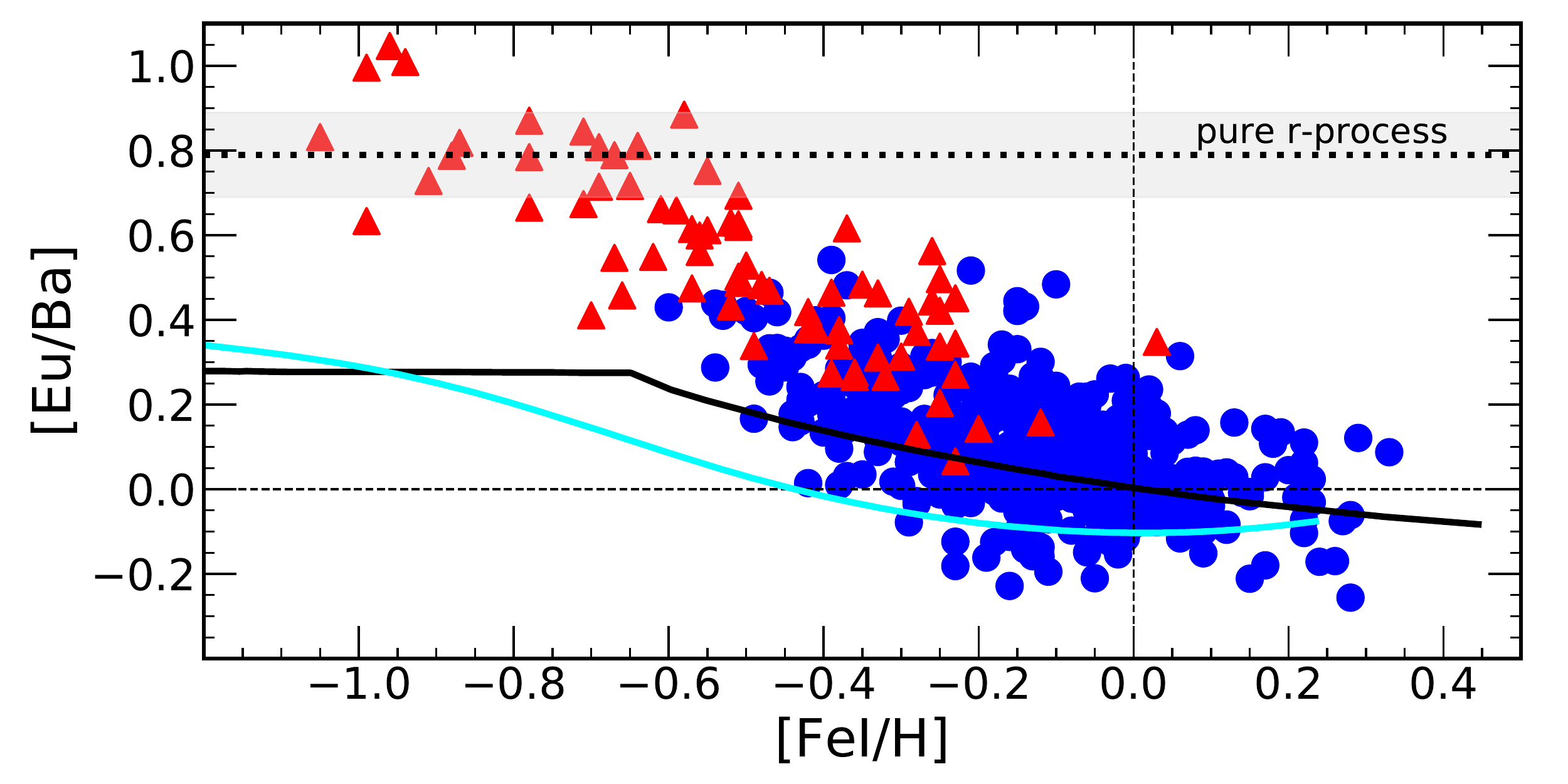}
            \caption{[Eu/Ba] ratio with respect to [Fe\,{\sc i}/H]. Symbols are the same as in Fig.~\ref{fig:hs_ls_feh}. The dotted line and the shadowed area of $\pm 0.1$~dex uncertainty represent a pure $r$-process ratio derived using the percentages of \citet{Bisterzo14} and the solar abundances of \citet{Grevesse07}.}
    \label{fig:euba}
\end{figure}

\begin{figure}
    \graphicspath{fig8.pdf} 
        \includegraphics[width=\columnwidth]{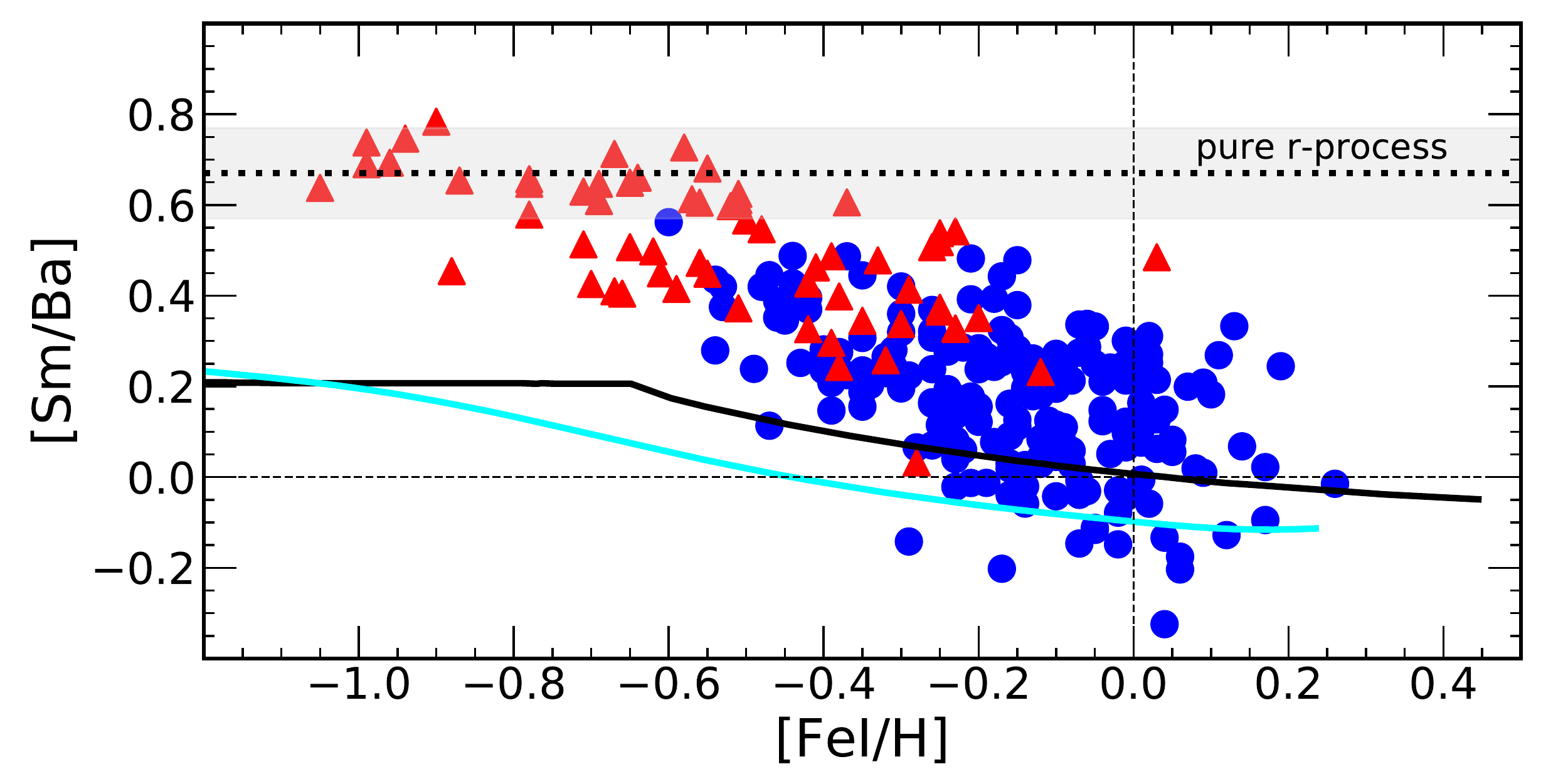}
            \caption{[Sm/Ba] ratio with respect to [Fe\,{\sc i}/H]. Symbols are the same as in Fig.~\ref{fig:euba}. }
    \label{fig:smba}
\end{figure}

The abundance patterns of $s$- and $r$-process dominated elements allow us to show how different star formation histories are in the Galactic discs. In Fig.~\ref{fig:euba} we compare abundance ratios of $r$-process dominated element europium and $s$-process dominated element barium [Eu/Ba] versus [Fe\,{\sc i}/H]  with the GCE model predictions for the thin disc by \citet{Prantzos18} and by \citet{Pagel97}, as well as with a pure $r$-process ratio derived using the percentages
of \citet{Bisterzo14} and the solar abundances of \citet{Grevesse07}. The GCE models of the thin disc clearly have to be further developed in order to account for the higher europium production and for other $r$-process dominated chemical elements such as samarium (Fig.~\ref{fig:smba}). 

For the evolution of the thick disc, \citet{Guiglion18} have recently raised the idea that the abundance ratios of [Eu/Ba] might be constant with metallicity. Other $r$-process dominated elements investigated in their study, gadolinium  and dysprosium, showed an abundance to barium ratios that decreased with increasing metallicity. The authors concluded that their observations clearly indicate a different nucleosynthesis history in the thick disc between Eu and Gd–Dy. In our study, we clearly see the decline of [Eu/Ba] ratio as a function of metallicity as in other studies (e.g. \citealt{Battistini16,Delgado17,Magrini18}). 

The [Eu/Ba] and [Sm/Ba] ratios are close to the pure $r$-process line for metal-poor thick-disc 
stars, meaning that the $r$-process was the only n-capture
process active at the beginning of the formation of the thick disc. 
 The difference between samarium-to-barium abundance ratios in the thin and thick discs is not as prominent as in the case of europium. This agrees with the relatively lower production of samarium in the $r$-process sites.

A comment could also be made about the [Eu/Ba] ratios in the super-solar metallicity stars. \citet{Delgado17} reported  that the [Eu/Mg] ratios in the thin-disc stars start to increase with increasing metallicity. This might be caused by the underestimation of barium abundances using the equivalent width method applied in this study or other methodological reasons. Another study by \citet{Trevisan14}, which was dedicated especially to the analysis of metal-rich stars, shows a quite flat but slightly decreasing trend until [Eu/Ba] of about $-0.2$~dex at [Fe/H] of about +0.5~dex. In our study, the stars span about [Fe/H] of +0.3~dex and show the same trend as in Trevisan et al. and other studies (e.g. \citealt{Battistini16, Magrini18}).

\subsection{[Eu/Mg] ratio}
\label{sec:EuMg}

\begin{figure}
    \graphicspath{fig9.pdf} 
        \includegraphics[width=\columnwidth]{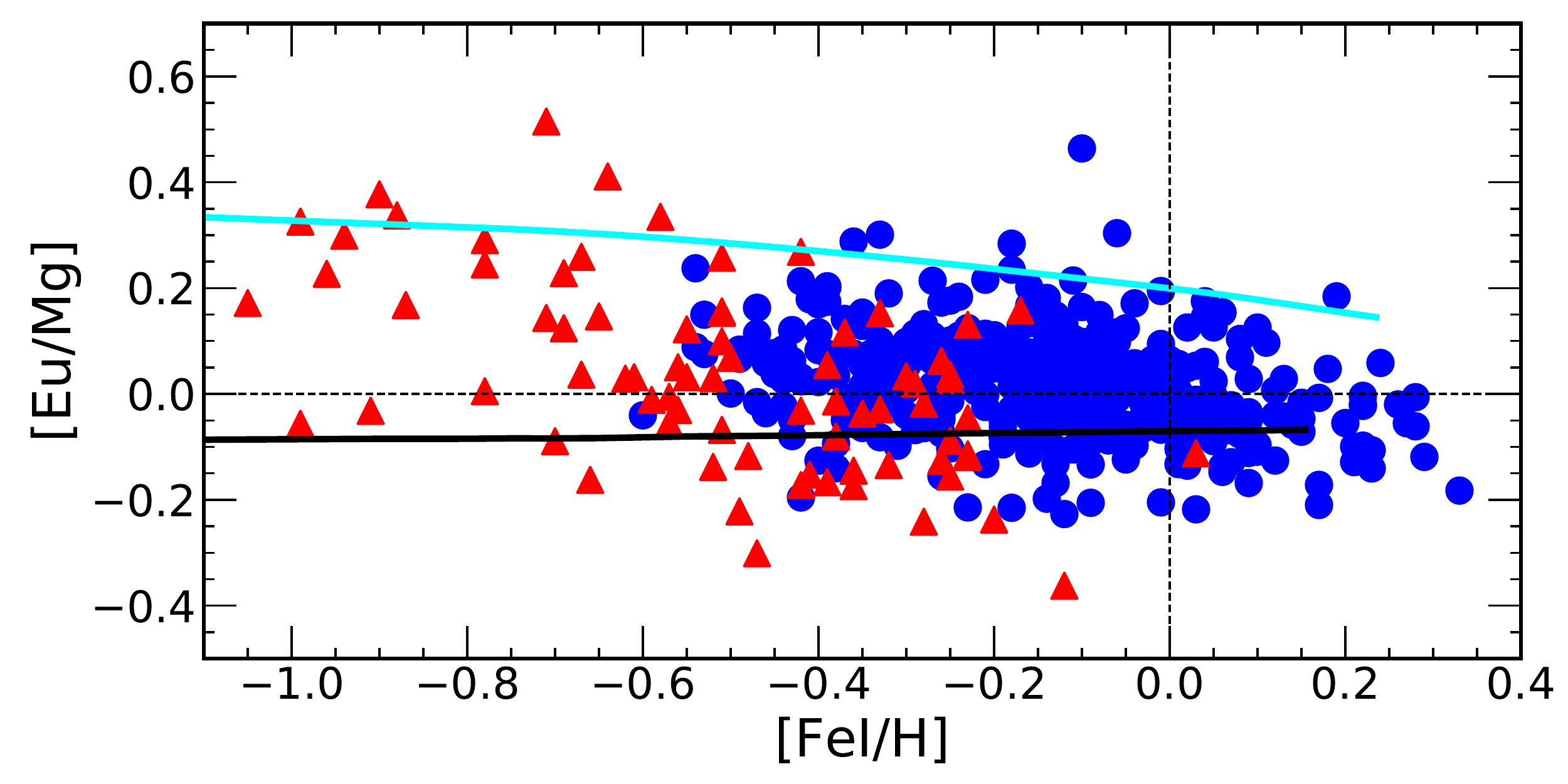}
            \caption{[Eu/Mg] ratio with respect to [Fe\,{\sc i}/H]. Symbols are the same as in Fig.~\ref{fig:hs_ls_feh}. }
    \label{fig:eumg}
\end{figure}

The study of magnesium and europium abundances in different galactic components is essential for determining the formation and evolution of the Galaxy.
\citet{Mashonkina_2001} studied 63 cool stars with metallicities ranging from $-2.20$~dex to 0.25~dex. In their Fig.~8, the authors pointed to a tendency of [Eu/Mg] ratios to decrease with increasing metallicity both for the thin- and thick-disc stars. They also found an overabundance of Eu relative to Mg in three halo stars. 
However, in 2003 they reanalysed the sample including 15 additional moderately metal-deficient stars \citep{Mashonkina_2003}, and an overabundance of Eu relative to Mg was determined only for two thick-disc and halo stars (see Fig.~5 in their paper). 
Furthermore, \citet{Delgado17} also found an almost flat [Eu/Mg] (see their Fig.~15), suggesting that these two elements receive an important contribution from supernovae progenitors of similar masses. However, \citet{Guiglion18} recently addressed the topic in the AMBRE project for a large sample of FGK Milky Way stars in a range of metallicities very similar to ours. They reported a decreasing [$r/\alpha$] trend for increasing metallicity, from which one of the main conclusions in their work was derived.

The europium-to-magnesium abundance ratios versus metallicity in our sample of stars are presented in 
Fig.~\ref{fig:eumg}. The magnesium abundances were taken from  \citetalias{Mikolaitis19} and  \citetalias{Tautvaisiene20}. We found that the [Eu/Mg] ratio in the investigated metallicity range tends to slightly decrease with increasing metallicity for the thin-disc stars. This is even more evident for the thick-disc stars. The abundances of Eu are higher when they are compared to the semi-empirical model by \citet{Pagel97}, which foresees the production of Eu by type~II supernova alone. The model by \citet{Prantzos18} extends far too high (mainly because of the Mg modelling uncertainties, see their Fig.~13).  

We have mentioned the negative trend of [$r/\alpha$] with increasing metallicity for the thin-disc stars found by \citet{Guiglion18} in the AMBRE project. Based on 25 thick-disc stars in the AMBRE sample, however, it was difficult to make a confident conclusion about the thick disc. Our sample of 76 thick-disc stars shows an even stronger decrease of [Eu/Mg], as well as [$r/\alpha$], with increasing metallicity than in the thin-disc stars. This indicates a different chemical evolution of these two Galactic components.  

\section{Abundance gradients with age}
\label{sect:age}

\begin{figure*}
    \graphicspath{ {fig10.pdf} }
        \includegraphics[width=\textwidth]{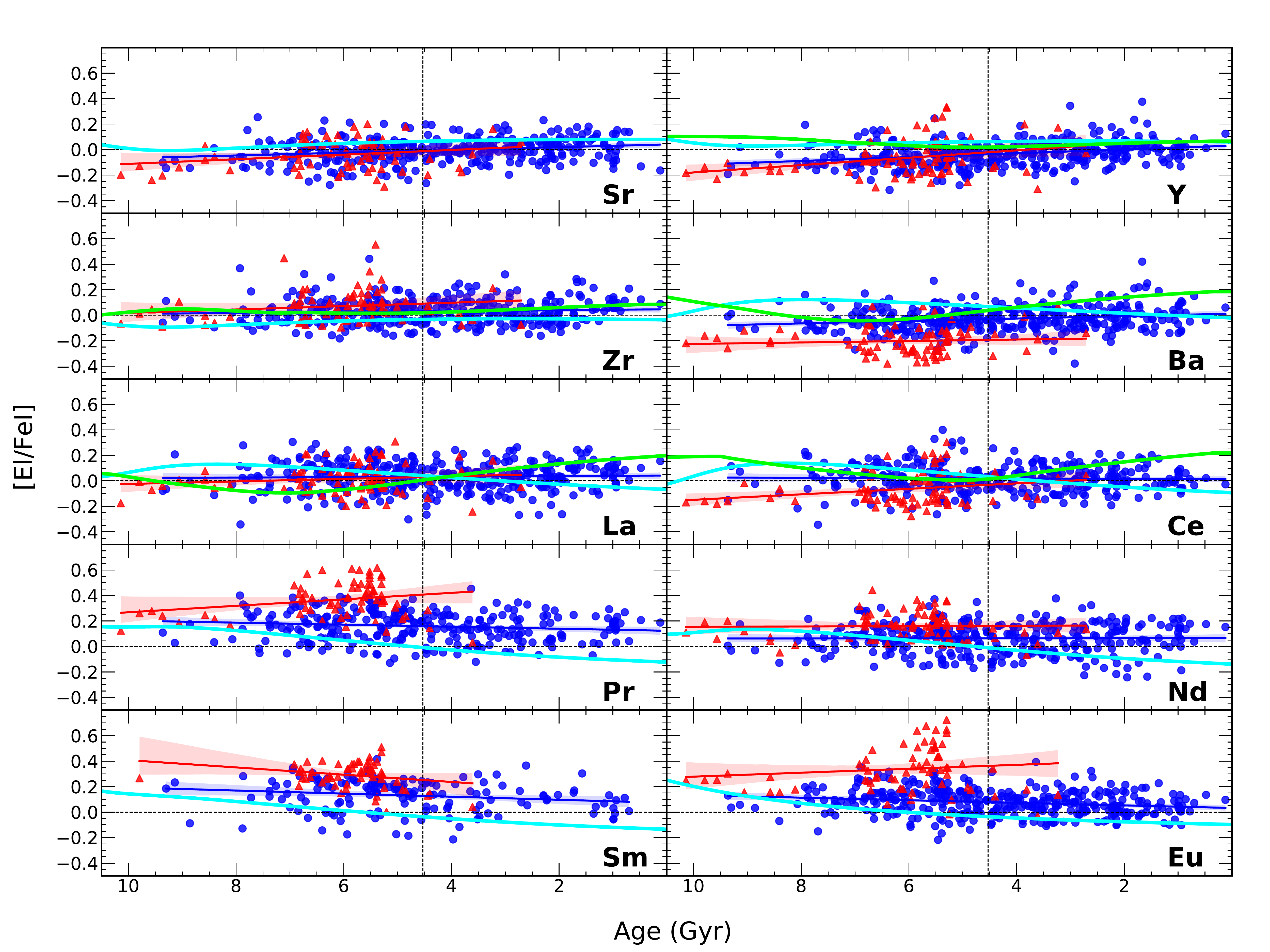}
        \caption{Elemental-to-iron abundance ratios as a function of age. The blue dots and red triangles represent the thin- and thick-disc stars of our study, respectively. The green lines represent the models by \citet{Maiorca12} and the cyan lines those by \citet{Prantzos18}. The blue lines are the linear fits for the thin-disc stars of this work and the red lines show the thick-disc stars, with a 95\% confidence interval for the  ordinary least-squares regressions. The vertical dashed lines indicate the adopted age of the Sun (4.53~Gyr).}
    \label{fig:abundance_age}
\end{figure*}

Fig.~\ref{fig:abundance_age} shows distributions of element-to-iron ratios as a function of stellar ages. The figure also includes the best linear fits to the [El/Fe\,{\sc i}] versus age and theoretical models by \citet{Prantzos18} and \citet{Maiorca12}.    

The determined best linear fits to the [El/Fe\,{\sc i}] and age relations are presented in Table~\ref{tab:slopes}. We provide the slopes, y-intercepts, Pearson correlation coefficients (PCC), and standard errors of the linear fits.

\begin{table*}
   \centering
        \caption{Best linear fits to the [El/Fe\,{\sc i}]$-$age distributions in dex\,Gyr$^{-1}$ and the Pearson correlation coefficients for the thin disc stars and comparison with other works.}
             \begin{threeparttable}
    \begin{tabular}{llrrrllc}
    \hline
    \hline
    El. & \multicolumn{2}{c} {Present work}  & \multicolumn{2}{c} {Battistini, Bensby}  &  \multicolumn{1}{c}  {Spina et al. 2018} & \multicolumn{2}{c} {Magrini et al. 2018}  \\
        & \multicolumn{2}{c}{thin disc}  & \multicolumn{2}{c}{thin disc} &   \multicolumn{1}{c}{solar twins} &  \multicolumn{2}{c}{age < 8~Gyr}      \\
         & \multicolumn{1}{c}{Slope} &\multicolumn{1}{c}{PCC}  & \multicolumn{1}{c}{Slope$^{\ast}$} &\multicolumn{1}{c}{PCC$^{\ast}$} & \multicolumn{1}{c}{Slope}  & \multicolumn{1}{c}{Slope$^{\ast\ast}$} & \multicolumn{1}{c}{PCC$^{\ast\ast}$}  \\
     \hline
    Sr & $-0.011\pm 0.003$  & $-0.22$ & $-0.023\pm0.009$ & $-0.40 $ & $-0.032\pm 0.002$  & &  \\
    Y  & $-0.015\pm0.002$  & $-0.30$ & $-0.011\pm0.002$ & $-0.35 $ & $-0.029\pm 0.002$  & $ -0.023\pm0.009$ & $-0.6$  \\
    Zr & $-0.004\pm0.003$  & $-0.07$ & $-0.003\pm0.006$ & $+0.11 $ & $-0.026\pm 0.002$  & $-0.038\pm 0.013$ & $-0.5$  \\
    Ba & $-0.010\pm0.003$  & $-0.19$ & $-0.023\pm0.003$ & $-0.43 $ & $-0.032\pm 0.002$ & $-0.027\pm 0.007 $ & $-0.8$  \\
    La & $-0.002\pm0.003$  & $-0.03$ & $+0.004\pm0.005$ & $+0.09 $ & $-0.023\pm 0.002$ & $ -0.005\pm0.015$ & $-0.2$  \\
    Ce & $+0.001\pm0.003$  & $+0.04$ & $-0.009\pm0.005$ & $-0.18 $ & $-0.024\pm 0.002$ & $-0.016\pm0.010 $ & $-0.5$  \\
    Pr & $+0.008\pm0.004$  & $+0.13$ & $ $ & $ $ & $-0.014\pm 0.003$&  &  \\
    Nd & $-0.000\pm0.003$ & $-0.01$ & $+0.006\pm0.004$ & $+0.14 $ & $-0.023\pm 0.002$&   &  \\
    Sm & $+0.014\pm0.006$  & $+0.19$ & $+0.010\pm0.006$ & $+0.22$ & $-0.008\pm 0.002$ & &  \\
    Eu & $+0.011\pm0.003$  & $+0.22$ & $+0.011\pm0.004$ & $+0.24 $ & $-0.007\pm 0.002$ & &  \\
   \hline
    \end{tabular}
    \label{tab:slopes}
%    \smallskip
\begin{tablenotes}
   \item[$^{\ast}$] Computed in this work for the potential thin-disc stars with age uncertainties better than 3~Gyr from data by \citet{Bensby14} and \citet{Battistini16}.
    \item[$^{\ast\ast}$] The slope was calculated including open clusters.
   \end{tablenotes}
      \end{threeparttable}
\end{table*}

For comparison, we also computed fits using data by \citet{Battistini16} for Sr, Zr, La, Ce, Nd, Sm, and Eu, and data by \citet{Bensby14} for Y and Ba. We took data of the thin-disc stars (TD/D < 0.5) with age determination uncertainties better than 3~Gyr. The determined slopes from our sample agree well with the slopes determined using the \citet{Battistini16} and \citet{Bensby14} studies, but there are more stars with higher barium abundances in the sample of \citet{Bensby14}, probably because this study included  more stars with larger mean galactocentric distances. 
In some of these young stars, which are often quite active, the high Ba abundances may be caused by the NLTE effects, which are under-accounted for the Ba\,{\sc ii} lines that emerge in the upper layers of stellar atmospheres (\citealt{Reddy17}).  

Table~\ref{tab:slopes} also shows slopes obtained for solar twins in the recent study by \citet{Spina18} and for the Galactic thin-disc stars with ages younger than 8~Gyr computed by \citet{Magrini18}. The authors investigated abundances of Y, Zr. Ba, La, and Ce using internal data of the $Gaia$-ESO Survey.

\begin{figure}

        \graphicspath{ {fig11.pdf} }
        \includegraphics[width=\columnwidth]{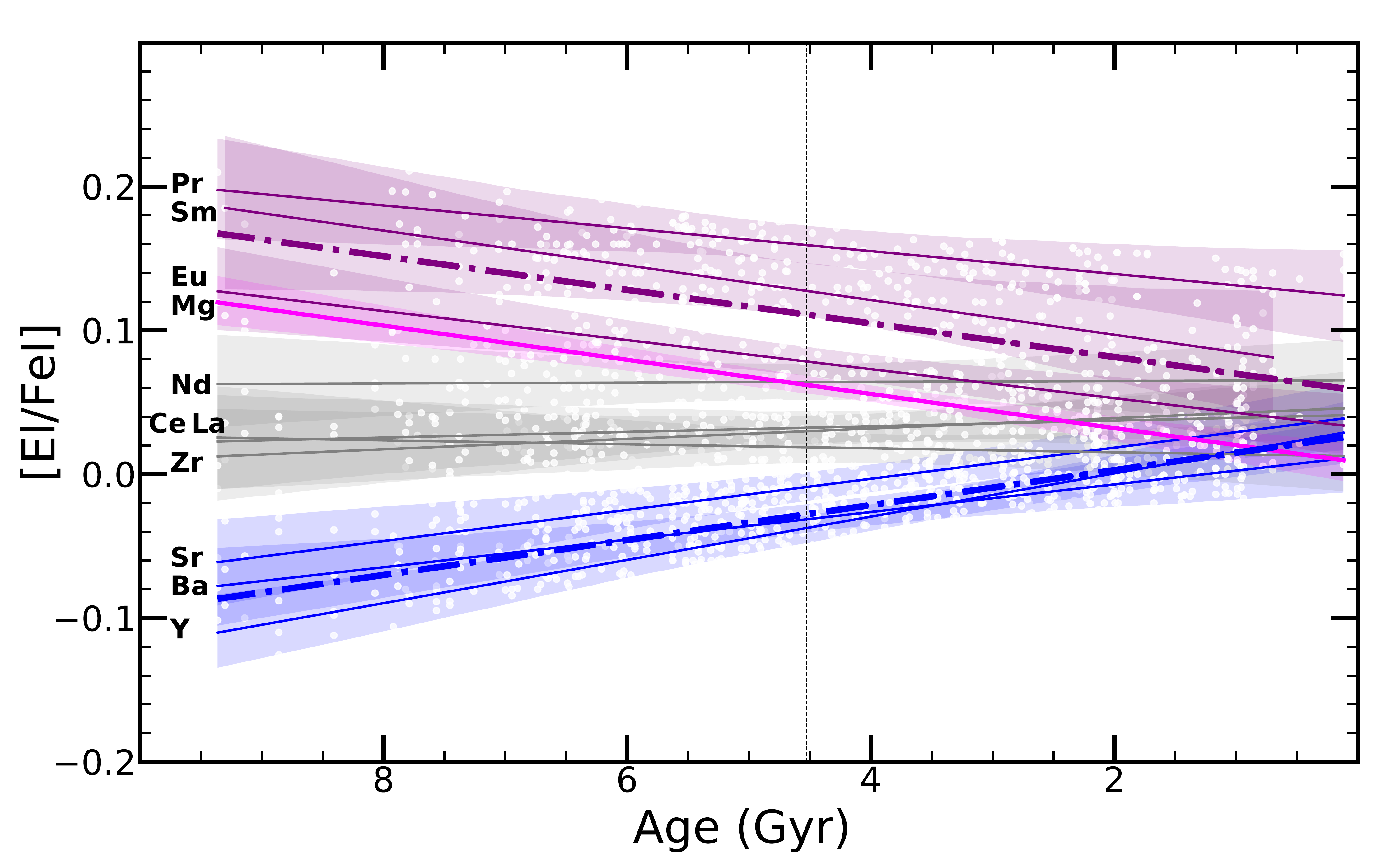}
    \caption{Compilation of [El/Fe{\sc I}] trends as a function of age for the thin-disc stars. The continuous purple lines represent the $r$-process dominated elements Eu, Sm, and Pr. The thick dash-dotted line is an averaged trend of these element-to-iron abundance ratios. The continuous blue lines represent the $s$-process dominated elements Y, Sr, and Ba. The thick dash-dotted line is an averaged trend of these element-to-iron abundance ratios. The grey continuous lines are for elements with negligible [El/Fe{\sc I}] age trends (Nd, Ce, La, and Zr). The continuous magenta line represents the [Mg/Fe{\sc I}] age correlation. The shadowed areas show the 95\% confidence interval for the regressions. The vertical dashed line marks the solar age. }
    \label{comilation_age_thin}
\end{figure}

In Fig.~\ref{comilation_age_thin} we compile all [El/Fe\,{\sc i}] trends as a function of age determined in our work. With increasing age, all the $s$-process dominated elements in the thin-disc stars of our sample show negative or flat trends (for [Sr/Fe], [Y/Fe], and [Ba/Fe], the downward trends being most clearly visible), while the $r$-process dominated elements have slightly positive or flat trends. 
These patterns agree with the GCE theory, which predicts a noticeable production of $s$-process dominated elements during the evolution of the Galactic disc. 

In Fig.~\ref{fig:abundance_age} we also compare the observational data with the models by \citet{Prantzos18} and \citet{Maiorca12}. The model by \citet{Maiorca12} closely follows our stars for Zr, but it has a slightly positive offset for Y at older ages. The model by \citet{Prantzos18} predicts quite flat trends for Sr, Y, and Zr from around 9~Gyr to the youngest objects, with a slightly positive offset for Sr and Y and a negative offset for Zr. 

Our [Ba, La, Ce/Fe] ratios are close to the chemical evolution model by \citet{Maiorca12} for the sub-solar ages. The younger stars display not so high abundance values as are predicted by the model, however. This is expected because the model was oriented to account for the higher abundances of Ba, La, and Ce that are observed in open clusters. The authors suggested that the observed enhancements might be produced
by nucleosynthesis in AGB stars of low mass ($M <1.5\,M_\odot$) if they release neutrons
from the $^{13}{\rm C}(\alpha,n)^{16}{\rm O}$ reaction in reservoirs larger by a factor of four than assumed in more massive AGB stars. In the current literature, the same assumptions
on the formation of the  $^{13}{\rm C}$ neutron source for all stars
between 1 and 3\,$M_\odot$ are reported. However, these elements are not so abundant in the field stars, and  \citet{Marsakov16}, for example, discussed on the basis of 90 open clusters the reasons  for the $s$-process dominated element overabundances in open clusters, which include high and elongated orbits, for instance.       

The model by \citet{Prantzos18} predicts a completely different behaviour for Ba, La, and Ce than the model by \citet{Maiorca12}. The model reaches its abundance peak at around 8~Gyr and then slightly decreases with decreasing age. Compared to our results, the model overestimates the abundances in the older stars and slightly underestimates the abundances in the youngest stars. 

As the [Ba/Fe] abundances in our work are very close to those of [La/Fe] and [Ce/Fe] and show close to solar abundances at young ages, we can state that our study does not support the barium puzzle anomaly (also opposed by \citealt{Reddy17}) and 
 does not require the additional intermediate process ($i$-process), as proposed by  \citep{Cowan77,Mishenina15}. 
   Stars with ages younger than 2~Gyr in our sample are dwarfs and unevolved giants. The mean uncertainty in their age determination is ($+1.17;-0.62$)~Gyr.

Praseodymium, neodymium, samarium, and europium have a significant input from $r$-process in their production. As $r$-process requires quite harsh conditions to occur, a production peak of such elements took place at earlier times in the Galaxy. 
For these elements the \citet{Prantzos18} model predicts a steady decline in [El/Fe] ratios with decreasing ages, but still steeper than our results suggest, especially for europium. 

\begin{figure}

        \graphicspath{ {fig12.pdf} }
        \includegraphics[width=\columnwidth]{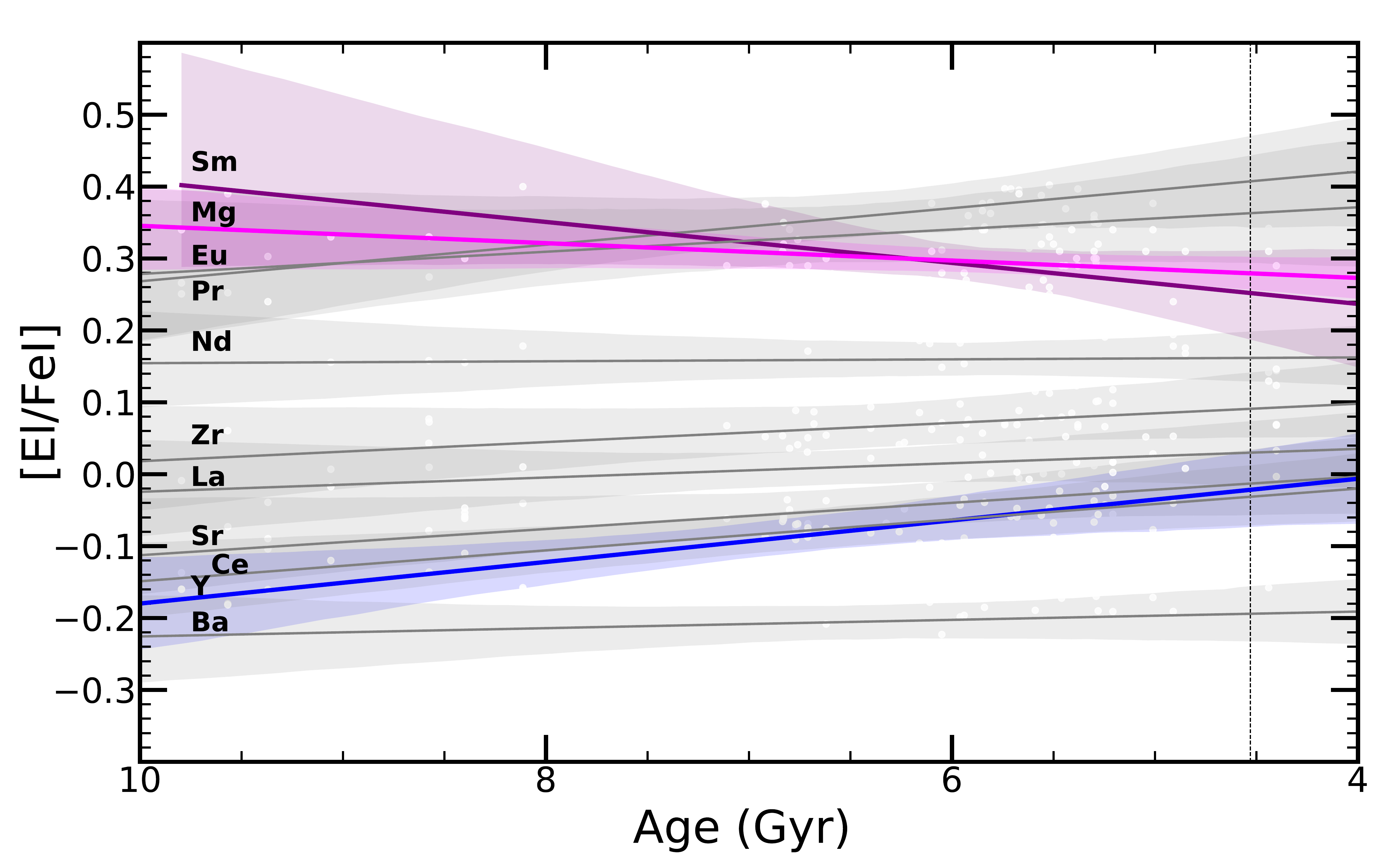}
    \caption{Compilation of [El/FeI] trends as a function of age for the thick-disc stars. The continuous purple line represents the $r$-process dominated element Sm, which alone shows a slight trend, but with a quite low accuracy. The continuous blue line represents the $s$-process dominated element Y, which has a more noticeable trend than the others. The grey continuous lines are for elements with negligible trends of [El/FeI] vs. age. The continuous magenta line represents the [Mg/FeI] age correlation. The shadowed areas show the 95\% confidence interval for the regressions. The vertical dashed line marks the solar age. }
    \label{comilation_age_thick}
\end{figure}

The 76 thick-disc stars of our sample show uncertain [El/Fe\,{\sc i}] trends with age (Fig.~\ref{comilation_age_thick}). 
Of the $r$-process dominated elements, only Sm shows a slightly positive trend. Its accuracy is quite low, however. In contrast, of the $s$-process dominated elements, only Y has a more noticeable but also quite uncertain trend. The [Mg/Fe\,{\sc i}] versus age slope is also presented for  comparison. We address the question of age dependence of the elemental abundance rations in the following section.     

\subsection{ [Y/Mg] as an age indicator for the thin-disc stars}

\begin{figure*}
        \graphicspath{ {fig13.pdf} }
        \includegraphics[width=\textwidth]{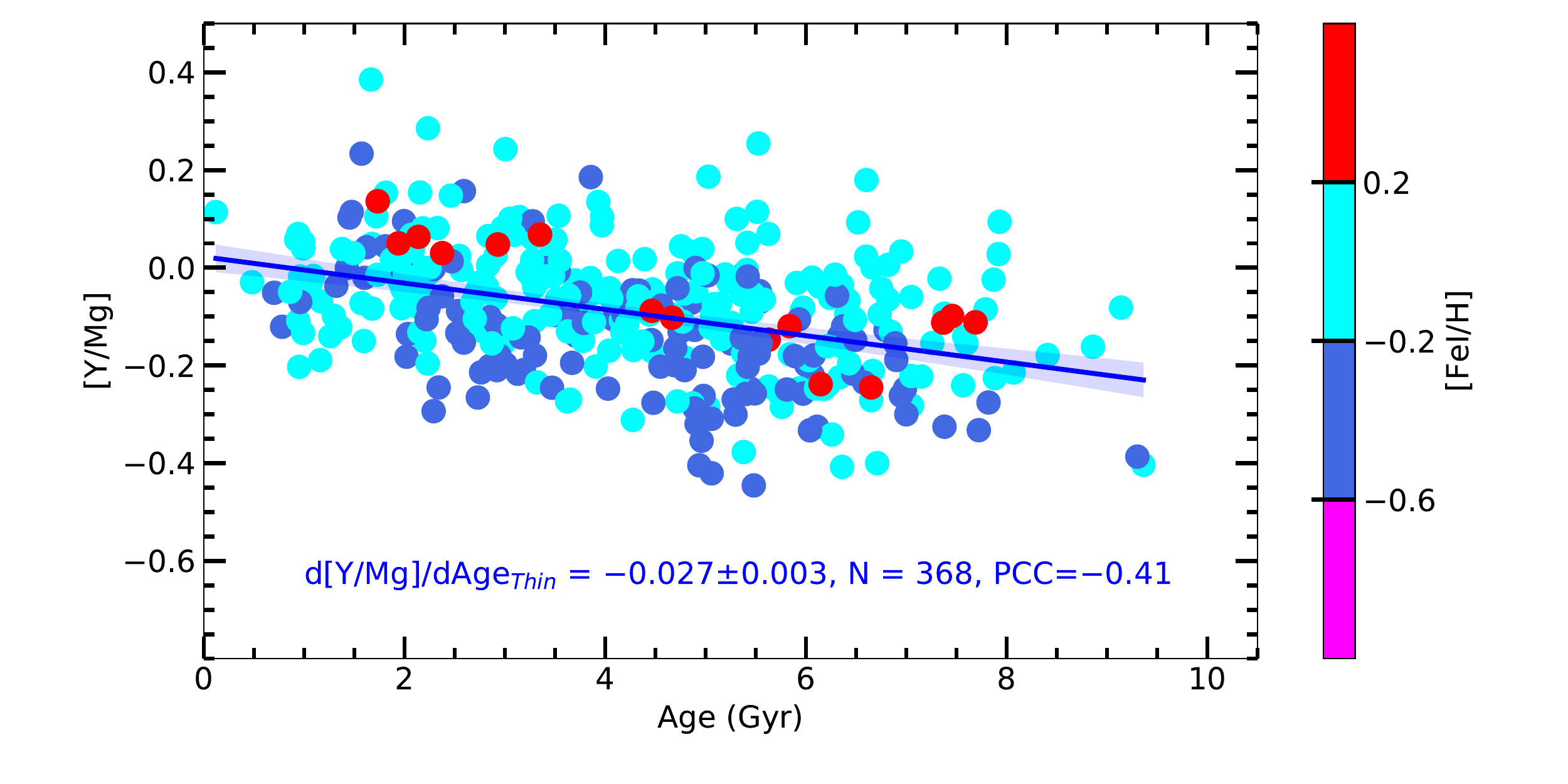}
    \caption{[Y/Mg] of the thin-disc stars colour-coded by metallicity as a function of age. The  continuous line is a fit to the sample of 368 stars. The shadowed area shows a 95\% confidence interval for the ordinary least-squares regression. } 
 \label{fig:YMg_age_thin}
\end{figure*}

\begin{figure*}
        \graphicspath{ {fig14.pdf} }
        \includegraphics[width=\textwidth]{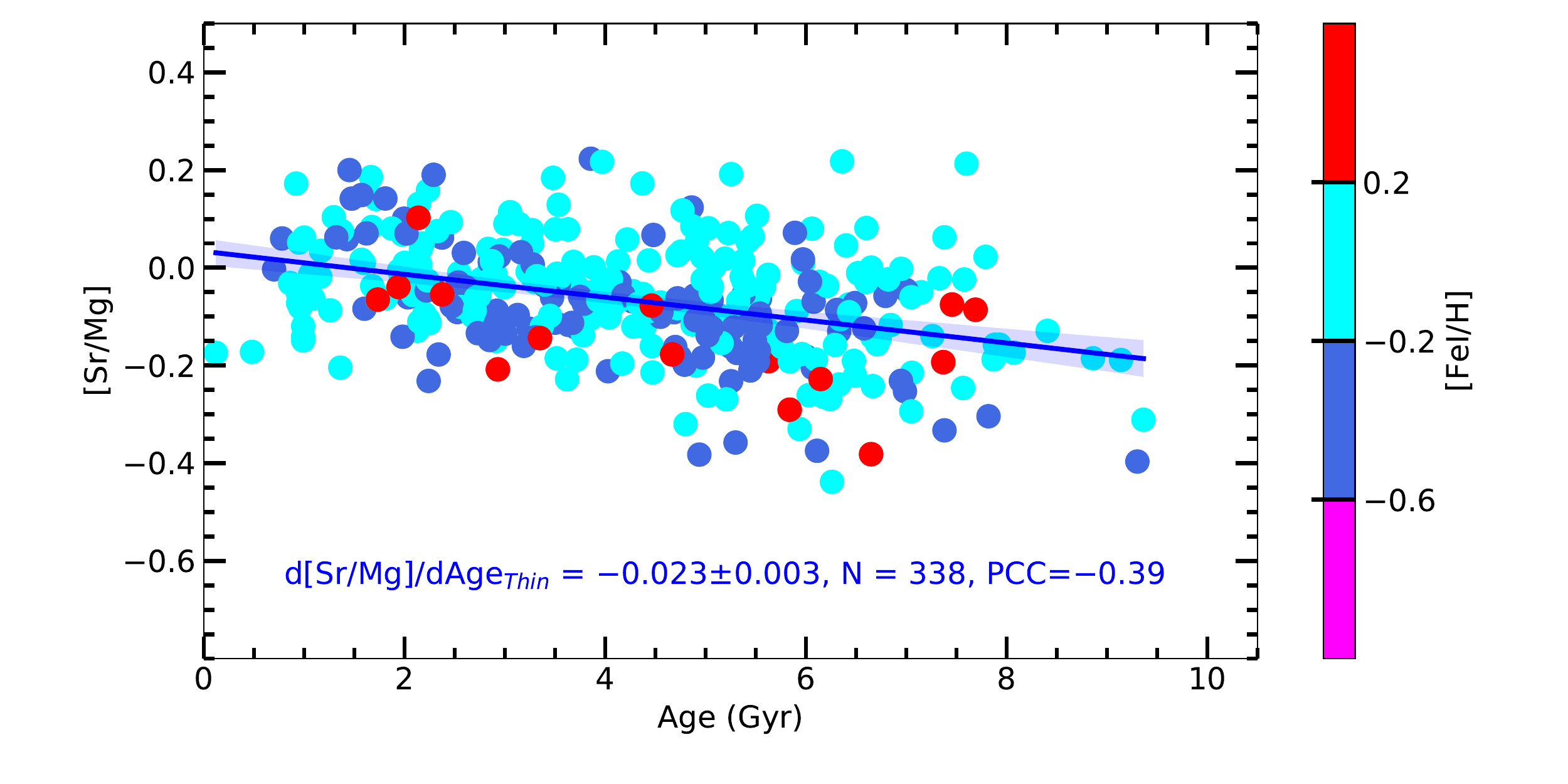}
    \caption{[Sr/Mg] of the thin-disc stars colour-coded by metallicity as a function of age. The  continuous line is a fit to the sample of 368 stars. The shadowed area shows the 95\% confidence interval for the ordinary least-squares regression. } 
 \label{fig:SrMg_age_thin}
\end{figure*}

As was proven previously, [Y/Mg] is a sensitive age indicator for solar twins  \citep{dasilva12,Nissen15,Nissen16,Spina16,TucciMaia16}. \citet{Feltzing17} showed that this relation depends on [Fe/H] and flattens for stars with lower than solar metallicities. They concluded that the [Y/Mg] versus age relation is unique to solar analogues.
However, \citet{Slumstrup17} confirmed that the empirical relation between [Y/Mg] and age as presented by \citet{Nissen16} also holds for the solar metallicity helium-core-burning giants. \citet{Nissen17} also confirmed that the abundances of the {\it Kepler} LEGACY stars, with asteroseismic ages, support the same  relation between [Y/Mg] and stellar age as was previously found for solar twins. The sub-sample of {\it Kepler} LEGACY stars in \citet{Nissen17} has a metallicity range similar to that of solar twins ($-0.15 < {\rm [Fe/H]} < +0.15$), but the range of effective temperatures is much wider ($5700 < T_{\rm eff} < 6400$~K). 

\citet{Nissen17} found that the relation [Y/Mg] versus age for the {\it Kepler} stars and solar twins has a slope of $-$0.035~dex\,Gyr$^{-1}$ with a [Y/Mg] intercept of 0.15~dex. 
The \citet{Nissen17} relation was based on a mixture of ages derived from two techniques, asteroseismology and isochrone fitting, but the slope coefficients are similar to those determined for solar twins by \citet{Nissen16}.
This relation was also calculated by \citet{Delgado19} for 
FGK solar-type dwarf stars from the HARPS-GTO program using Padova and Yonsei-Yale isochrones and Hipparcos and Gaia parallaxes. The authors obtained a result for the slope of $-$0.041~dex\,Gyr$^{-1}$, for the solar twins. \citet{Casali20} recently studied how the slope varies with metallicity for solar twins in the Gaia-ESO samples of open clusters and field stars, analysing this relation in four different regions of metallicity. They found a slope of $-0.018$~dex\,Gyr$^{-1}$ for ${\rm [Fe/H]} > 0.1$, $-0.040$ for $-0.1 < {\rm [Fe/H]} < +0.1$, $-0.042$ for $-0.3 < {\rm [Fe/H]} <-0.1$, and $-0.038$ for $-0.5 < {\rm [Fe/H]} < -0.3$.

In our study, we took abundances of magnesium and the thin-to-thick disc separation of the investigated stars from  \citetalias{Mikolaitis19} and \citetalias{Tautvaisiene20} into account, and from the thin-disc sample of 371 stars with the mean <$T_{\rm eff}> = 5503\pm{56}$~K, <log\,$g$> = 3.51~$\pm{0.26}$, <[Fe/H]> = $-0.09\pm{0.11}$, and the UniDAM ages spanning from 0.1 to 9~Gyrs, calculated the following [Y/Mg] age relation with ${\rm PCC}=-0.41$: 

\begin{equation}
{\rm [Y/Mg]_{Thin}}=0.022\,(\pm0.015)-0.027\,(\pm 0.003)\cdot{\rm age [Gyr]}.
\label{eq:YM_age}
\end{equation}

The [Y/Mg] data and the computed slope with respect to age are shown in Fig.~\ref{fig:YMg_age_thin}. 
The slope and intercept are almost the same as those of \citet{Titarenko19}, who investigated the AMBRE project sample of 325 turn-off thin-disc stars in the solar neighbourhood.

\begin{figure}

        \graphicspath{ {fig15.pdf} }
        \includegraphics[width=\columnwidth]{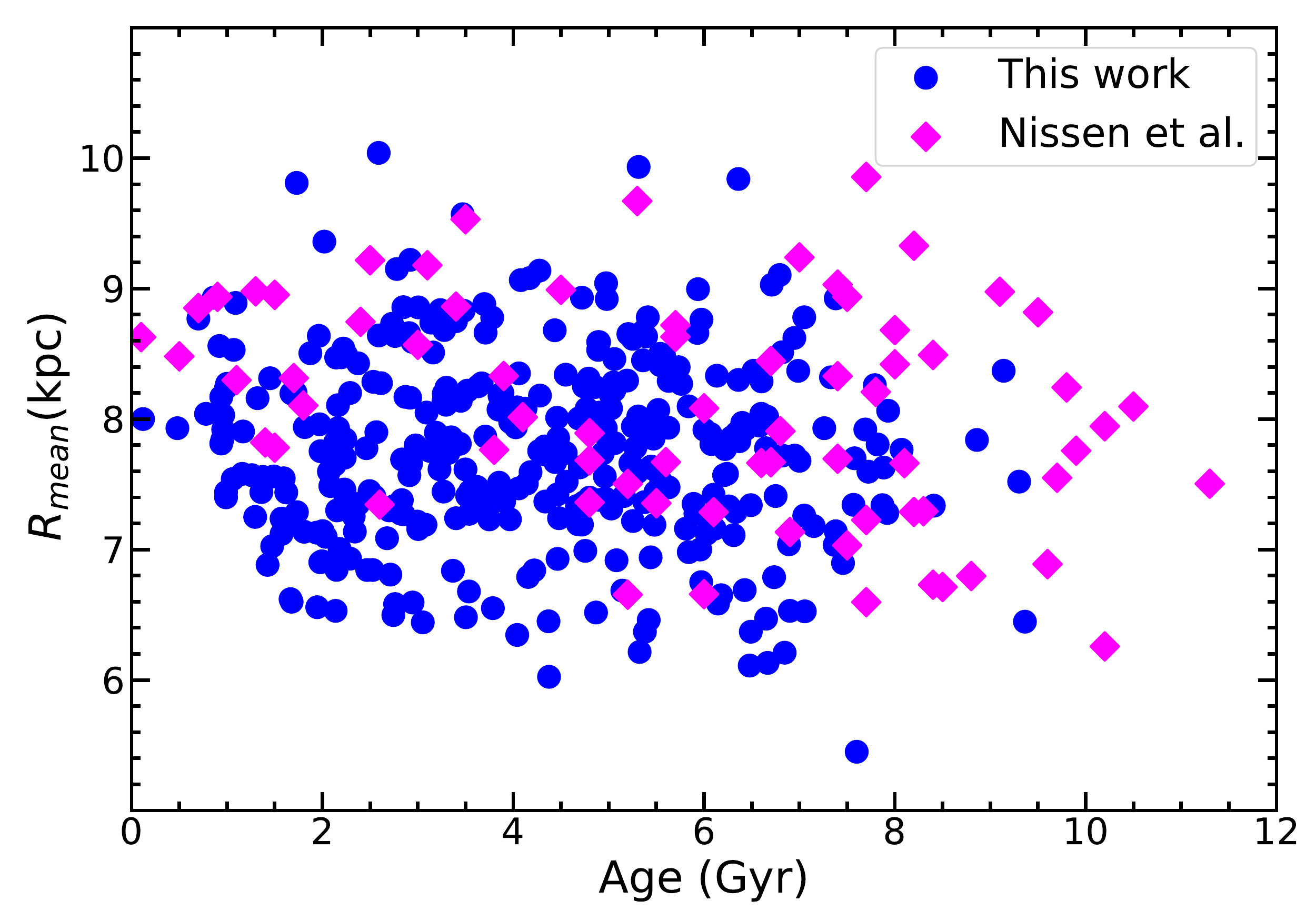}
    \caption{$R_{\rm mean}$ and age relation for the sample of thin-disc stars investigated in this work (blue dots) and in \citet{Nissen20} (magenta diamonds). This figure shows that in the \citet{Nissen20} sample, young stars predominantly have larger $R_{\rm mean}$ . This causes a larger intercept in the [Y/Mg] and age relation in \citet{Nissen20} than in this work. }
    \label{fig:our_Nissen_thin}
\end{figure}

When we compare the [Y/Mg] versus age slope obtained in this work to the slope obtained by \citet{Nissen20}, we see a difference of about 0.15~dex in the intercepts even for stars of the same metallicity.  In search for the reason, we decided to compute the mean galactocentric distances for the Nissen et al. sample stars and to compare them with those of our sample. 
Fig.~\ref{fig:our_Nissen_thin} shows that in the \citet{Nissen20} sample, young stars predominantly have larger $R_{\rm mean}$ . This may cause a larger intercept in the [Y/Mg] and age relation compared to our study and that of \citet{Titarenko19}. Therefore we follow \citet{Casali20} and  \citet{Magrini21} in concluding that the relations between ages and abundance ratios obtained from samples of stars located in a limited region of the Galaxy cannot be translated into general relations valid for the whole disc.  It is worth mentioning that differences in methods used for computing relations and accounting for uncertainties in different studies may also play a role while comparing the results.

We also tested the [Y/Al] ratio as a function of age for the thin disc, as a potential age indicator and obtained 

\begin{equation}
{\rm [Y/Al]_{Thin}}=0.067\,(\pm 0.015)-0.029\,(\pm 0.003)\cdot{\rm age [Gyr]}  
\label{eq:YAl_age}
\end{equation}
with ${\rm PCC}=-0.42$, which is comparable with the [Y/Mg]-age relation.  We also targeted other low-scatter light $s$-process dominated elements as age indicators. Sr also appears to have a good sensitivity to stellar age: 
\begin{equation}
{\rm [Sr/Mg]_{Thin}}= 0.033\,(\pm 0.014)-0.023\,(\pm 0.003)\cdot{\rm age [Gyr]} 
\label{eq:YAl_age}
\end{equation}
 with ${\rm PPC} =-0.39$, which makes the [Sr/Mg] abundance ratio a promising age indicator  (displayed in Fig.~\ref{fig:SrMg_age_thin}). Very recently, \citet{Nissen20} also tested the sensitivity of Sr in solar-type stars, obtaining a result very similar to that of Y, which is in agreement with our results.  \citet{Nissen20} used the same single Sr\,{\sc i} 4607.34~{\AA} line as in our study.

\subsection{ [Y/Mg] as age indicator for the thick-disc stars}

\begin{figure*}
        \graphicspath{ {fig16.pdf} }
        \includegraphics[width=\textwidth]{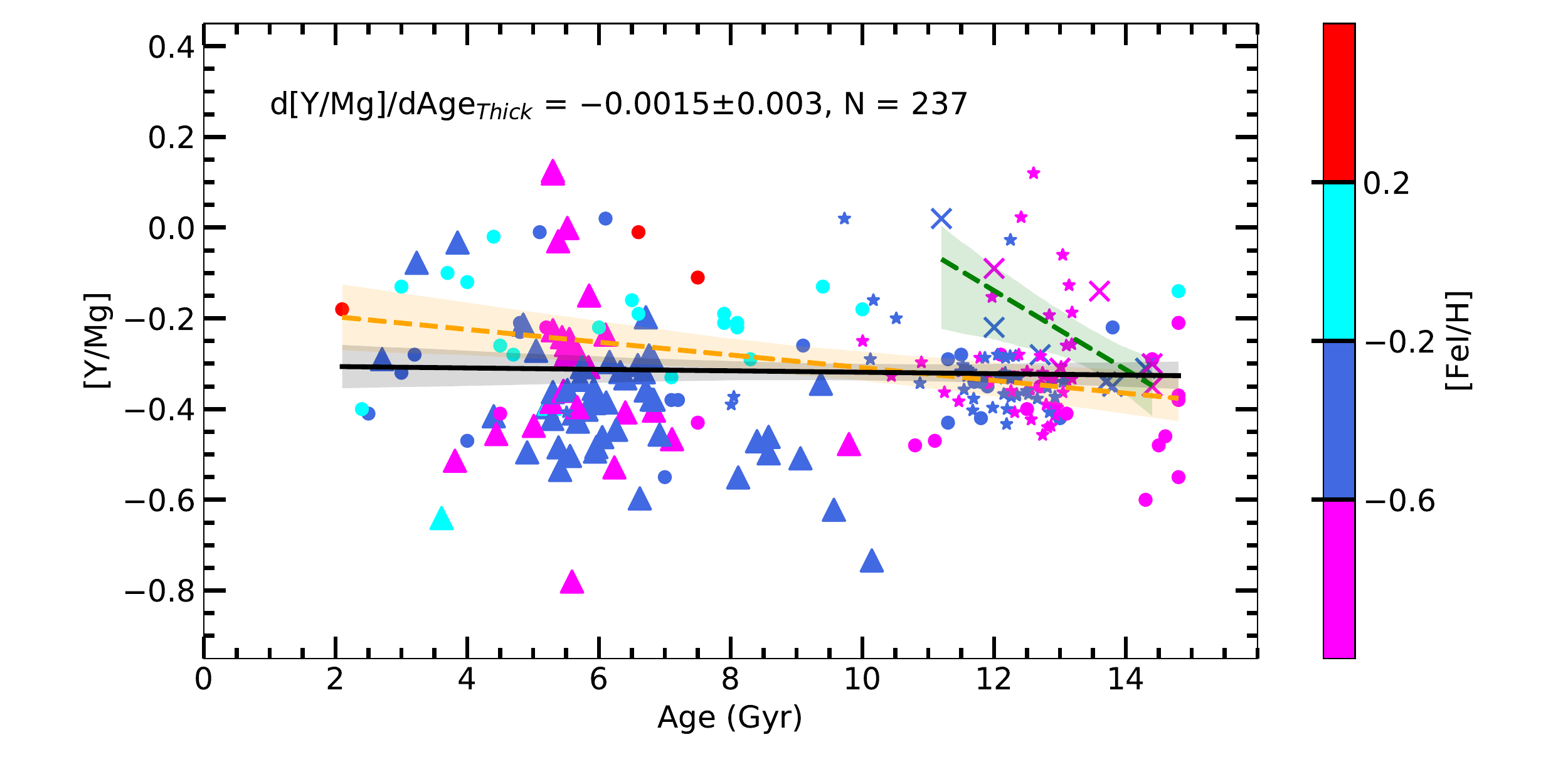}
    \caption{[Y/Mg] of the thick-disc stars colour-coded by metallicity as a function of age. Our sample stars (triangles) are suplemented by data from \citet{Bensby14} (dots), \citet{DelgadoMena18} and \citet{Adibekyan12} (stars), and \citet{Titarenko19} (crosses). The black continuous line is a fit to the entire sample of 237 stars. For a comparison, the dashed green line represents the trend taken from \citet{Titarenko19}, who computed it for the AMBRE sample of 11 thick-disc stars; and the dashed yellow line represents the trend that we computed using Bensby et al. data alone covering the entire investigated age interval. The shadowed areas show the 95\% confidence interval for the ordinary least-squares regressions.  } 
 \label{fig:YMg_age_thick}
\end{figure*}

Our thick-disc sample contains 76 stars with estimated ages. They have a mean $<T_{\rm eff}> = 4715\pm{77}$~K, $<{\rm log}\,g>=2.49~\pm{0.25}$ and $<{\rm [Fe/H]}>=-0.46~\pm{0.12}$, and ages from about 3 to 10~Gyr. 
Using this sample, we computed a steeper (than the thin-disc stars) negative slope of $-0.041\pm 0.013$~dex\,Gyr$^{-1}$ with a [Y/Mg] intercept of $-0.115\pm{0.079}$~dex, and  ${\rm PPC} =-0.35$. The [Y/Mg]-age slope given by \citet{Titarenko19} for the AMBRE sample of 11 thick-disc turn-off stars with ages from about 11.5 to 14.5~Gyrs strongly disagrees with our result. 

Because of this disagreement, we decided to increase the sample of 76 thick-disc stars by including data from other studies that provided abundances of Y and Mg, as well as ages and separated components of the Galactic disc. Consequently, we included 62 stars (with age uncertainties $<3$~Gyr) from \citet{Bensby14}, 88 stars (with age uncertainties $<3$~Gyr) from \citet{Delgado17} and \citet{Adibekyan12}, and 11 stars from \citet{Titarenko19}. The final sample consisted of 237 stars that cover different ranges of metallicity and age.

In Fig.~\ref{fig:YMg_age_thick} we show the almost negligible [Y/Mg] correlation with age (${\rm PCC}=-0.04$) for the compilation of 237 thick-disc stars: 

\begin{equation}
{\rm [Y/Mg]_{Thick}}=-0.303\,(\pm0.027)-0.002\,(\pm 0.003)\cdot{\rm age [Gyr]}.
\label{eq:YM_age_thick}
\end{equation}

\begin{table*}
\centering
        \caption{Best linear fits to the [El/Fe\,{\sc i}]$-R_{\rm mean}$ and $|z_{\rm max}|$ distributions in dex\,kpc$^{-1}$  for the thin-disc stars.}
        \label{tab:rmeanslopes}
        \begin{tabular}{lccrccr} % 4 columns, alignment for each
                \hline
                \hline
                El. & $R_{\rm mean}$ slope  & y-intercept & $\chi^2$ & $|z_{\rm max}|$ slope & y-intercept & $\chi^2$ \\
%                \\
%        &  dex\,Gyr$^{-1}$ & dex  &  \\
                        \hline
                Sr & $+0.002\pm0.007$ & $-0.032\pm 0.052$ & $+0.67$ & $-0.080\pm0.024$ & $+0.005\pm 0.007$ & $+0.65$  \\ 
                Y & $-0.005\pm0.006$ & $+0.014\pm 0.047$ & $+1.00$ & $-0.082\pm0.024$ & $-0.010\pm 0.006$ & $+0.92$ \\
%               Zr\,{\sc i} & $+0.001\pm0.008$ &$+0.021$ & $+0.01$ & $-0.001\pm0.021$ & $+0.027$ & $-0.002$ \\
                Zr & $+0.020\pm0.007$ &$-0.126\pm 0.052$  & $+0.49$ & $+0.023\pm0.023$ & $+0.026\pm 0.007$ & $+0.50$\\
                Ba & $-0.013\pm 0.006$ & $+0.072\pm 0.045$ & $+1.09$ & $-0.052\pm 0.024$ & $-0.015\pm 0.007$ & $+1.09$ \\ 
                La & $+0.025\pm0.007$ & $-0.170\pm 0.054$ & $+0.75$ & $+0.030\pm0.025$ & $+0.018\pm 0.008$ & $+0.77$ \\
                Ce & $+0.015\pm0.007$ & $-0.105\pm 0.054$ & $+0.69$ & $-0.034\pm0.027$ & $+0.021\pm 0.008$ & $+0.70$ \\
                Pr & $+0.045\pm0.009$ & $-0.218\pm 0.063$ & $+1.33$ & $+0.094\pm 0.035$ & $+0.124\pm 0.011$ & $+1.41$ \\
                Nd & $+0.034\pm0.008$ & $-0.218\pm 0.063$ & $+1.16$ & $+0.070\pm0.034$ & $+0.028\pm 0.010$ & $+1.20$ \\
                Sm & $+0.022\pm0.011$ & $-0.057\pm 0.089$ & $+1.10$ & $+0.071\pm0.039$ & $+0.100\pm 0.014$ & $+1.10$ \\
                Eu & $+0.023\pm0.006$ & $-0.098\pm 0.049$ & $+0.53$ & $+0.123\pm0.023$ & $+0.048\pm 0.007$ & $+0.51$\\
        \hline
        \end{tabular}
\end{table*}

\begin{table*}
\centering
        \caption{Best linear fits to the [El/Fe\,{\sc i}]$-R_{\rm mean}$ and $|z_{\rm max}|$ distributions in dex\,kpc$^{-1}$ for the thick-disc stars.}
        \label{tab:rmeanslopesthick}
        \begin{tabular}{lccrccr} % 4 columns, alignment for each
                \hline
                \hline
                El. & $R_{\rm mean}$ slope  & y-intercept & $\chi^2$ & $|z_{\rm max}|$ slope & y-intercept & $\chi^2$ \\
                        \hline
                Sr & $+0.016\pm 0.011$ & $-0.203\pm 0.086$ & $+1.09$ & $-0.107\pm 0.038$ & $-0.029\pm 0.020$ & $+1.01$  \\ 
                Y  & $+0.006\pm 0.012$ & $-0.128\pm 0.092$ & $+1.53$ & $-0.062\pm 0.040$ & $-0.059\pm 0.020$ & $+1.49$ \\
                Zr & $-0.001\pm 0.014$ & $+0.060\pm 0.114$ & $+0.60$ & $-0.087\pm 0.046$ & $+0.091\pm 0.024$ & $+0.58$ \\
                Ba & $-0.005\pm 0.010$ & $-0.137\pm 0.079$ & $+1.10$ & $-0.080\pm 0.033$ & $-0.142\pm 0.017$ & $+1.03$ \\ 
                La & $-0.002\pm 0.012$ & $+0.039\pm 0.096$ & $+0.86$ & $-0.034\pm 0.042$ & $+0.040\pm 0.021$ & $+0.85$ \\
                Ce & $-0.002\pm 0.013$ & $-0.038\pm 0.102$ & $+0.90$ & $-0.062\pm 0.045$ & $-0.034\pm 0.021$ & $+0.88$ \\
                Pr & $+0.024\pm 0.013$ & $+0.136\pm 0.099$ & $+1.20$ & $-0.046\pm 0.046$ & $+0.340\pm 0.022$ & $+1.24$ \\
                Nd & $+0.001\pm 0.010$ & $+0.133\pm 0.078$ & $+0.61$ & $+0.021\pm 0.036$ & $+0.132\pm 0.017$ & $+0.61$ \\
                Sm & $+0.001\pm 0.014$ & $+0.245\pm 0.109$ & $+0.67$ & $-0.009\pm 0.050$ & $+0.256\pm 0.023$ & $+0.67$ \\
                Eu & $-0.004\pm 0.018$ & $+0.332\pm 0.144$ & $+1.38$ & $-0.012\pm 0.061$ & $+0.302\pm 0.031$ & $+1.38$\\
        \hline
        \end{tabular}
\end{table*}

 In Fig.~\ref{fig:YMg_age_thick}  we show the trend taken from \citet{Titarenko19} for comparison, who computed this for the AMBRE sample of 11 thick-disc stars, which are all older. This slope disagrees with the results of other old thick-disc stars investigated by \citet{Delgado17}, \citet{Bensby14}, and \citet{Adibekyan12}. Having in mind that the \citet{Bensby14} sample contains thick-disc stars of a wide range of ages, we computed and displayed a slope based on the Bensby et al. stellar sample alone. This slope is also quite shallow (${\rm [Y/Mg]}=-0.168\,(\pm0.040)-0.014\,(\pm 0.004)\cdot{\rm age [Gyr]})$.

Thus, our study shows that [Y/Mg] can be used as a clock to find stellar ages for the thin-disc stars. The chemical evolution of the thick disc is different, however, and it seems that the [Y/Mg] ratios cannot be used to evaluate the age. Hopefully, the number of thick-disc stars with determined n-capture element abundances will soon increase and a comprehensive study of the thick-disc evolution will proceed.

\section{Radial and vertical abundance gradients} 
\label{sect:5}

 \begin{figure*}
    \graphicspath{ {fig17.pdf} }
        \includegraphics[width=\textwidth]{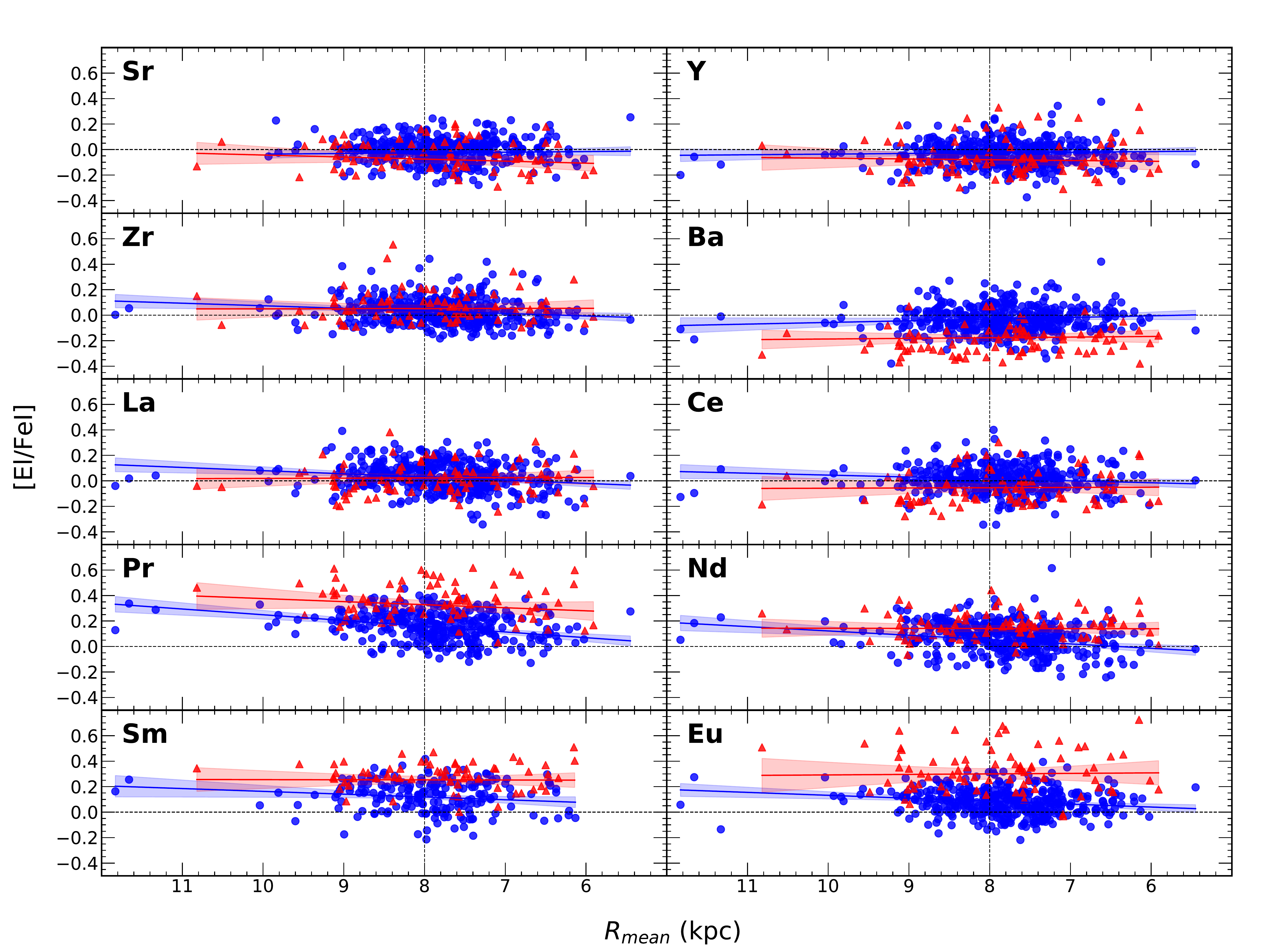}
                \caption{Elemental-to-iron abundance ratios as a function of the mean galactocentric distances. The blue dots and red triangles represent the thin- and thick-disc stars of our study, respectively.  The blue lines are the linear fits for the thin-disc stars and the red lines show the thick-disc stars. The uncertainties in [El/Fe] and $R_{\rm mean}$ were taken into account while computing the slopes. The 95\% confidence intervals for the slopes are shadowed. The dotted lines correspond to the solar mean galactocentric distance. }
    \label{fig:abundance_rmean}
\end{figure*}

\begin{figure*}
    \graphicspath{ {fig18.pdf} }
        \includegraphics[width=\textwidth]{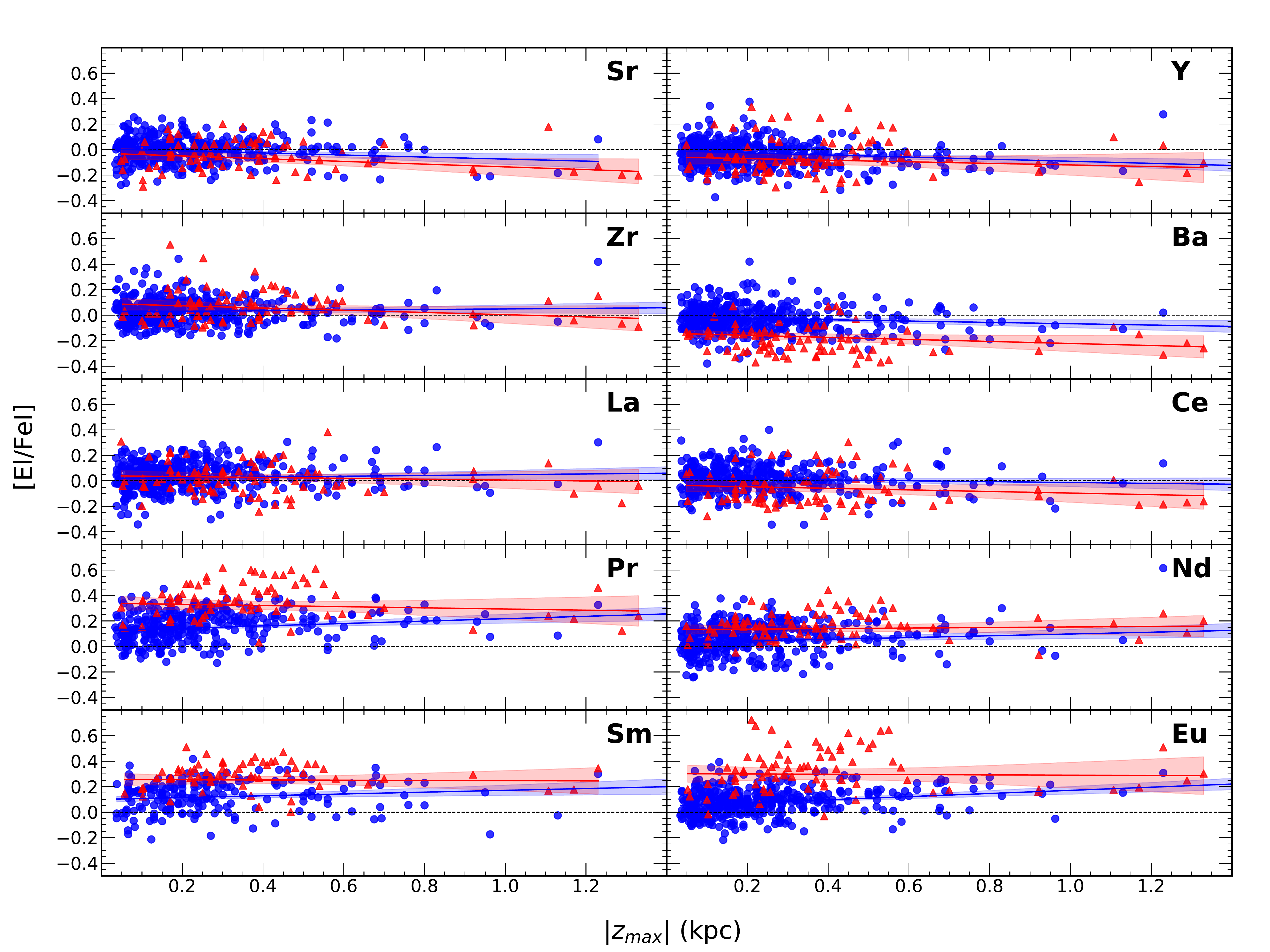}
                \caption{Elemental-to-iron abundance ratios as a function of the $|z_{\rm max}|$.  Symbols are the same as in Fig.~\ref{fig:abundance_rmean}.}
    \label{fig:abundance_zmax}
\end{figure*}

 The radial and vertical abundance [El/Fe\,{\sc i}] gradients were calculated as a function  of the mean galactocentric distance ($R_ {\rm mean}$) and the maximum height from the Galactic plane ($|z_{\rm max}|$). 
  $R_ {\rm mean}$ is a better indicator of the stellar birth place than an in situ$ $ galactocentric distance. It was used in several previous studies (e.g. \citealt{Edvardsson93,Rocha-Pinto04,Trevisan14,Adibekyan16,Mikolaitis19}), and could be used more widely, especially while investigating older stars.
 The target selection strategy in the selected fields was to observe all stars in the selected magnitude and colour range. In this sense, the sample is homogeneous and minimally biased. The targeted sample is magnitude ($V < 8$~mag) and colour $((B-V)>0.39)$ limited, as described in \citetalias{Mikolaitis19} and  \citetalias{Tautvaisiene20} and Sect.~\ref{sect:2}. The final sample consists of 506 F, G, and K stars distributed in wide atmospheric parameters and age space. The different populations of stars with different metallicities and ages might introduce a bias in the derived abundance radial and vertical gradients. We therefore divided our sample into thin- and thick-disc subsamples in order to avoid the potential bias in analysing gradients.

The resulting values of the radial and vertical abundance gradients for the thin- and thick-disc  stars are listed in Table~\ref{tab:rmeanslopes} and Table~\ref{tab:rmeanslopesthick}, respectively.   The individual uncertainties of the x- and y-axis data were taken into account when we computed the linear fits with the orthogonal distance regression
method using the Kapteyn package (\citealt{KapteynPackage}). 

In Fig.~\ref{fig:abundance_rmean} we show the radial distribution of [El/Fe\,{\sc i}] abundances. The sample stars span the range  5.5~<~$R_{\rm mean}$~<~10~kpc, except for one star at 11.8~kpc. 

For the thin-disc stars, the radial  [El/Fe] abundance gradients are negligible for the $s$-process dominated elements and become slightly positive for the elements with stronger $r$-process domination. The same slightly positive radial [El/Fe] gradients as for the $r$-process dominated elements  in our study were found for $\alpha$-elements (except odd ones) in \citetalias{Mikolaitis19} as well as in \citet{Li18}. 

Studies of the radial n-capture-to-iron abundance gradients are very scarce so far. We can only search for a broad agreement of our results with several studies of abundance gradients with galactocentric distances ($R_{\sc gc}$). \citet{daSilva16} studied n-capture elements across the Galactic thin disc based on Cepheid variables. Because the Cepheids are young stars, their $R_{\sc gc}$ may be rather close to their birthplaces and $R_{\sc mean}$. da\,Silva et al. supplemented their sample of 111 Cepheids with 324 more stars from other studies and found that the [Y/Fe] distribution is flat throughout the entire disc. In our study, we confirm this finding not only based on the whole thin-disc sample of stars and on a subsample of younger $\leq 4$~Gyr stars, but also add another light $s$-process dominated element strontium.  Like in our study,     da\,Silva et al. also obtained positive [El/Fe] radial gradients for La, Ce, Nd, and Eu. The slopes are rather similar. For [Eu/Fe], they differ just by 0.002~dex\,kpc$^{-1}$. More recently, \citet{Luck18b} also investigated the gradients of n-capture element abundance-to-iron ratios with respect to $R_{\sc gc}$ for a sample of 435 Cepheids. It is interesting to note that the [Ba/Fe] versus $R_{\sc gc}$ slope according to this Cepheid sample is also negative, as in our study. [Ba/Fe] is the only n-capture element-to-iron ratio with a negative radial gradient in our sample of stars and in \citet{Luck18b}.  \citet{Overbeek16} investigated trends of Pr, Nd, and Eu to Fe abundance ratios with respect to $R_{\sc gc}$ using 23 open clusters. As in our study, they found that these elements have positive linear trends with galactocentric radius (the linear regression slopes are of about +0.04~dex\,kpc$^{-1}$). They also suggested that the [El/Fe] relation of Pr and Nd, but not Eu, with the galactocentric radius may not be linear because the [El/Fe] of these elements appears to be enhanced around 10~kpc and drop around 12~kpc. Because only a small number of stars lie at these large radial distances, we cannot address this question. For the thick-disc stars, the radial abundance-to-iron slopes are negligible, as was found for  $\alpha$-process elements by \citet{Li18}, even though the production sites of $\alpha$-elements  and $s$-processes dominated elements are quite different.   

In Fig.~\ref{fig:abundance_zmax} we present the vertical distribution of the [El/Fe\,{\sc i}] abundances. The analysed stars are in the range of 0~<~$|z_{\rm max}|$~<~1.3~kpc, except for one star at 3.73~kpc. 

For the thin-disc stars, the vertical abundance gradients are predominately negative for the $s$-process dominated elements and become slightly positive for the $r$-process dominated elements. For the thick-disc stars, 
the vertical  [El/Fe] gradients for the investigated n-capture elements are negative and become flatter for the $r$-process dominated elements. The y-intercepts are noticeably larger for the $r$-process dominated elements. 

 The vertical [El/Fe] ratio gradients have recently been studied on rather large samples of thin- and thick-disc stars (e.g. \citealt{Duong18,Li18,Yan19}), and for the thin disc in \citetalias{Mikolaitis19}. Although only $\alpha$-process elements were investigated in these studies, we can still compare these trends with the $r$-process dominated element trends in our study because the abundance-to-iron ratios of $\alpha$-elements and $r$-process dominated elements are both affected by yields of type\,II supernovae.  It was found that the vertical [$\alpha$/Fe] trends are  positive in the thin disc and negligible in the thick disc. This agrees with the trends of the $r$-process dominated elements in our study.

 Neutron-capture elements are the least well understood in terms of nucleosynthesis and
formation environments. These elements can provide much information about the Galactic star formation and enrichment,
but observational data are limited, especially for the thick disc.

\section{Summary and conclusions}
\label{sect:summary}

With the aim of testing trends of n-capture element-to-iron abundance ratios versus stellar metallicity, age, mean galactocentric distance, and maximum height from the Galactic plane, we have analysed high-resolution spectra for a sample of 506 F, G, and K spectral-type solar vicinity ($V < 8$~mag) dwarf, subgiant, and giant stars. The investigated stars are located in the continuous viewing zone of the TESS space mission and in two neighbouring preliminary PLATO space mission fields. 

With this work we complement the studies we carried out in \citetalias{Mikolaitis18}, \citetalias{Mikolaitis19}, and \citetalias{Tautvaisiene20}, where we provided stellar parameters and abundances of 24 chemical species, which together with those of the present study reach 34 elements, calculated using the same method and instrumentation. The elemental abundance values of the n-capture elements were determined in our work by a differential line-by-line synthesis of spectral lines using MARCS one-dimensional model atmospheres and accounting for the hyperfine-structure effects. 

The approximate ranges of atmospheric parameters covered by the stars are the following: $3800 < T_{\rm eff} < 6900$~K, $0.9 < {\rm log}\,g < 4.7$, and $-1.0 < {\rm [Fe/ H]} < +0.5$. The stellar ages span from 0.1 to 10~Gyr, $R_{\rm mean}$ are between 5.5 and 11.8~kpc, and $|z_{\rm max}|$ are between 0.03 and 3.74~kpc. 

The sample contains 424 stars attributed to the Galactic thin disc and 82 stars that belong to the thick disc. 
We determined abundances of Sr (453 stars), Y (506 stars), Zr\,{\sc i} (307 stars), Zr\,{\sc ii} (476 stars), Ba (504 stars), La (504 stars), Ce (467 stars), Pr (402 stars), Nd (466 stars), Sm (257 stars), and Eu (489 stars). 

We compared the observational results with the recent Galactic chemical evolution model by \citet{Prantzos18}.
The detailed evaluation of [El/Fe\,{\sc i}] versus [Fe\,{\sc i}/H] and age trends, and heavy-to-light $s$-process dominated element abundance ratios versus [Fe\,{\sc i}/H] in particular showed that 
the contribution from the LIMS has to be further investigated and developed. The model agreement with the observations might be better if the convex shape of the model had its maximum not at the [Fe\,{\sc i}/H] of $-0.4$~dex, but at about $-0.2$~dex,  and would start accounting the LIMS input at higher metallicities (about $-0.7$~dex).

Abundance correlations with age slopes for the thin-disc stars are slightly negative or near zero for the majority of $s$-process dominated elements. The light $s$-process elements strontium and yttrium stand out by having a strong negative abundance
correlation with age. For the $r$-process dominated elements, the abundance slopes with age become positive.
Similar slopes were also computed in our work using data by \citet{Bensby14} and \citet{Battistini16}, whereas for a sample of solar twins (\citealt{Spina18}), all slopes of the same elements were negative. Further studies of the europium abundance versus age gradients deserve special attention. 

The neighbouring Galactic thin-disc stars investigated in our study do not show the barium abundance anomaly, like it was found also by \citet{Reddy17} and \citet{Marsakov16}. 
 They do not require the additional $i$-process as proposed by  \citet{Cowan77} or \citet{Mishenina15} to act in the production of barium alone.   

The relation obtained from our sample of 424 thin-disc stars is ${\rm[Y/Mg]}= 0.022(\pm0.015)-0.027(\pm 0.003)\cdot{\rm age [Gyr]}$. This slope is very similar to the one obtained by \citet{Titarenko19} for the 325 turn-off thin-disc stars in the AMBRE project.

The thick-disc sample of 76 stars with a mean ${\rm <[Fe/H]}>=-0.51\pm{0.16}$ might imply a steeper negative slope of [Y/Mg] and age relation. When we increased the sample with data from other studies for the thick disc to 237 stars including larger ages, however, we concluded that the [Y/Mg] and age slope is almost negligible. This proves different chemical evolution histories of the Galactic thin and thick disc components and shows that [Y/Mg]  can be used as a clock to determine the age only for the thin-disc stars. 

Our sample of 76 thick-disc stars clearly shows a visible decrease in [Eu/Mg] and [$r/\alpha$] with increasing metallicity compared to the thin-disc stars. This indicates a different chemical evolution of these two Galactic components.  

We used the advantage of new possibilities that were opened by the $Gaia$ space mission \citep{Gaia16,Gaia18} to determine accurate  stellar locations in the Galaxy. We computed slopes of n-capture element abundances with respect to the mean galactocentric distances and distances from the Galactic plane. The mean galactocentric distances of stars are much more informative in indicating stellar birthplaces than the {\it \textup{in situ}} galactocentric distances. 

The best linear fits to the [El/Fe\,{\sc i}]$-R_{\rm mean}$ and $|z_{\rm max}|$ relations  (Table~\ref{tab:rmeanslopes}) show that the radial abundance gradients in the thin disc are negligible for the light $s$-process dominated elements and become positive for the elements with stronger $r$-process domination. 
The vertical abundance gradients are negative for the light $s$-process dominated elements and become positive for the $r$-process dominated elements. For the thick disc, the the radial abundance-to-iron slopes are negligible, and the vertical slopes are predominantly negative. 

The Galactic chemical evolution models still have to be further developed by investigating the roles of yields from rotating massive stars and low- and intermediate-mass stars in producing $s$-process dominated elements. The relative roles of neutron star mergers and core-collapse supernova in producing $r$-process dominated elements also await further investigations. Observational data are still scarce for the thick-disc stars. More observational data should be collected for stars of different ages and Galactic locations. The first results from the Gaia-ESO Spectroscopic survey \citep{Magrini18}, the APOGEE Survey \citep{Cunha17}, the R-Alliance \citep{Hansen18,Sakari18},
the on-going  GALAH \citep{Martell17,Griffith19} and up-coming WEAVE \citep{Dalton16} and 4MOST  \citep{deJong2019} surveys as well as smaller projects show great potential in bringing new useful observational results. We continue observations in the SPFOT survey as well  and plan to compare our expanded observational results with the new GCE models by \citet{Grisoni20}, which appear to fit the thick-disc chemical evolution rather well. With increasing numbers of stars with accurate asteroseismic ages and masses determined from the currently working NASA TESS and the upcoming ESA PLATO space missions, the detailed stellar composition results obtained in the current work and other studies will be very useful for further investigations of Galaxy formation and evolution and for characterising newly found exoplanets.

\section*{Acknowledgements}
We acknowledge funding from the Research Council of Lithuania (LMTLT) (Grant No. LAT-08/2016) and the European
Social Fund via the LMTLT grant No. 09.3.3-LMT-K-712-01-0103).
This work has made use of data from the European Space Agency (ESA) mission
{\it Gaia} (\url{https://www.cosmos.esa.int/gaia}), processed by the {\it Gaia}
Data Processing and Analysis Consortium (DPAC,
\url{https://www.cosmos.esa.int/web/gaia/dpac/consortium}). Funding for the DPAC
has been provided by national institutions, in particular the institutions
participating in the {\it Gaia} Multilateral Agreement. We gratefully acknowledge Laura Magrini and Nikos Prantzos for providing the GCE models, Alexey Mints for his help with UniDAM, Mark Taylor for his help with TopCat,  and Martin Vogelaar for his advises with the Kapteyn Package. 
We thank the anonymous referee, whose constructive review helped to improve this paper comprehensively.
This research made use of the Stilts and Topcat \citep{2005ASPC..347...29T,2006ASPC..351..666T}, Astropy\footnote{http://www.astropy.org} \citep{2018AJ....156..123A}, {\it galpy} \citep{Bovy15} and UniDAM \citep{2017A&A...604A.108M} astronomical tools. We have made extensive use of the NASA ADS and SIMBAD databases.

\bibliographystyle{aa} % style aa.bst

\bibliography{biblio}        % PLATO.bib is the name of our database

\begin{thebibliography}{144}
\expandafter\ifx\csname natexlab\endcsname\relax\def\natexlab#1{#1}\fi

\bibitem[{{Abbott} {et~al.}(2017){Abbott}, {Abbott}, {Abbott}, {Acernese},
  {Ackley}, {Adams}, {Adams}, {Addesso}, {Adhikari}, {Adya}, \&
  et~al.}]{Abbott17}
{Abbott}, B.~P., {Abbott}, R., {Abbott}, T.~D., {et~al.} 2017, \apjl, 848, L12

\bibitem[{{Adibekyan} {et~al.}(2018){Adibekyan}, {de Laverny}, {Recio-Blanco},
  {Sousa}, {Delgado-Mena}, {Kordopatis}, {Ferreira}, {Santos}, {Hakobyan}, \&
  {Tsantaki}}]{Adibekyan18}
{Adibekyan}, V., {de Laverny}, P., {Recio-Blanco}, A., {et~al.} 2018, \aap,
  619, A130

\bibitem[{{Adibekyan} {et~al.}(2016){Adibekyan}, {Delgado-Mena}, {Figueira},
  {Sousa}, {Santos}, {Gonz{\'a}lez Hern{\'a}ndez}, {Minchev}, {Faria},
  {Israelian}, {Harutyunyan}, {Su{\'a}rez-Andr{\'e}s}, \&
  {Hakobyan}}]{Adibekyan16}
{Adibekyan}, V., {Delgado-Mena}, E., {Figueira}, P., {et~al.} 2016, \aap, 592,
  A87

\bibitem[{{Adibekyan} {et~al.}(2012){Adibekyan}, {Sousa}, {Santos}, {Delgado
  Mena}, {Gonz{\'a}lez Hern{\'a}ndez}, {Israelian}, {Mayor}, \&
  {Khachatryan}}]{Adibekyan12}
{Adibekyan}, V.~Z., {Sousa}, S.~G., {Santos}, N.~C., {et~al.} 2012, \aap, 545,
  A32

\bibitem[{{Alvarez} \& {Plez}(1998)}]{1998A&A...330.1109A}
{Alvarez}, R. \& {Plez}, B. 1998, \aap, 330, 1109

\bibitem[{{Arlandini} {et~al.}(1999){Arlandini}, {K{\"a}ppeler}, {Wisshak},
  {Gallino}, {Lugaro}, {Busso}, \& {Straniero}}]{Arlandini99}
{Arlandini}, C., {K{\"a}ppeler}, F., {Wisshak}, K., {et~al.} 1999, \apj, 525,
  886

\bibitem[{{Asplund} {et~al.}(2009){Asplund}, {Grevesse}, {Sauval}, \&
  {Scott}}]{Asplund09}
{Asplund}, M., {Grevesse}, N., {Sauval}, A.~J., \& {Scott}, P. 2009, \araa, 47,
  481

\bibitem[{{Astropy Collaboration} {et~al.}(2018){Astropy Collaboration},
  {Price-Whelan}, {Sip{\H o}cz}, {G{\"u}nther}, {Lim}, {Crawford}, {Conseil},
  {Shupe}, {Craig}, {Dencheva}, {Ginsburg}, {VanderPlas}, {Bradley},
  {P{\'e}rez-Su{\'a}rez}, {de Val-Borro}, {Aldcroft}, {Cruz}, {Robitaille},
  {Tollerud}, {Ardelean}, {Babej}, {Bach}, {Bachetti}, {Bakanov}, {Bamford},
  {Barentsen}, {Barmby}, {Baumbach}, {Berry}, {Biscani}, {Boquien}, {Bostroem},
  {Bouma}, {Brammer}, {Bray}, {Breytenbach}, {Buddelmeijer}, {Burke},
  {Calderone}, {Cano Rodr{\'{\i}}guez}, {Cara}, {Cardoso}, {Cheedella},
  {Copin}, {Corrales}, {Crichton}, {D'Avella}, {Deil}, {Depagne}, {Dietrich},
  {Donath}, {Droettboom}, {Earl}, {Erben}, {Fabbro}, {Ferreira}, {Finethy},
  {Fox}, {Garrison}, {Gibbons}, {Goldstein}, {Gommers}, {Greco}, {Greenfield},
  {Groener}, {Grollier}, {Hagen}, {Hirst}, {Homeier}, {Horton}, {Hosseinzadeh},
  {Hu}, {Hunkeler}, {Ivezi{\'c}}, {Jain}, {Jenness}, {Kanarek}, {Kendrew},
  {Kern}, {Kerzendorf}, {Khvalko}, {King}, {Kirkby}, {Kulkarni}, {Kumar},
  {Lee}, {Lenz}, {Littlefair}, {Ma}, {Macleod}, {Mastropietro}, {McCully},
  {Montagnac}, {Morris}, {Mueller}, {Mumford}, {Muna}, {Murphy}, {Nelson},
  {Nguyen}, {Ninan}, {N{\"o}the}, {Ogaz}, {Oh}, {Parejko}, {Parley}, {Pascual},
  {Patil}, {Patil}, {Plunkett}, {Prochaska}, {Rastogi}, {Reddy Janga},
  {Sabater}, {Sakurikar}, {Seifert}, {Sherbert}, {Sherwood-Taylor}, {Shih},
  {Sick}, {Silbiger}, {Singanamalla}, {Singer}, {Sladen}, {Sooley},
  {Sornarajah}, {Streicher}, {Teuben}, {Thomas}, {Tremblay}, {Turner},
  {Terr{\'o}n}, {van Kerkwijk}, {de la Vega}, {Watkins}, {Weaver}, {Whitmore},
  {Woillez}, {Zabalza}, \& {Astropy Contributors}}]{2018AJ....156..123A}
{Astropy Collaboration}, {Price-Whelan}, A.~M., {Sip{\H o}cz}, B.~M., {et~al.}
  2018, \aj, 156, 123

\bibitem[{{Bailer-Jones} {et~al.}(2018){Bailer-Jones}, {Rybizki}, {Fouesneau},
  {Mantelet}, \& {Andrae}}]{Bailer-Jones2018}
{Bailer-Jones}, C.~A.~L., {Rybizki}, J., {Fouesneau}, M., {Mantelet}, G., \&
  {Andrae}, R. 2018, VizieR Online Data Catalog, I/347

\bibitem[{{Battistini} \& {Bensby}(2016)}]{Battistini16}
{Battistini}, C. \& {Bensby}, T. 2016, \aap, 586, A49

\bibitem[{{Bensby} {et~al.}(2014){Bensby}, {Feltzing}, \& {Oey}}]{Bensby14}
{Bensby}, T., {Feltzing}, S., \& {Oey}, M.~S. 2014, \aap, 562, A71

\bibitem[{{Bi{\'e}mont} {et~al.}(2011){Bi{\'e}mont}, {Blagoev}, {Engstr{\"o}m},
  {Hartman}, {Lundberg}, {Malcheva}, {Nilsson}, {Whitehead}, {Palmeri}, \&
  {Quinet}}]{Biemont11}
{Bi{\'e}mont}, {\'E}., {Blagoev}, K., {Engstr{\"o}m}, L., {et~al.} 2011,
  \mnras, 414, 3350

\bibitem[{{Biemont} {et~al.}(1981){Biemont}, {Grevesse}, {Hannaford}, \&
  {Lowe}}]{Biemont81}
{Biemont}, E., {Grevesse}, N., {Hannaford}, P., \& {Lowe}, R.~M. 1981, \apj,
  248, 867

\bibitem[{{Bisterzo} {et~al.}(2015){Bisterzo}, {Gallino}, {K{\"a}ppeler},
  {Wiescher}, {Imbriani}, {Straniero}, {Cristallo}, {G{\"o}rres}, \&
  {deBoer}}]{Bisterzo15}
{Bisterzo}, S., {Gallino}, R., {K{\"a}ppeler}, F., {et~al.} 2015, \mnras, 449,
  506

\bibitem[{{Bisterzo} {et~al.}(2014){Bisterzo}, {Travaglio}, {Gallino},
  {Wiescher}, \& {K{\"a}ppeler}}]{Bisterzo14}
{Bisterzo}, S., {Travaglio}, C., {Gallino}, R., {Wiescher}, M., \&
  {K{\"a}ppeler}, F. 2014, \apj, 787, 10

\bibitem[{{Bisterzo} {et~al.}(2017){Bisterzo}, {Travaglio}, {Wiescher},
  {K{\"a}ppeler}, \& {Gallino}}]{Bisterzo17}
{Bisterzo}, S., {Travaglio}, C., {Wiescher}, M., {K{\"a}ppeler}, F., \&
  {Gallino}, R. 2017, \apj, 835, 97

\bibitem[{{Bovy}(2015)}]{Bovy15}
{Bovy}, J. 2015, \apjs, 216, 29

\bibitem[{{Burbidge} {et~al.}(1957){Burbidge}, {Burbidge}, {Fowler}, \&
  {Hoyle}}]{Burbidge57}
{Burbidge}, E.~M., {Burbidge}, G.~R., {Fowler}, W.~A., \& {Hoyle}, F. 1957,
  Reviews of Modern Physics, 29, 547

\bibitem[{{Busso} {et~al.}(1999){Busso}, {Gallino}, \& {Wasserburg}}]{Busso99}
{Busso}, M., {Gallino}, R., \& {Wasserburg}, G.~J. 1999, \araa, 37, 239

\bibitem[{{Casali} {et~al.}(2020){Casali}, {Spina}, {Magrini}, {Karakas},
  {Kobayashi}, {Casey}, {Feltzing}, {Van der Swaelmen}, {Tsantaki},
  {Jofr{\'e}}, {Bragaglia}, {Feuillet}, {Bensby}, {Biazzo}, {Gonneau},
  {Tautvai{\v{s}}ien{\.{e}}}, {Baratella}, {Roccatagliata}, {Pancino}, {Sousa},
  {Adibekyan}, {Martell}, {Bayo}, {Jackson}, {Jeffries}, {Gilmore}, {Randich},
  {Alfaro}, {Koposov}, {Korn}, {Recio-Blanco}, {Smiljanic}, {Franciosini},
  {Hourihane}, {Monaco}, {Morbidelli}, {Sacco}, {Worley}, \&
  {Zaggia}}]{Casali20}
{Casali}, G., {Spina}, L., {Magrini}, L., {et~al.} 2020, \aap, 639, A127

\bibitem[{{Clayton} \& {Rassbach}(1967)}]{Clayton67}
{Clayton}, D.~D. \& {Rassbach}, M.~E. 1967, \apj, 148, 69

\bibitem[{{Coc} \& {Vangioni}(2017)}]{Coc17}
{Coc}, A. \& {Vangioni}, E. 2017, International Journal of Modern Physics E,
  26, 1741002

\bibitem[{{Corliss} \& {Bozman}(1962)}]{Corliss62}
{Corliss}, C.~H. \& {Bozman}, W.~R. 1962, {Experimental transition
  probabilities for spectral lines of seventy elements; derived from the NBS
  Tables of spectral-line intensities}

\bibitem[{{C{\^o}t{\'e}} {et~al.}(2019){C{\^o}t{\'e}}, {Eichler}, {Arcones},
  {Hansen}, {Simonetti}, {Frebel}, {Fryer}, {Pignatari}, {Reichert},
  {Belczynski}, \& {Matteucci}}]{Cote19}
{C{\^o}t{\'e}}, B., {Eichler}, M., {Arcones}, A., {et~al.} 2019, \apj, 875, 106

\bibitem[{{C{\^o}t{\'e}} {et~al.}(2018){C{\^o}t{\'e}}, {Fryer}, {Belczynski},
  {Korobkin}, {Chru{\'s}li{\'n}ska}, {Vassh}, {Mumpower}, {Lippuner},
  {Sprouse}, {Surman}, \& {Wollaeger}}]{Cote18}
{C{\^o}t{\'e}}, B., {Fryer}, C.~L., {Belczynski}, K., {et~al.} 2018, \apj, 855,
  99

\bibitem[{{Cowan} \& {Rose}(1977)}]{Cowan77}
{Cowan}, J.~J. \& {Rose}, W.~K. 1977, \apj, 212, 149

\bibitem[{{Cowan} {et~al.}(2019){Cowan}, {Sneden}, {Lawler}, {Aprahamian},
  {Wiescher}, {Langanke}, {Mart{\'\i}nez-Pinedo}, \& {Thielemann}}]{Cowan19}
{Cowan}, J.~J., {Sneden}, C., {Lawler}, J.~E., {et~al.} 2019, arXiv e-prints,
  arXiv:1901.01410

\bibitem[{{Cowley} \& {Corliss}(1983)}]{Cowley83}
{Cowley}, C.~R. \& {Corliss}, C.~H. 1983, \mnras, 203, 651

\bibitem[{{Cristallo} {et~al.}(2015){Cristallo}, {Abia}, {Straniero}, \&
  {Piersanti}}]{Cristallo15}
{Cristallo}, S., {Abia}, C., {Straniero}, O., \& {Piersanti}, L. 2015, \apj,
  801, 53

\bibitem[{{Cunha} {et~al.}(2017){Cunha}, {Smith}, {Hasselquist}, {Souto},
  {Shetrone}, {Allende Prieto}, {Bizyaev}, {Frinchaboy},
  {Garc{\'\i}a-Hern{\'a}ndez}, {Holtzman}, {Johnson}, {J{\H{o}}nsson},
  {Majewski}, {M{\'e}sz{\'a}ros}, {Nidever}, {Pinsonneault}, {Schiavon},
  {Sobeck}, {Skrutskie}, {Zamora}, {Zasowski}, \&
  {Fern{\'a}ndez-Trincado}}]{Cunha17}
{Cunha}, K., {Smith}, V.~V., {Hasselquist}, S., {et~al.} 2017, \apj, 844, 145

\bibitem[{{Cutri} \& {et al.}(2014)}]{Cutri14}
{Cutri}, R.~M. \& {et al.} 2014, VizieR Online Data Catalog, II/328

\bibitem[{{da Silva} {et~al.}(2016){da Silva}, {Lemasle}, {Bono}, {Genovali},
  {McWilliam}, {Cristallo}, {Bergemann}, {Buonanno}, {Fabrizio}, {Ferraro},
  {Fran{\c{c}}ois}, {Iannicola}, {Inno}, {Laney}, {Kudritzki}, {Matsunaga},
  {Nonino}, {Primas}, {Przybilla}, {Romaniello}, {Th{\'e}venin}, \&
  {Urbaneja}}]{daSilva16}
{da Silva}, R., {Lemasle}, B., {Bono}, G., {et~al.} 2016, \aap, 586, A125

\bibitem[{{Da Silva} {et~al.}(2012){Da Silva}, {Porto de Mello}, {Milone}, {da
  Silva}, {Ribeiro}, \& {Rocha-Pinto}}]{dasilva12}
{Da Silva}, R., {Porto de Mello}, G.~F., {Milone}, A.~C., {et~al.} 2012, \aap,
  542, A84

\bibitem[{{Dalton} {et~al.}(2016){Dalton}, {Trager}, {Abrams}, {Bonifacio},
  {Aguerri}, {Middleton}, {Benn}, {Dee}, {Say{\`e}de}, {Lewis}, {Pragt},
  {Pico}, {Walton}, {Rey}, {Allende Prieto}, {Pe{\~n}ate}, {Lhome},
  {Ag{\'o}cs}, {Alonso}, {Terrett}, {Brock}, {Gilbert}, {Schallig}, {Ridings},
  {Guinouard}, {Verheijen}, {Tosh}, {Rogers}, {Lee}, {Steele}, {Stuik},
  {Tromp}, {Jask{\'o}}, {Carrasco}, {Farcas}, {Kragt}, {Lesman}, {Kroes},
  {Mottram}, {Bates}, {Rodriguez}, {Gribbin}, {Delgado}, {Herreros}, {Martin},
  {Cano}, {Navarro}, {Irwin}, {Lewis}, {Gonzalez Solares}, {Murphy}, {Worley},
  {Bassom}, {O'Mahoney}, {Bianco}, {Zurita}, {ter Horst}, {Molinari}, {Lodi},
  {Guerra}, {Martin}, {Vallenari}, {Salasnich}, {Baruffolo}, {Jin}, {Hill},
  {Smith}, {Drew}, {Poggianti}, {Pieri}, {Dominquez Palmero}, \&
  {Farina}}]{Dalton16}
{Dalton}, G., {Trager}, S., {Abrams}, D.~C., {et~al.} 2016, in Society of
  Photo-Optical Instrumentation Engineers (SPIE) Conference Series, Vol. 9908,
  \procspie, 99081G

\bibitem[{{Davidson} {et~al.}(1992){Davidson}, {Snoek}, {Volten}, \&
  {Doenszelmann}}]{Davidson92}
{Davidson}, M.~D., {Snoek}, L.~C., {Volten}, H., \& {Doenszelmann}, A. 1992,
  \aap, 255, 457

\bibitem[{{de Jong} {et~al.}(2019){de Jong}, {Agertz}, {Berbel}, {Aird},
  {Alexander}, {Amarsi}, {Anders}, {Andrae}, {Ansarinejad}, {Ansorge},
  {Antilogus}, {Anwand-Heerwart}, {Arentsen}, {Arnadottir}, {Asplund}, {Auger},
  {Azais}, {Baade}, {Baker}, {Baker}, {Balbinot}, {Baldry}, {Banerji},
  {Barden}, {Barklem}, {Barth{\'e}l{\'e}my-Mazot}, {Battistini}, {Bauer},
  {Bell}, {Bellido-Tirado}, {Bellstedt}, {Belokurov}, {Bensby}, {Bergemann},
  {Bestenlehner}, {Bielby}, {Bilicki}, {Blake}, {Bland-Hawthorn}, {Boeche},
  {Boland}, {Boller}, {Bongard}, {Bongiorno}, {Bonifacio}, {Boudon}, {Brooks},
  {Brown}, {Brown}, {Br{\"u}ggen}, {Brynnel}, {Brzeski}, {Buchert},
  {Buschkamp}, {Caffau}, {Caillier}, {Carrick}, {Casagrande}, {Case}, {Casey},
  {Cesarini}, {Cescutti}, {Chapuis}, {Chiappini}, {Childress}, {Christlieb},
  {Church}, {Cioni}, {Cluver}, {Colless}, {Collett}, {Comparat}, {Cooper},
  {Couch}, {Courbin}, {Croom}, {Croton}, {Daguis{\'e}}, {Dalton}, {Davies},
  {Davis}, {de Laverny}, {Deason}, {Dionies}, {Disseau}, {Doel}, {D{\"o}scher},
  {Driver}, {Dwelly}, {Eckert}, {Edge}, {Edvardsson}, {Youssoufi}, {Elhaddad},
  {Enke}, {Erfanianfar}, {Farrell}, {Fechner}, {Feiz}, {Feltzing}, {Ferreras},
  {Feuerstein}, {Feuillet}, {Finoguenov}, {Ford}, {Fotopoulou}, {Fouesneau},
  {Frenk}, {Frey}, {Gaessler}, {Geier}, {Fusillo}, {Gerhard}, {Giannantonio},
  {Giannone}, {Gibson}, {Gillingham}, {Gonz{\'a}lez-Fern{\'a}ndez},
  {Gonzalez-Solares}, {Gottloeber}, {Gould}, {Grebel}, {Gueguen}, {Guiglion},
  {Haehnelt}, {Hahn}, {Hansen}, {Hartman}, {Hauptner}, {Hawkins}, {Haynes},
  {Haynes}, {Heiter}, {Helmi}, {Aguayo}, {Hewett}, {Hinton}, {Hobbs}, {Hoenig},
  {Hofman}, {Hook}, {Hopgood}, {Hopkins}, {Hourihane}, {Howes}, {Howlett},
  {Huet}, {Irwin}, {Iwert}, {Jablonka}, {Jahn}, {Jahnke}, {Jarno}, {Jin},
  {Jofre}, {Johl}, {Jones}, {J{\"o}nsson}, {Jordan}, {Karovicova}, {Khalatyan},
  {Kelz}, {Kennicutt}, {King}, {Kitaura}, {Klar}, {Klauser}, {Kneib}, {Koch},
  {Koposov}, {Kordopatis}, {Korn}, {Kosmalski}, {Kotak}, {Kovalev}, {Kreckel},
  {Kripak}, {Krumpe}, {Kuijken}, {Kunder}, {Kushniruk}, {Lam}, {Lamer},
  {Laurent}, {Lawrence}, {Lehmitz}, {Lemasle}, {Lewis}, {Li}, {Lidman}, {Lind},
  {Liske}, {Lizon}, {Loveday}, {Ludwig}, {McDermid}, {Maguire}, {Mainieri},
  {Mali}, {Mandel}, {Mandel}, {Mannering}, {Martell}, {Martinez Delgado},
  {Matijevic}, {McGregor}, {McMahon}, {McMillan}, {Mena}, {Merloni}, {Meyer},
  {Michel}, {Micheva}, {Migniau}, {Minchev}, {Monari}, {Muller}, {Murphy},
  {Muthukrishna}, {Nandra}, {Navarro}, {Ness}, {Nichani}, {Nichol}, {Nicklas},
  {Niederhofer}, {Norberg}, {Obreschkow}, {Oliver}, {Owers}, {Pai},
  {Pankratow}, {Parkinson}, {Paschke}, {Paterson}, {Pecontal}, {Parry},
  {Phillips}, {Pillepich}, {Pinard}, {Pirard}, {Piskunov}, {Plank},
  {Pl{\"u}schke}, {Pons}, {Popesso}, {Power}, {Pragt}, {Pramskiy}, {Pryer},
  {Quattri}, {Queiroz}, {Quirrenbach}, {Rahurkar}, {Raichoor}, {Ramstedt},
  {Rau}, {Recio-Blanco}, {Reiss}, {Renaud}, {Revaz}, {Rhode}, {Richard},
  {Richter}, {Rix}, {Robotham}, {Roelfsema}, {Romaniello}, {Rosario},
  {Rothmaier}, {Roukema}, {Ruchti}, {Rupprecht}, {Rybizki}, {Ryde}, {Saar},
  {Sadler}, {Sahl{\'e}n}, {Salvato}, {Sassolas}, {Saunders}, {Saviauk},
  {Sbordone}, {Schmidt}, {Schnurr}, {Scholz}, {Schwope}, {Seifert}, {Shanks},
  {Sheinis}, {Sivov}, {Sk{\'u}lad{\'o}ttir}, {Smartt}, {Smedley}, {Smith},
  {Smith}, {Sorce}, {Spitler}, {Starkenburg}, {Steinmetz}, {Stilz}, {Storm},
  {Sullivan}, {Sutherland}, {Swann}, {Tamone}, {Taylor}, {Teillon}, {Tempel},
  {ter Horst}, {Thi}, {Tolstoy}, {Trager}, {Traven}, {Tremblay}, {Tresse},
  {Valentini}, {van de Weygaert}, {van den Ancker}, {Veljanoski}, {Venkatesan},
  {Wagner}, {Wagner}, {Walcher}, {Waller}, {Walton}, {Wang}, {Winkler},
  {Wisotzki}, {Worley}, {Worseck}, {Xiang}, {Xu}, {Yong}, {Zhao}, {Zheng},
  {Zscheyge}, \& {Zucker}}]{deJong2019}
{de Jong}, R.~S., {Agertz}, O., {Berbel}, A.~A., {et~al.} 2019, The Messenger,
  175, 3

\bibitem[{{Delgado Mena} {et~al.}(2018){Delgado Mena}, {Adibekyan}, {Figueira},
  {Gonz{\'a}lez Hern{\'a}ndez}, {Santos}, {Tsantaki}, {Sousa}, {Faria}, {Su{\'a
  }rez-Andr{\'e}s}, \& {Israelian}}]{DelgadoMena18}
{Delgado Mena}, E., {Adibekyan}, V.~Z., {Figueira}, P., {et~al.} 2018,
  Publications of the Astronomical Society of the Pacific, 130, 094202

\bibitem[{{Delgado Mena} {et~al.}(2019){Delgado Mena}, {Moya}, {Adibekyan},
  {Tsantaki}, {Gonz{\'a}lez Hern{\'a}ndez}, {Israelian}, {Davies}, {Chaplin},
  {Sousa}, {Ferreira}, \& {Santos}}]{Delgado19}
{Delgado Mena}, E., {Moya}, A., {Adibekyan}, V., {et~al.} 2019, \aap, 624, A78

\bibitem[{{Delgado Mena} {et~al.}(2017){Delgado Mena}, {Tsantaki}, {Adibekyan},
  {Sousa}, {Santos}, {Gonz{\'a}lez Hern{\'a}ndez}, \& {Israelian}}]{Delgado17}
{Delgado Mena}, E., {Tsantaki}, M., {Adibekyan}, V.~Z., {et~al.} 2017, \aap,
  606, A94

\bibitem[{{Den Hartog} {et~al.}(2003){Den Hartog}, {Lawler}, {Sneden}, \&
  {Cowan}}]{DenHartog03}
{Den Hartog}, E.~A., {Lawler}, J.~E., {Sneden}, C., \& {Cowan}, J.~J. 2003,
  \apjs, 148, 543

\bibitem[{{D'Orazi} {et~al.}(2009){D'Orazi}, {Magrini}, {Randich}, {Galli},
  {Busso}, \& {Sestito}}]{D'Orazi09}
{D'Orazi}, V., {Magrini}, L., {Randich}, S., {et~al.} 2009, \apjl, 693, L31

\bibitem[{{Duong} {et~al.}(2018){Duong}, {Freeman}, {Asplund}, {Casagrande},
  {Buder}, {Lind}, {Ness}, {Bland-Hawthorn}, {De Silva}, {D'Orazi}, {Kos},
  {Lewis}, {Lin}, {Martell}, {Schlesinger}, {Sharma}, {Simpson}, {Zucker},
  {Zwitter}, {Anguiano}, {Da Costa}, {Hyde}, {Horner}, {Kafle}, {Nataf},
  {Reid}, {Stello}, {Ting}, \& {Wyse}}]{Duong18}
{Duong}, L., {Freeman}, K.~C., {Asplund}, M., {et~al.} 2018, \mnras, 476, 5216

\bibitem[{{Edvardsson} {et~al.}(1993){Edvardsson}, {Andersen}, {Gustafsson},
  {Lambert}, {Nissen}, \& {Tomkin}}]{Edvardsson93}
{Edvardsson}, B., {Andersen}, J., {Gustafsson}, B., {et~al.} 1993, \aap, 500,
  391

\bibitem[{{Feltzing} {et~al.}(2017){Feltzing}, {Howes}, {McMillan}, \&
  {Stonkut{\.e}}}]{Feltzing17}
{Feltzing}, S., {Howes}, L.~M., {McMillan}, P.~J., \& {Stonkut{\.e}}, E. 2017,
  \mnras, 465, L109

\bibitem[{{Forsberg} {et~al.}(2019){Forsberg}, {J{\"o}nsson}, {Ryde}, \&
  {Matteucci}}]{Forsberg19}
{Forsberg}, R., {J{\"o}nsson}, H., {Ryde}, N., \& {Matteucci}, F. 2019, \aap,
  631, A113

\bibitem[{{Freiburghaus} {et~al.}(1999){Freiburghaus}, {Rosswog}, \&
  {Thielemann}}]{Freiburghaus99}
{Freiburghaus}, C., {Rosswog}, S., \& {Thielemann}, F.-K. 1999, \apjl, 525,
  L121

\bibitem[{{Gaia Collaboration} {et~al.}(2018){Gaia Collaboration}, {Brown},
  {Vallenari}, {Prusti}, {de Bruijne}, {Babusiaux}, {Bailer-Jones}, {Biermann},
  {Evans}, {Eyer}, \& et~al.}]{Gaia18}
{Gaia Collaboration}, {Brown}, A.~G.~A., {Vallenari}, A., {et~al.} 2018, \aap,
  616, A1

\bibitem[{{Gaia Collaboration} {et~al.}(2016){Gaia Collaboration}, {Prusti},
  {de Bruijne}, {Brown}, {Vallenari}, {Babusiaux}, {Bailer-Jones}, {Bastian},
  {Biermann}, {Evans}, \& et~al.}]{Gaia16}
{Gaia Collaboration}, {Prusti}, T., {de Bruijne}, J.~H.~J., {et~al.} 2016,
  \aap, 595, A1

\bibitem[{{Gallino} {et~al.}(1998){Gallino}, {Arlandini}, {Busso}, {Lugaro},
  {Travaglio}, {Straniero}, {Chieffi}, \& {Limongi}}]{Gallino98}
{Gallino}, R., {Arlandini}, C., {Busso}, M., {et~al.} 1998, \apj, 497, 388

\bibitem[{{Gratton} \& {Sneden}(1994)}]{Gratton94}
{Gratton}, R.~G. \& {Sneden}, C. 1994, \aap, 287, 927

\bibitem[{{Grevesse} {et~al.}(2007){Grevesse}, {Asplund}, \&
  {Sauval}}]{Grevesse07}
{Grevesse}, N., {Asplund}, M., \& {Sauval}, A.~J. 2007, \ssr, 130, 105

\bibitem[{{Griffith} {et~al.}(2019){Griffith}, {Johnson}, \&
  {Weinberg}}]{Griffith19}
{Griffith}, E., {Johnson}, J.~A., \& {Weinberg}, D.~H. 2019, \apj, 886, 84

\bibitem[{{Grisoni} {et~al.}(2020){Grisoni}, {Cescutti}, {Matteucci},
  {Forsberg}, {J{\"o}nsson}, \& {Ryde}}]{Grisoni20}
{Grisoni}, V., {Cescutti}, G., {Matteucci}, F., {et~al.} 2020, \mnras, 492,
  2828

\bibitem[{{Guiglion} {et~al.}(2018){Guiglion}, {de Laverny}, {Recio-Blanco}, \&
  {Prantzos}}]{Guiglion18}
{Guiglion}, G., {de Laverny}, P., {Recio-Blanco}, A., \& {Prantzos}, N. 2018,
  \aap, 619, A143

\bibitem[{{Gustafsson} {et~al.}(2008){Gustafsson}, {Edvardsson}, {Eriksson},
  {J{\o}rgensen}, {Nordlund}, \& {Plez}}]{Gustafsson08}
{Gustafsson}, B., {Edvardsson}, B., {Eriksson}, K., {et~al.} 2008, \aap, 486,
  951

\bibitem[{{Hansen} {et~al.}(2018){Hansen}, {Holmbeck}, {Beers}, {Placco},
  {Roederer}, {Frebel}, {Sakari}, {Simon}, \& {Thompson}}]{Hansen18}
{Hansen}, T.~T., {Holmbeck}, E.~M., {Beers}, T.~C., {et~al.} 2018, \apj, 858,
  92

\bibitem[{{Haynes} \& {Kobayashi}(2019)}]{Haynes19}
{Haynes}, C.~J. \& {Kobayashi}, C. 2019, \mnras, 483, 5123

\bibitem[{{Heiter} {et~al.}(2021){Heiter}, {Lind}, {Bergemann}, {Asplund},
  {Mikolaitis}, {Barklem}, {Masseron}, {de Laverny}, {Magrini}, {Edvardsson},
  {J{\"o}nsson}, {Pickering}, {Ryde}, {Bayo Ar{\'a}n}, {Bensby}, {Casey},
  {Feltzing}, {Jofr{\'e}}, {Korn}, {Pancino}, {Damiani}, {Lanzafame}, {Lardo},
  {Monaco}, {Morbidelli}, {Smiljanic}, {Worley}, {Zaggia}, {Randich}, \&
  {Gilmore}}]{Heiter2021}
{Heiter}, U., {Lind}, K., {Bergemann}, M., {et~al.} 2021, \aap, 645, A106

\bibitem[{Ivarsson {et~al.}(2001)Ivarsson, Litzen, \& Wahlgren}]{Ivarsson01}
Ivarsson, S., Litzen, U., \& Wahlgren, G.~M. 2001, Physica Scripta, 64, 455

\bibitem[{{Jacobson} \& {Friel}(2013)}]{Jacobson13}
{Jacobson}, H.~R. \& {Friel}, E.~D. 2013, \aj, 145, 107

\bibitem[{{J{\"o}nsson} {et~al.}(2017){J{\"o}nsson}, {Ryde}, {Nordlander},
  {Pehlivan Rhodin}, {Hartman}, {J{\"o}nsson}, \& {Eriksson}}]{Jonsson17}
{J{\"o}nsson}, H., {Ryde}, N., {Nordlander}, T., {et~al.} 2017, \aap, 598, A100

\bibitem[{{Jurgenson} {et~al.}(2016){Jurgenson}, {Fischer}, {McCracken},
  {Sawyer}, {Giguere}, {Szymkowiak}, {Santoro}, \& {Muller}}]{Jurgenson16}
{Jurgenson}, C., {Fischer}, D., {McCracken}, T., {et~al.} 2016, Journal of
  Astronomical Instrumentation, 5, 1650003

\bibitem[{{Jurgenson} {et~al.}(2014){Jurgenson}, {Fischer}, {McCracken},
  {Stoll}, {Szymkowiak}, {Giguere}, {Santoro}, \&
  {Muller}}]{2014SPIE.9147E..7FJ}
{Jurgenson}, C.~A., {Fischer}, D.~A., {McCracken}, T.~M., {et~al.} 2014, in
  \procspie, Vol. 9147, Ground-based and Airborne Instrumentation for Astronomy
  V, 91477F

\bibitem[{{Kappeler} {et~al.}(1989){Kappeler}, {Beer}, \&
  {Wisshak}}]{Kappeler89}
{Kappeler}, F., {Beer}, H., \& {Wisshak}, K. 1989, Reports on Progress in
  Physics, 52, 945

\bibitem[{{K{\"a}ppeler} {et~al.}(2011){K{\"a}ppeler}, {Gallino}, {Bisterzo},
  \& {Aoki}}]{Kappeler11}
{K{\"a}ppeler}, F., {Gallino}, R., {Bisterzo}, S., \& {Aoki}, W. 2011, Reviews
  of Modern Physics, 83, 157

\bibitem[{{Karakas}(2014)}]{Karakas14}
{Karakas}, A.~I. 2014, in IAU Symposium, Vol. 298, Setting the scene for Gaia
  and LAMOST, ed. S.~{Feltzing}, G.~{Zhao}, N.~A. {Walton}, \& P.~{Whitelock},
  142--153

\bibitem[{{Katz} {et~al.}(2018){Katz}, {Sartoretti}, {Cropper}, {Panuzzo},
  {Seabroke}, {Viala}, {Benson}, {Blomme}, {Jasniewicz}, {Jean-Antoine},
  {Huckle}, {Smith}, {Baker}, {Crifo}, {Damerdji}, {David}, {Dolding},
  {Fr{\'e}mat}, {Gosset}, {Guerrier}, {Guy}, {Haigron}, {Jan{\ss}en},
  {Marchal}, {Plum}, {Soubiran}, {Th{\'e}venin}, {Ajaj}, {Allende Prieto},
  {Babusiaux}, {Boudreault}, {Chemin}, {Delle Luche}, {Fabre}, {Gueguen},
  {Hambly}, {Lasne}, {Meynadier}, {Pailler}, {Panem}, {Royer}, {Tauran},
  {Zurbach}, {Zwitter}, {Arenou}, {Bossini}, {Gomez}, {Lemaitre}, {Leclerc},
  {Morel}, {Munari}, {Turon}, {Vallenari}, \& {{\v Z}erjal}}]{Katz18}
{Katz}, D., {Sartoretti}, P., {Cropper}, M., {et~al.} 2018, arXiv e-prints
  [\eprint[arXiv]{1804.09372}]

\bibitem[{{Kobayashi} {et~al.}(2020){Kobayashi}, {Karakas}, \&
  {Lugaro}}]{Kobayashi20}
{Kobayashi}, C., {Karakas}, A.~I., \& {Lugaro}, M. 2020, \apj, 900, 179

\bibitem[{{Korotin} {et~al.}(2015){Korotin}, {Andrievsky}, {Hansen}, {Caffau},
  {Bonifacio}, {Spite}, {Spite}, \& {Fran{\c{c}}ois}}]{Korotin15}
{Korotin}, S.~A., {Andrievsky}, S.~M., {Hansen}, C.~J., {et~al.} 2015, \aap,
  581, A70

\bibitem[{{Lamb} {et~al.}(1977){Lamb}, {Howard}, {Truran}, \& {Iben}}]{Lamb77}
{Lamb}, S.~A., {Howard}, W.~M., {Truran}, J.~W., \& {Iben}, Jr., I. 1977, \apj,
  217, 213

\bibitem[{{Lawler} {et~al.}(2001{\natexlab{a}}){Lawler}, {Bonvallet}, \&
  {Sneden}}]{Lawler01La}
{Lawler}, J.~E., {Bonvallet}, G., \& {Sneden}, C. 2001{\natexlab{a}}, \apj,
  556, 452

\bibitem[{{Lawler} {et~al.}(2006){Lawler}, {Den Hartog}, {Sneden}, \&
  {Cowan}}]{Lawler06}
{Lawler}, J.~E., {Den Hartog}, E.~A., {Sneden}, C., \& {Cowan}, J.~J. 2006,
  \apjs, 162, 227

\bibitem[{{Lawler} {et~al.}(2009){Lawler}, {Sneden}, {Cowan}, {Ivans}, \& {Den
  Hartog}}]{Lawler09}
{Lawler}, J.~E., {Sneden}, C., {Cowan}, J.~J., {Ivans}, I.~I., \& {Den Hartog},
  E.~A. 2009, \apjs, 182, 51

\bibitem[{{Lawler} {et~al.}(2001{\natexlab{b}}){Lawler}, {Wickliffe}, {den
  Hartog}, \& {Sneden}}]{Lawler01Eu}
{Lawler}, J.~E., {Wickliffe}, M.~E., {den Hartog}, E.~A., \& {Sneden}, C.
  2001{\natexlab{b}}, \apj, 563, 1075

\bibitem[{{Li} {et~al.}(2018){Li}, {Zhao}, {Zhai}, \& {Jia}}]{Li18}
{Li}, C., {Zhao}, G., {Zhai}, M., \& {Jia}, Y. 2018, \apj, 860, 53

\bibitem[{{Li} {et~al.}(2007){Li}, {Chatelain}, {Holt}, {Rehse}, {Rosner}, \&
  {Scholl}}]{Li07}
{Li}, R., {Chatelain}, R., {Holt}, R.~A., {et~al.} 2007, \physscr, 76, 577

\bibitem[{{Liu} {et~al.}(2020){Liu}, {Shi}, \& {Wu}}]{Liu20}
{Liu}, S., {Shi}, J., \& {Wu}, Z. 2020, \apj, 896, 64

\bibitem[{{Luck}(2017)}]{Luck17}
{Luck}, R.~E. 2017, \aj, 153, 21

\bibitem[{{Luck}(2018{\natexlab{a}})}]{Luck18a}
{Luck}, R.~E. 2018{\natexlab{a}}, \aj, 155, 111

\bibitem[{{Luck}(2018{\natexlab{b}})}]{Luck18b}
{Luck}, R.~E. 2018{\natexlab{b}}, \aj, 156, 171

\bibitem[{{Luri} {et~al.}(2018){Luri}, {Brown}, {Sarro}, {Arenou},
  {Bailer-Jones}, {Castro-Ginard}, {de Bruijne}, {Prusti}, {Babusiaux}, \&
  {Delgado}}]{Luri18}
{Luri}, X., {Brown}, A.~G.~A., {Sarro}, L.~M., {et~al.} 2018, \aap, 616, A9

\bibitem[{{Magrini} {et~al.}(2018){Magrini}, {Spina}, {Randich}, {Friel},
  {Kordopatis}, {Worley}, {Pancino}, {Bragaglia}, {Donati}, {Tautvai{\v
  s}ien{\.e}}, {Bagdonas}, {Delgado-Mena}, {Adibekyan}, {Sousa},
  {Jim{\'e}nez-Esteban}, {Sanna}, {Roccatagliata}, {Bonito}, {Sbordone},
  {Duffau}, {Gilmore}, {Feltzing}, {Jeffries}, {Vallenari}, {Alfaro}, {Bensby},
  {Francois}, {Koposov}, {Korn}, {Recio-Blanco}, {Smiljanic}, {Bayo},
  {Carraro}, {Casey}, {Costado}, {Damiani}, {Franciosini}, {Frasca},
  {Hourihane}, {Jofr{\'e}}, {de Laverny}, {Lewis}, {Masseron}, {Monaco},
  {Morbidelli}, {Prisinzano}, {Sacco}, \& {Zaggia}}]{Magrini18}
{Magrini}, L., {Spina}, L., {Randich}, S., {et~al.} 2018, \aap, 617, A106

\bibitem[{{Magrini} {et~al.}(2021){Magrini}, {Vescovi}, {Casali}, {Cristallo},
  {Viscasillas V{\'a}zquez}, {Cescutti}, {Spina}, {Van Der Swaelmen}, \&
  {Randich}}]{Magrini21}
{Magrini}, L., {Vescovi}, D., {Casali}, G., {et~al.} 2021, \aap, 646, L2

\bibitem[{{Maiorca} {et~al.}(2012){Maiorca}, {Magrini}, {Busso}, {Randich},
  {Palmerini}, \& {Trippella}}]{Maiorca12}
{Maiorca}, E., {Magrini}, L., {Busso}, M., {et~al.} 2012, \apj, 747, 53

\bibitem[{{Marsakov} {et~al.}(2016){Marsakov}, {Gozha}, {Koval'}, \&
  {Shpigel'}}]{Marsakov16}
{Marsakov}, V.~A., {Gozha}, M.~L., {Koval'}, V.~V., \& {Shpigel'}, L.~V. 2016,
  Astronomy Reports, 60, 61

\bibitem[{{Martell} {et~al.}(2017){Martell}, {Sharma}, {Buder}, {Duong},
  {Schlesinger}, {Simpson}, {Lind}, {Ness}, {Marshall}, {Asplund},
  {Bland-Hawthorn}, {Casey}, {De Silva}, {Freeman}, {Kos}, {Lin}, {Zucker},
  {Zwitter}, {Anguiano}, {Bacigalupo}, {Carollo}, {Casagrande}, {Da Costa},
  {Horner}, {Huber}, {Hyde}, {Kafle}, {Lewis}, {Nataf}, {Navin}, {Stello},
  {Tinney}, {Watson}, \& {Wittenmyer}}]{Martell17}
{Martell}, S.~L., {Sharma}, S., {Buder}, S., {et~al.} 2017, \mnras, 465, 3203

\bibitem[{{Mashonkina} \& {Gehren}(2001)}]{Mashonkina_2001}
{Mashonkina}, L. \& {Gehren}, T. 2001, \aap, 376, 232

\bibitem[{{Mashonkina} {et~al.}(2003){Mashonkina}, {Gehren}, {Travaglio}, \&
  {Borkova}}]{Mashonkina_2003}
{Mashonkina}, L., {Gehren}, T., {Travaglio}, C., \& {Borkova}, T. 2003, \aap,
  397, 275

\bibitem[{Matteucci {et~al.}(2014)Matteucci, Romano, Arcones, Korobkin, \&
  Rosswog}]{Matteucci14}
Matteucci, F., Romano, D., Arcones, A., Korobkin, O., \& Rosswog, S. 2014,
  Monthly Notices of the Royal Astronomical Society, 438, 2177–2185

\bibitem[{{Meggers} {et~al.}(1975){Meggers}, {Corliss}, \&
  {Scribner}}]{Meggers75}
{Meggers}, W.~F., {Corliss}, C.~H., \& {Scribner}, B.~F. 1975, {Tables of
  spectral-line intensities. Part I, II\_- arranged by elements.}

\bibitem[{{Miglio} {et~al.}(2017){Miglio}, {Chiappini}, {Mosser}, {Davies},
  {Freeman}, {Girardi}, {Jofr{\'e}}, {Kawata}, {Rendle}, {Valentini},
  {Casagrande}, {Chaplin}, {Gilmore}, {Hawkins}, {Holl}, {Appourchaux},
  {Belkacem}, {Bossini}, {Brogaard}, {Goupil}, {Montalb{\'a}n}, {Noels},
  {Anders}, {Rodrigues}, {Piotto}, {Pollacco}, {Rauer}, {Allende Prieto},
  {Avelino}, {Babusiaux}, {Barban}, {Barbuy}, {Basu}, {Baudin}, {Benomar},
  {Bienaym{\'e}}, {Binney}, {Bland-Hawthorn}, {Bressan}, {Cacciari},
  {Campante}, {Cassisi}, {Christensen-Dalsgaard}, {Combes}, {Creevey}, {Cunha},
  {Jong}, {Laverny}, {Degl'Innocenti}, {Deheuvels}, {Depagne}, {Ridder}, {Di
  Matteo}, {Di Mauro}, {Dupret}, {Eggenberger}, {Elsworth}, {Famaey},
  {Feltzing}, {Garc{\'{\i}}a}, {Gerhard}, {Gibson}, {Gizon}, {Haywood},
  {Handberg}, {Heiter}, {Hekker}, {Huber}, {Ibata}, {Katz}, {Kawaler},
  {Kjeldsen}, {Kurtz}, {Lagarde}, {Lebreton}, {Lund}, {Majewski}, {Marigo},
  {Martig}, {Mathur}, {Minchev}, {Morel}, {Ortolani}, {Pinsonneault}, {Plez},
  {Prada Moroni}, {Pricopi}, {Recio-Blanco}, {Reyl{\'e}}, {Robin}, {Roxburgh},
  {Salaris}, {Santiago}, {Schiavon}, {Serenelli}, {Sharma}, {Silva Aguirre},
  {Soubiran}, {Steinmetz}, {Stello}, {Strassmeier}, {Ventura}, {Ventura},
  {Walton}, \& {Worley}}]{Miglio17}
{Miglio}, A., {Chiappini}, C., {Mosser}, B., {et~al.} 2017, Astronomische
  Nachrichten, 338, 644

\bibitem[{{Mikolaitis} {et~al.}(2019){Mikolaitis}, {Drazdauskas}, {Minkevi{\v
  c}i{\= u}t{\. e}}, {Stonkut{\. e}}, {Tautvai{\v s}ien{\. e}}, {Klebonas},
  {Bagdonas}, {Pak{\v s}tien{\. e}}, \& {Janulis}}]{Mikolaitis19}
{Mikolaitis}, {\v S.}., {Drazdauskas}, A., {Minkevi{\v c}i{\= u}t{\. e}}, R.,
  {et~al.} 2019, A\&A, doi:10.1051/0004-6361/201835004, (Paper II)

\bibitem[{{Mikolaitis} {et~al.}(2018){Mikolaitis}, {Tautvai{\v s}ien{\.e}},
  {Drazdauskas}, {Minkevi{\v c}i{\= u}t{\.e}}, {Klebonas}, {Bagdonas}, {Pak{\v
  s}tien{\.e}}, \& {Janulis}}]{Mikolaitis18}
{Mikolaitis}, {\v S}., {Tautvai{\v s}ien{\.e}}, G., {Drazdauskas}, A., {et~al.}
  2018, \pasp, 130, 074202 (Paper I)

\bibitem[{{Miles} \& {Wiese}(1969)}]{Miles69}
{Miles}, B.~M. \& {Wiese}, W.~L. 1969, Atomic Data, 1, 1

\bibitem[{{Mints} \& {Hekker}(2017)}]{2017A&A...604A.108M}
{Mints}, A. \& {Hekker}, S. 2017, \aap, 604, A108

\bibitem[{{Mishenina} {et~al.}(2015){Mishenina}, {Pignatari}, {Carraro},
  {Kovtyukh}, {Monaco}, {Korotin}, {Shereta}, {Yegorova}, \&
  {Herwig}}]{Mishenina15}
{Mishenina}, T., {Pignatari}, M., {Carraro}, G., {et~al.} 2015, \mnras, 446,
  3651

\bibitem[{{Mishenina} {et~al.}(2019){Mishenina}, {Pignatari}, {Gorbaneva},
  {Bisterzo}, {Travaglio}, {Thielemann}, \& {Soubiran}}]{Mishenina19}
{Mishenina}, T., {Pignatari}, M., {Gorbaneva}, T., {et~al.} 2019, \mnras, 484,
  3846

\bibitem[{{Mishenina} {et~al.}(2013){Mishenina}, {Pignatari}, {Korotin},
  {Soubiran}, {Charbonnel}, {Thielemann}, {Gorbaneva}, \&
  {Basak}}]{Mishenina13}
{Mishenina}, T.~V., {Pignatari}, M., {Korotin}, S.~A., {et~al.} 2013, \aap,
  552, A128

\bibitem[{{Nishimura} {et~al.}(2006){Nishimura}, {Kotake}, {Hashimoto},
  {Yamada}, {Nishimura}, {Fujimoto}, \& {Sato}}]{Nishimura06}
{Nishimura}, S., {Kotake}, K., {Hashimoto}, M.-a., {et~al.} 2006, \apj, 642,
  410

\bibitem[{{Nissen}(2015)}]{Nissen15}
{Nissen}, P.~E. 2015, \aap, 579, A52

\bibitem[{{Nissen}(2016)}]{Nissen16}
{Nissen}, P.~E. 2016, \aap, 593, A65

\bibitem[{{Nissen} {et~al.}(2020){Nissen}, {Christensen-Dalsgaard},
  {Mosumgaard}, {Silva Aguirre}, {Spitoni}, \& {Verma}}]{Nissen20}
{Nissen}, P.~E., {Christensen-Dalsgaard}, J., {Mosumgaard}, J.~R., {et~al.}
  2020, \aap, 640, A81

\bibitem[{{Nissen} {et~al.}(2017){Nissen}, {Silva Aguirre},
  {Christensen-Dalsgaard}, {Collet}, {Grundahl}, \& {Slumstrup}}]{Nissen17}
{Nissen}, P.~E., {Silva Aguirre}, V., {Christensen-Dalsgaard}, J., {et~al.}
  2017, \aap, 608, A112

\bibitem[{{Overbeek} {et~al.}(2016){Overbeek}, {Friel}, \&
  {Jacobson}}]{Overbeek16}
{Overbeek}, J.~C., {Friel}, E.~D., \& {Jacobson}, H.~R. 2016, \apj, 824, 75

\bibitem[{{Pagel} \& {Tautvaisiene}(1997)}]{Pagel97}
{Pagel}, B.~E.~J. \& {Tautvaisiene}, G. 1997, \mnras, 288, 108

\bibitem[{{Parkinson}(1976)}]{Parkinson76}
{Parkinson}, J.~H. 1976, Philosophical Transactions of the Royal Society of
  London Series A, 281, 375

\bibitem[{{Peters}(1968)}]{Peters68}
{Peters}, J.~G. 1968, \apj, 154, 225

\bibitem[{{Pignatari} {et~al.}(2010){Pignatari}, {Gallino}, {Heil}, {Wiescher},
  {K{\"a}ppeler}, {Herwig}, \& {Bisterzo}}]{Pignatari10}
{Pignatari}, M., {Gallino}, R., {Heil}, M., {et~al.} 2010, \apj, 710, 1557

\bibitem[{{Prantzos} {et~al.}(2020){Prantzos}, {Abia}, {Cristallo}, {Limongi},
  \& {Chieffi}}]{Prantzos20}
{Prantzos}, N., {Abia}, C., {Cristallo}, S., {Limongi}, M., \& {Chieffi}, A.
  2020, \mnras, 491, 1832

\bibitem[{{Prantzos} {et~al.}(2018){Prantzos}, {Abia}, {Limongi}, {Chieffi}, \&
  {Cristallo}}]{Prantzos18}
{Prantzos}, N., {Abia}, C., {Limongi}, M., {Chieffi}, A., \& {Cristallo}, S.
  2018, \mnras, 476, 3432

\bibitem[{{Rauer} {et~al.}(2014){Rauer}, {Catala}, {Aerts}, {Appourchaux},
  {Benz}, {Brandeker}, {Christensen-Dalsgaard}, {Deleuil}, {Gizon}, {Goupil},
  {G{\"u}del}, {Janot-Pacheco}, {Mas-Hesse}, {Pagano}, {Piotto}, {Pollacco},
  {Santos}, {Smith}, {Su{\'a}rez}, {Szab{\'o}}, {Udry}, {Adibekyan}, {Alibert},
  {Almenara}, {Amaro-Seoane}, {Eiff}, {Asplund}, {Antonello}, {Barnes},
  {Baudin}, {Belkacem}, {Bergemann}, {Bihain}, {Birch}, {Bonfils}, {Boisse},
  {Bonomo}, {Borsa}, {Brand{\~a}o}, {Brocato}, {Brun}, {Burleigh}, {Burston},
  {Cabrera}, {Cassisi}, {Chaplin}, {Charpinet}, {Chiappini}, {Church},
  {Csizmadia}, {Cunha}, {Damasso}, {Davies}, {Deeg}, {D{\'{\i}}az}, {Dreizler},
  {Dreyer}, {Eggenberger}, {Ehrenreich}, {Eigm{\"u}ller}, {Erikson}, {Farmer},
  {Feltzing}, {de Oliveira Fialho}, {Figueira}, {Forveille}, {Fridlund},
  {Garc{\'{\i}}a}, {Giommi}, {Giuffrida}, {Godolt}, {Gomes da Silva},
  {Granzer}, {Grenfell}, {Grotsch-Noels}, {G{\"u}nther}, {Haswell}, {Hatzes},
  {H{\'e}brard}, {Hekker}, {Helled}, {Heng}, {Jenkins}, {Johansen},
  {Khodachenko}, {Kislyakova}, {Kley}, {Kolb}, {Krivova}, {Kupka}, {Lammer},
  {Lanza}, {Lebreton}, {Magrin}, {Marcos-Arenal}, {Marrese}, {Marques},
  {Martins}, {Mathis}, {Mathur}, {Messina}, {Miglio}, {Montalban}, {Montalto},
  {Monteiro}, {Moradi}, {Moravveji}, {Mordasini}, {Morel}, {Mortier},
  {Nascimbeni}, {Nelson}, {Nielsen}, {Noack}, {Norton}, {Ofir}, {Oshagh},
  {Ouazzani}, {P{\'a}pics}, {Parro}, {Petit}, {Plez}, {Poretti}, {Quirrenbach},
  {Ragazzoni}, {Raimondo}, {Rainer}, {Reese}, {Redmer}, {Reffert},
  {Rojas-Ayala}, {Roxburgh}, {Salmon}, {Santerne}, {Schneider}, {Schou},
  {Schuh}, {Schunker}, {Silva-Valio}, {Silvotti}, {Skillen}, {Snellen}, {Sohl},
  {Sousa}, {Sozzetti}, {Stello}, {Strassmeier}, {{\v S}vanda}, {Szab{\'o}},
  {Tkachenko}, {Valencia}, {Van Grootel}, {Vauclair}, {Ventura}, {Wagner},
  {Walton}, {Weingrill}, {Werner}, {Wheatley}, \&
  {Zwintz}}]{2014ExA....38..249R}
{Rauer}, H., {Catala}, C., {Aerts}, C., {et~al.} 2014, Experimental Astronomy,
  38, 249

\bibitem[{{Reddy} \& {Lambert}(2017)}]{Reddy17}
{Reddy}, A.~B.~S. \& {Lambert}, D.~L. 2017, \apj, 845, 151

\bibitem[{{Ricker} {et~al.}(2015){Ricker}, {Winn}, {Vanderspek}, {Latham},
  {Bakos}, {Bean}, {Berta-Thompson}, {Brown}, {Buchhave}, {Butler}, {Butler},
  {Chaplin}, {Charbonneau}, {Christensen-Dalsgaard}, {Clampin}, {Deming},
  {Doty}, {De Lee}, {Dressing}, {Dunham}, {Endl}, {Fressin}, {Ge}, {Henning},
  {Holman}, {Howard}, {Ida}, {Jenkins}, {Jernigan}, {Johnson}, {Kaltenegger},
  {Kawai}, {Kjeldsen}, {Laughlin}, {Levine}, {Lin}, {Lissauer}, {MacQueen},
  {Marcy}, {McCullough}, {Morton}, {Narita}, {Paegert}, {Palle}, {Pepe},
  {Pepper}, {Quirrenbach}, {Rinehart}, {Sasselov}, {Sato}, {Seager},
  {Sozzetti}, {Stassun}, {Sullivan}, {Szentgyorgyi}, {Torres}, {Udry}, \&
  {Villasenor}}]{Ricker15}
{Ricker}, G.~R., {Winn}, J.~N., {Vanderspek}, R., {et~al.} 2015, Journal of
  Astronomical Telescopes, Instruments, and Systems, 1, 014003

\bibitem[{{Rocha-Pinto} {et~al.}(2004){Rocha-Pinto}, {Flynn}, {Scalo},
  {H{\"a}nninen}, {Maciel}, \& {Hensler}}]{Rocha-Pinto04}
{Rocha-Pinto}, H.~J., {Flynn}, C., {Scalo}, J., {et~al.} 2004, \aap, 423, 517

\bibitem[{{Sakari} {et~al.}(2018){Sakari}, {Placco}, {Farrell}, {Roederer},
  {Wallerstein}, {Beers}, {Ezzeddine}, {Frebel}, {Hansen}, {Holmbeck},
  {Sneden}, {Cowan}, {Venn}, {Davis}, {Matijevi{\v{c}}}, {Wyse},
  {Bland-Hawthorn}, {Chiappini}, {Freeman}, {Gibson}, {Grebel}, {Helmi},
  {Kordopatis}, {Kunder}, {Navarro}, {Reid}, {Seabroke}, {Steinmetz}, \&
  {Watson}}]{Sakari18}
{Sakari}, C.~M., {Placco}, V.~M., {Farrell}, E.~M., {et~al.} 2018, \apj, 868,
  110

\bibitem[{{Sch{\"o}nrich} \& {Weinberg}(2019)}]{Schonrich19}
{Sch{\"o}nrich}, R.~A. \& {Weinberg}, D.~H. 2019, \mnras, 487, 580

\bibitem[{{Shen} {et~al.}(2015){Shen}, {Cooke}, {Ramirez-Ruiz}, {Madau},
  {Mayer}, \& {Guedes}}]{Shen15}
{Shen}, S., {Cooke}, R.~J., {Ramirez-Ruiz}, E., {et~al.} 2015, \apj, 807, 115

\bibitem[{{Siegel} {et~al.}(2019){Siegel}, {Barnes}, \& {Metzger}}]{Siegel19}
{Siegel}, D.~M., {Barnes}, J., \& {Metzger}, B.~D. 2019, \nat, 569, 241

\bibitem[{{Skrutskie} {et~al.}(2006){Skrutskie}, {Cutri}, {Stiening},
  {Weinberg}, {Schneider}, {Carpenter}, {Beichman}, {Capps}, {Chester},
  {Elias}, {Huchra}, {Liebert}, {Lonsdale}, {Monet}, {Price}, {Seitzer},
  {Jarrett}, {Kirkpatrick}, {Gizis}, {Howard}, {Evans}, {Fowler}, {Fullmer},
  {Hurt}, {Light}, {Kopan}, {Marsh}, {McCallon}, {Tam}, {Van Dyk}, \&
  {Wheelock}}]{Skrutskie06}
{Skrutskie}, M.~F., {Cutri}, R.~M., {Stiening}, R., {et~al.} 2006, \aj, 131,
  1163

\bibitem[{{Slumstrup} {et~al.}(2017){Slumstrup}, {Grundahl}, {Brogaard},
  {Thygesen}, {Nissen}, {Jessen-Hansen}, {Van Eylen}, \&
  {Pedersen}}]{Slumstrup17}
{Slumstrup}, D., {Grundahl}, F., {Brogaard}, K., {et~al.} 2017, \aap, 604, L8

\bibitem[{{Sneden} {et~al.}(2008){Sneden}, {Cowan}, \& {Gallino}}]{Sneden08}
{Sneden}, C., {Cowan}, J.~J., \& {Gallino}, R. 2008, \araa, 46, 241

\bibitem[{{Sneden} {et~al.}(2009){Sneden}, {Lawler}, {Cowan}, {Ivans}, \& {Den
  Hartog}}]{Sneden2009}
{Sneden}, C., {Lawler}, J.~E., {Cowan}, J.~J., {Ivans}, I.~I., \& {Den Hartog},
  E.~A. 2009, \apjs, 182, 80

\bibitem[{{Sneden}(1973)}]{1973PhDT.......180S}
{Sneden}, C.~A. 1973, PhD thesis, The University of Texas at Austin.

\bibitem[{{Spina} {et~al.}(2018){Spina}, {Mel{\'e}ndez}, {Karakas}, {dos
  Santos}, {Bedell}, {Asplund}, {Ram{\'{\i}}rez}, {Yong}, {Alves-Brito},
  {Bean}, \& {Dreizler}}]{Spina18}
{Spina}, L., {Mel{\'e}ndez}, J., {Karakas}, A.~I., {et~al.} 2018, \mnras, 474,
  2580

\bibitem[{{Spina} {et~al.}(2016){Spina}, {Mel{\'e}ndez}, {Karakas},
  {Ram{\'{\i}}rez}, {Monroe}, {Asplund}, \& {Yong}}]{Spina16}
{Spina}, L., {Mel{\'e}ndez}, J., {Karakas}, A.~I., {et~al.} 2016, \aap, 593,
  A125

\bibitem[{{Spina} {et~al.}(2020){Spina}, {Nordlander}, {Casey}, {Bedell},
  {D'Orazi}, {Mel{\'e}ndez}, {Karakas}, {Desidera}, {Baratella}, {Yana
  Galarza}, \& {Casali}}]{Spina20}
{Spina}, L., {Nordlander}, T., {Casey}, A.~R., {et~al.} 2020, \apj, 895, 52

\bibitem[{{Sullivan} {et~al.}(2015){Sullivan}, {Winn}, {Berta-Thompson},
  {Charbonneau}, {Deming}, {Dressing}, {Latham}, {Levine}, {McCullough},
  {Morton}, {Ricker}, {Vanderspek}, \& {Woods}}]{Sullivan15}
{Sullivan}, P.~W., {Winn}, J.~N., {Berta-Thompson}, Z.~K., {et~al.} 2015, \apj,
  809, 77

\bibitem[{{Surman} {et~al.}(2008){Surman}, {McLaughlin}, {Ruffert}, {Janka}, \&
  {Hix}}]{Surman08}
{Surman}, R., {McLaughlin}, G.~C., {Ruffert}, M., {Janka}, H.-T., \& {Hix},
  W.~R. 2008, \apjl, 679, L117

\bibitem[{Tautvai{\v{s}}ien{\.{e}} {et~al.}(2020)Tautvai{\v{s}}ien{\.{e}},
  Mikolaitis, Drazdauskas, Stonkut{\.{e}}, Minkevi{\v{c}}i{\={u}}t{\.{e}},
  Kjeldsen, Brogaard, von Essen, Grundahl, Pak{\v{s}}tien{\.{e}}, Bagdonas, \&
  V{\'{a}}zquez}]{Tautvaisiene20}
Tautvai{\v{s}}ien{\.{e}}, G., Mikolaitis, {\v{S}}., Drazdauskas, A., {et~al.}
  2020, The Astrophysical Journal Supplement Series, 248, 19

\bibitem[{{Taylor}(2005)}]{2005ASPC..347...29T}
{Taylor}, M.~B. 2005, in Astronomical Society of the Pacific Conference Series,
  Vol. 347, Astronomical Data Analysis Software and Systems XIV, ed.
  P.~{Shopbell}, M.~{Britton}, \& R.~{Ebert}, 29

\bibitem[{{Taylor}(2006)}]{2006ASPC..351..666T}
{Taylor}, M.~B. 2006, in Astronomical Society of the Pacific Conference Series,
  Vol. 351, Astronomical Data Analysis Software and Systems XV, ed.
  C.~{Gabriel}, C.~{Arviset}, D.~{Ponz}, \& S.~{Enrique}, 666

\bibitem[{{Terlouw} \& {Vogelaar}(2014)}]{KapteynPackage}
{Terlouw}, J.~P. \& {Vogelaar}, M.~G.~R. 2014, {Kapteyn Package, version
  2.3b3}, {Kapteyn Astronomical Institute}, Groningen, available from
  \url{http://www.astro.rug.nl/software/kapteyn/}

\bibitem[{{Thielemann} {et~al.}(2017){Thielemann}, {Eichler}, {Panov}, \&
  {Wehmeyer}}]{Thielemann17}
{Thielemann}, F.~K., {Eichler}, M., {Panov}, I.~V., \& {Wehmeyer}, B. 2017,
  Annual Review of Nuclear and Particle Science, 67, 253

\bibitem[{{Titarenko} {et~al.}(2019){Titarenko}, {Recio-Blanco}, {de Laverny},
  {Hayden}, \& {Guiglion}}]{Titarenko19}
{Titarenko}, A., {Recio-Blanco}, A., {de Laverny}, P., {Hayden}, M., \&
  {Guiglion}, G. 2019, \aap, 622, A59

\bibitem[{{Travaglio} {et~al.}(2004){Travaglio}, {Gallino}, {Arnone}, {Cowan},
  {Jordan}, \& {Sneden}}]{Travaglio04}
{Travaglio}, C., {Gallino}, R., {Arnone}, E., {et~al.} 2004, \apj, 601, 864

\bibitem[{{Travaglio} {et~al.}(2001){Travaglio}, {Gallino}, {Busso}, \&
  {Gratton}}]{Travaglio01}
{Travaglio}, C., {Gallino}, R., {Busso}, M., \& {Gratton}, R. 2001, \apj, 549,
  346

\bibitem[{{Trevisan} \& {Barbuy}(2014)}]{Trevisan14}
{Trevisan}, M. \& {Barbuy}, B. 2014, \aap, 570, A22

\bibitem[{{Trippella} {et~al.}(2016){Trippella}, {Busso}, {Palmerini},
  {Maiorca}, \& {Nucci}}]{Trippella16}
{Trippella}, O., {Busso}, M., {Palmerini}, S., {Maiorca}, E., \& {Nucci}, M.~C.
  2016, \apj, 818, 125

\bibitem[{{Tucci Maia} {et~al.}(2016){Tucci Maia}, {Ram{\'{\i}}rez},
  {Mel{\'e}ndez}, {Bedell}, {Bean}, \& {Asplund}}]{TucciMaia16}
{Tucci Maia}, M., {Ram{\'{\i}}rez}, I., {Mel{\'e}ndez}, J., {et~al.} 2016,
  \aap, 590, A32

\bibitem[{{Watson} {et~al.}(2019){Watson}, {Hansen}, {Selsing}, {Koch},
  {Malesani}, {Andersen}, {Fynbo}, {Arcones}, {Bauswein}, {Covino}, {Grado},
  {Heintz}, {Hunt}, {Kouveliotou}, {Leloudas}, {Levan}, {Mazzali}, \&
  {Pian}}]{Watson19}
{Watson}, D., {Hansen}, C.~J., {Selsing}, J., {et~al.} 2019, \nat, 574, 497

\bibitem[{{Woosley} {et~al.}(1994){Woosley}, {Wilson}, {Mathews}, {Hoffman}, \&
  {Meyer}}]{Woosley94}
{Woosley}, S.~E., {Wilson}, J.~R., {Mathews}, G.~J., {Hoffman}, R.~D., \&
  {Meyer}, B.~S. 1994, \apj, 433, 229

\bibitem[{{Yan} {et~al.}(2019){Yan}, {Du}, {Liu}, {Li}, {Shi}, {Chen}, {Ma}, \&
  {Wu}}]{Yan19}
{Yan}, Y., {Du}, C., {Liu}, S., {et~al.} 2019, \apj, 880, 36

\bibitem[{{Yong} {et~al.}(2012){Yong}, {Carney}, \& {Friel}}]{Yong12}
{Yong}, D., {Carney}, B.~W., \& {Friel}, E.~D. 2012, \aj, 144, 95

\bibitem[{{Zhao} {et~al.}(2016){Zhao}, {Mashonkina}, {Yan}, {Alexeeva},
  {Kobayashi}, {Pakhomov}, {Shi}, {Sitnova}, {Tan}, {Zhang}, {Zhang}, {Zhou},
  {Bolte}, {Chen}, {Li}, {Liu}, \& {Zhai}}]{Zhao16}
{Zhao}, G., {Mashonkina}, L., {Yan}, H.~L., {et~al.} 2016, \apj, 833, 225

\end{thebibliography}

%\newpage
%\begin{appendix}

%\section{}

%\end{appendix}

\end{document}